\documentclass[aps,prb,amsmath,amssymb,superscriptaddress,showpacs,floatfix,longbibliography]{revtex4-1}

\makeatother
\usepackage[utf8]{inputenc}
\usepackage[dvipsnames]{xcolor}
\usepackage{amsmath}
\usepackage{amssymb}
\usepackage{graphicx}
\usepackage{bm}
\usepackage{float}
\usepackage{physics}
\usepackage{lipsum}
\usepackage{afterpage}
\usepackage{threeparttable}
\usepackage{xfrac}
\usepackage{color}
\usepackage[colorlinks=true,bookmarks=false,citecolor=blue,linkcolor=red,hyperfootnotes=true,urlcolor=blue]{hyperref}

\newcommand{\qe}{{\sc Quantum ESPRESSO}}
\newcommand{\ocean}{{\sc ocean}}
\newcommand{\exciting}{{\texttt {exciting}}}
\newcommand{\xs}{{\sc xspectra}}
\newcommand{\bs}{\mathbf}

\newcommand{\beginappendix}{%
        \appendix
        \setcounter{table}{0}
        \renewcommand{\thetable}{B\arabic{table}}%
        \setcounter{figure}{0}
        \renewcommand{\thefigure}{B\arabic{figure}}%
     }

\begin{document}

\title{Multi-code Benchmark on Simulated Ti K-edge X-ray Absorption Spectra of Ti-O Compounds}

\author{Fanchen Meng}
\affiliation{ 
Center for Functional Nanomaterials,
Brookhaven National Laboratory, Upton, New York 11973,
United States
}%

\author{Benedikt Maurer}
\affiliation{ 
Institut f\"ur Physik and IRIS Adlershof, Humboldt-Universit\"at zu Berlin, Berlin Germany
}%

\author{Fabian Peschel}
\affiliation{ 
Institut f\"ur Physik and IRIS Adlershof, Humboldt-Universit\"at zu Berlin, Berlin Germany
}%

\author{Sencer Selcuk}
\affiliation{ 
Center for Functional Nanomaterials,
Brookhaven National Laboratory, Upton, New York 11973,
United States
}%

\author{Mark Hybertsen}
\affiliation{ 
Center for Functional Nanomaterials,
Brookhaven National Laboratory, Upton, New York 11973,
United States
}%

\author{Xiaohui Qu}
\affiliation{ 
Center for Functional Nanomaterials,
Brookhaven National Laboratory, Upton, New York 11973,
United States
}%

\author{Christian Vorwerk}
\email{vorwerk@uchicago.edu}
\altaffiliation[Also at ]{Institut f\"ur Physik and IRIS Adlershof, Humboldt-Universit\"at zu Berlin, Berlin Germany}%Lines break automatically or can be forced with \\
\affiliation{ 
Pritzker School of Molecular Engineering, University of Chicago, Chicago, Illinois 60637, United States
}%

\author{Claudia Draxl}
\email{claudia.draxl@physik.hu-berlin.de}
\altaffiliation[Also at ]{European Theoretical Spectroscopy Facility (ETSF)}%Lines break automatically or can be forced with \\
\affiliation{ 
Institut f\"ur Physik and IRIS Adlershof, Humboldt-Universit\"at zu Berlin, Berlin Germany
}%

\author{John Vinson}
\email{john.vinson@nist.gov}
\affiliation{ Material Measurement Laboratory, National Institute of Standards and Technology, Gaithersburg, Maryland 20899, United States
}%

\author{Deyu Lu}
\email{dlu@bnl.gov}
\affiliation{ 
Center for Functional Nanomaterials,
Brookhaven National Laboratory, Upton, New York 11973,
United States
}
\date{\today}

\begin{abstract}
X-ray absorption spectroscopy (XAS) is an element-specific materials characterization technique that is sensitive to structural and electronic properties. First-principles simulated XAS has been widely used as a powerful tool to interpret experimental spectra and draw physical insights. Recently, there has also been growing interest in building computational XAS databases to enable data analytics and machine learning applications. However, there are non-trivial differences among commonly used XAS simulation codes, both in underlying theoretical  formalism and in technical implementation. Reliable and reproducible computational XAS databases require systematic benchmark studies. In this work, we benchmarked Ti K-edge XAS simulations of ten representative Ti-O binary compounds, which we refer to as the Ti-O-10 dataset, using three state-of-the-art codes: \xs{}, \ocean{} and \exciting{}. We systematically studied the convergence behavior with respect to the input parameters and developed a workflow to automate and standardize the calculations to ensure converged spectra. Our benchmark comparison considers a 35 eV spectral range starting from the K-edge onset, representative of widely used near-edge spectra. Quantitative comparison over this range is based on Spearman's rank correlation score ($r_{sp}$). Our results show: (1) the two Bethe-Salpeter equation (BSE) codes (\ocean{} and \exciting{}) have excellent agreement with an average $r_{sp}$ of 0.998; (2) good agreement is obtained between the core-hole potential code (\xs{}) and BSE codes (\ocean{} and \exciting{}) with an average $r_{sp}$ of 0.990, and this smaller $r_{sp}$ reflects the noticeable differences in the main edge spectral shape that can be primarily attributed to the difference in the strength of the screened core-hole potential; (3) simulations from both methods overall reproduce well the main experimental spectral features of rutile and anatase, and the different treatments of the screened core-hole potential have visible impact on pre-edge intensities and the peak ratio of the main edge; (4) there exist moderate
differences in the relative edge alignment of the three codes with a standard deviation of about 0.2 eV, which arise from multiple contributions including the frozen core approximation, final state effects, and different approximations used for the self-energy correction. Our benchmark study provides important standards for first-principles XAS simulations with broad impact in data-driven XAS analysis.
\end{abstract}

\maketitle

\section{Introduction}
X-ray absorption spectroscopy (XAS) probes the excitations that promote deeply bound core electrons into the unoccupied states. Due to the energy separation between core levels of different elements and the small spatial extent of core orbitals, XAS is element-specific and sensitive to the local chemical environment around the absorbing atoms. As a first approximation the spectra reflect details of the unoccupied density of states localized on the site of the absorbing atom, although spectra can be heavily modified by the presence of the positively charged core hole. Particularly, the low-lying excitation region extending about 30 eV above the onset of absorption from a specific core level, known as x-ray absorption near-edge structure (XANES)~\cite{rehr2000theoretical}, has been routinely used to extract the local chemical characteristics of the absorption site, such as local symmetry, type of hybridization, charge state, spin state, and bond distortions~\cite{yamamoto2008assignment}. 
Physically, local structural and chemical characteristics that modify the local unoccupied orbitals can result in measurable changes to the XANES. 

The rich information contained in XANES leads to its extensive use for materials characterization in condensed matter physics,\cite{saitoh1995electronic,prendergast2006x} materials science,\cite{zhang2017multi} chemistry,\cite{li2015complex} and biology.\cite{yano2009x}
%, excitations from a deeply bound core level to low-lying conduction states within about 20~eV of the conduction band minimum, have long been used as a probe of the local structure of materials. 
%XAS can be used to observe changes in coordination, charge, or chemical bonds. 
Despite these broad applications, interpreting XANES spectra is non-trivial because the observed spectral features represent a complex convolution of the atomic and electronic structures. Standard XANES analysis relies on fitting the measured spectra with empirical fingerprints collected from experimental standards with known chemical formulae and atomic arrangements.\cite{yamamoto2008assignment} However, the empirical fingerprint approach is limited by the chemical and configuration space spanned by available experimental standards. Typically the available data for a target element are restricted to simple crystals and small molecules.  Therefore, it remains challenging to analyze XANES spectra of complex materials (e.g., surfaces, interfaces, nanostructures and amorphous materials) and structural evolution during physical and chemical processes (e.g., phase transitions and chemical reactions).

%In the past few decades, there has been remarkable progress in first-principles XAS spectroscopy. 
First-principles XANES simulations have made remarkable progress in the past few decades. Due to their predictive power, first-principles calculations provide a concrete connection between XANES spectra and the underlying atomic and electronic structures. This makes them a powerful tool for spectral interpretation.\cite{rehr2000theoretical,de2008core} In many cases, simulated XANES spectra yield excellent agreement with measurements on key spectral features, e.g., number of peaks, peak positions and peak height ratios, thus supporting quantitative assignment of the spectra.\cite{Haverkort2012, PhysRevLett.118.096402, tang2022many}  Recently, there is a growing interest in data-driven XANES analysis leveraging machine learning (ML).\cite{doi:10.1063/5.0049111,timoshenko2019inverting} First-principles simulations of XANES play a significant role in this new paradigm. In practice, the first step in a data-driven approach is to construct a database containing atomic structures and the corresponding simulated XANES spectra for either curated configuration spaces (e.g., metal clusters~\cite{doi:10.1021/acs.jpclett.7b02364}, bi-metallic clusters~\cite{marcella2020neural, liu2021probing}, metal oxide clusters~\cite{liu2019mapping}, small molecules with structural distortions~\cite{guda2021understanding}, transition metal compounds~\cite{li2019deep}, amorphous materials~\cite{aarva2019understandingI,aarva2019understandingII}, catalysts~\cite{routh2021latent,xiang2022solving}, or interfaces~\cite{trejo2019elucidating}) or chemical spaces derived from public structure databases (e.g., a wide range of small molecules at their equilibrium structures~\cite{carbone2020machine,tetef2021unsupervised} or transition metal oxides~\cite{carbone2019classification,Torrisi2020,zheng2020random,rankine2020deep, yan2019ultrathin}). Multiple data-analytics methods can be applied subsequently, including computational spectral fingerprints~\cite{yan2019ultrathin,guda2021understanding,aarva2019understandingI,aarva2019understandingII,kuban2022I,kuban2022II}, ML surrogates to predict spectrum from structure (i.e., the forward problem)~\cite{carbone2020machine,rankine2020deep,rankine2022accurate,Ghose2023} and ML classifiers to extract physical descriptors from spectra (i.e., the inverse problem)~\cite{doi:10.1021/acs.jpclett.7b02364,li2019deep,marcella2020neural, liu2021probing,liu2019mapping,guda2021understanding,routh2021latent,xiang2022solving,trejo2019elucidating,tetef2021unsupervised,carbone2019classification,Torrisi2020,zheng2020random}. Bridging from models trained on simulated datasets to experiment can be challenging. Although promising results were reported in special cases when ML classifiers trained on synthetic data were directly applied to experimental data~\cite{doi:10.1021/acs.jpclett.7b02364,marcella2020neural, liu2021probing,marcella2020neural, liu2021probing,guda2021understanding}, in general the systematic error between theory and experiment needs to be carefully addressed.

In comparison to the traditional first-principles modeling approach, the data-driven approach has the advantage that the generated data are findable, accessible, interoperable, reusable (known as the FAIR data principles) and expandable. This allows researchers to identify important trends from the full energy range of XANES of diverse materials spaces in an unbiased way. As a result, the data-driven approach is playing a more and more important role in XANES analysis, especially for high-throughput structure identification and real-time spectral interpretation pertinent to autonomous experimentation. A critical first step in this paradigm is building high-quality XANES databases that expand a wide energy range of 30 to 40 eV, in contrast to many existing studies that focus on features in a narrow energy range. Despite several existing simulated XANES databases~\cite{MP-Kedges,MP-Ledges,Shibata2022}, many more need to be constructed to cover the vast chemical space and materials diversity in practical applications. However, there are several caveats in the simulated XANES database construction.

First, one needs to choose from many available XANES simulation codes. Density functional theory (DFT)-based or many-body perturbation theory-based XANES simulation methods are a favorable choice due to the interpretability associated with band structure theory. It is also feasible to use these methods to build a database of XANES spectra. Because of different theoretical methods (e.g., treatment of the core-hole final-state effects) and numerical implementations (e.g., basis sets) employed, it is important to systematically understand the performance of different codes on representative benchmark systems both among themselves and against experiment. In addition, the continued improvement of x-ray measurements (e.g., increased resolution and decreased noise and sample damage) highlights the need for quantitative theoretical predictions. Systematic and quantitative comparisons between codes are necessary to elucidate the nature of discrepancies, which can originate from different approximations in the theory and different numerical implementations~\cite{shirley2021core,draxl2021}. Such systematic comparison has, to our knowledge, not previously been attempted, though comparisons with limited scope have been carried out for the O K edge~\cite{rehr2005final,PhysRevLett.118.096402}, C K edge~\cite{PhysRevB.95.115112}, Mg K edge~\cite{rehr2005final} and Ti K edge~\cite{zhang2017multi,yan2019ultrathin}. In part, the lack of such benchmark studies is due to the complexity of accounting for different approximations and settings in multiple codes or methods, which can be non-trivial even for domain experts, making it difficult to establish reliable comparisons. The computational cost associated with multiple codes and multiple choices of parameters is another limiting factor.
%the difficulty in establishing quantitative measures of agreement between x-ray spectra. , and .
%It is non-trivial to obtain fully converged XAS results with DFT-based methods, due to the lack of the systematic understanding of the convergence behavior of key input parameters. This issue has become a roadblock hampering the reliability and reproducibility of computational 
%\cite{doi:10.1021/acs.jpclett.7b02364,doi:10.1063/5.0049111,Torrisi2020} 

Second, the quality of the simulated XANES spectra strongly depends on the numerical convergence of multiple key parameters. While some parameters are generic to the excited state calculations, many are code specific, such as the choice of pseudopotentials. We emphasize that spectral data quality is essential to data-driven XANES analysis, because it directly affects the performance of the downstream data analytics applications. In other words, robust control for parameter choice is essential to the utility of spectral databases, in particular to avoid misleading or unphysical trends in subsequent analysis. Establishing the know-how to achieve the convergence in multiple codes requires a collective effort from multiple research groups.  Significant benchmark studies have been performed for ground state DFT codes~\cite{lejaeghere2016reproducibility} and \emph{GW} codes~\cite{van2015gw,caruso2016benchmark}. However, similar cross-code benchmark studies are still needed in the computational x-ray spectroscopy field.

Last, in order to generate large spectral databases from high-throughput calculations, it is mandatory to have an automated workflow that requires little human oversight. This workflow needs to provide system-independent default parameters and, more importantly, educated guesses of system-dependent parameters based on well-established trends from benchmark studies. Such a workflow not only provides a common standard to obtain reliable XANES simulation data, but also lowers the barrier of XANES simulation for non-experts. In addition, a well tested workflow avoids repetitive work in determining input parameters on similar systems and thus saves research time and computational resources. Due to the complexities associated with the first two issues discussed above, a multi-code XANES workflow does not exist.

A meaningful benchmark against experimental data requires a set of highly reliable measurements with quantified estimations of errors and uncertainties as well as a quantification of the often uncontrolled approximations used in the theoretical approaches. Comparison to measured data from the literature is confounded by differences between published measurements arising from such factors as calibration, instrumental resolution, sample validation and preparation, self-absorption corrections, and other issues. Theoretical approaches adopt a number of widely used approximations, the effects of which can be large compared to the differences between codes as well as significant for comparison to measured spectral features. Case studies illustrate the impact of many common approximations, including neglect of the role of vibrations~\cite{PhysRevB.90.205207,10.1063/1.4856835,PhysRevB.81.115125,PhysRevB.81.115125,PhysRevB.98.014111,PhysRevB.98.214305} and different technical approximations in first-principles computational approaches~\cite{10.1063/5.0030493,PhysRevB.94.035163,doi:10.1073/pnas.2201258119, PhysRevB.101.245119,DEGROOT2021147061,deGroot_2009}.  In view of these well-known issues, we have phased our study.  In this paper, we focus on first establishing the level of agreement and uniformity between standard XANES simulation methods with only a brief comparison with experiment. A benchmark against a broader set of measured data will be the focus of a future study.

Specifically, in this study, we compare simulated XANES spectra among multiple codes that implement state-of-the-art first-principles theoretical approaches. We carry out a quantitative comparison among three popular codes, \ocean{}~\cite{vinson2011bethe,ocean-3}, \exciting{}~\cite{gulans2014exciting,vorwerk2017addressing,vorwerk2019implementation}, and \xs{}~\cite{taillefumier2002x,gougoussis2009first}, for calculating XANES spectra. These codes follow two differing theoretical approaches and rely on different implementations and approximations, as will be discussed in the next section. We benchmark Ti K-edge XANES for ten representative Ti$_x$O$_y$ compounds, which we refer to as the Ti-O-10 dataset. We have developed scripts to automatically generate consistent input files from the crystal structure, establish defaults for general input parameters, and carry out convergence tests on several key parameters. We find overall good agreement in the calculated spectra among the three codes over the spectral range typically considered in XANES, despite differences in theoretical approximations and technical implementations. We identify the main reasons for the differences that are observed, and we discuss the comparison to experiment for the widely studied examples of TiO$_2$ in the rutile and anatase structure.
%-----------------------------------------------------------------------------
\section{Theoretical framework}
%-----------------------------------------------------------------------------
%-----------------------------------------------------------------------------
\subsection{X-ray Absorption Spectroscopy} \label{sec:xas}
%-----------------------------------------------------------------------------
The x-ray absorption cross section can be calculated from Fermi's golden rule according to~\cite{de2008core}
\begin{equation} \label{eq:goldenrule}
  \sigma(\omega)=4\pi^2 \alpha\, \omega\, \sum_f \left | M_{0,f} \right |^2  \delta(E_f-E_0-\omega),
\end{equation}
where $E_0$ and $E_f$ are the total energies of the many-body initial state $|\Psi_0 \rangle$ and final state $|\Psi_f\rangle$. $\alpha$ is the fine structure constant, and $\omega$ is the x-ray energy. Unless otherwise specified, we use atomic units throughout the rest of the paper. $M_{0,f}=\langle \Psi_f |\hat{O} | \Psi_0 \rangle $ is the transition matrix element with $\hat{O}$ the transition operator. Under the electric field of the photon beam, the dipole and quadrupole terms are given by $\hat{O}=\bs{e}\cdot \bs{r}+i/2 (\bs{e}\cdot \bs{r}) (\bs{q}\cdot \bs{r})$, where $\bs{e}$ and $\bs{q}$ are the polarization vector and
the wave vector of the photon beam, and $\bs{r}$ is the position of the electron. Within the scope of this work, we will only consider the dominating contribution from the dipole term except where noted below.  Evaluating XANES spectra at the ground state or the final state Hamiltonian are referred to as the initial or finial state rule, respectively. Under the initial state rule, XANES spectra correspond to excitation of independent particles, while under the final state rule, the many-body effects of interacting particles (such as core hole screening) are included.

Practical calculations build on specific approximations. In practice, approximations need to be made to evaluate Eq.~\ref{eq:goldenrule}. Here we start with the single particle picture based on the Kohn-Sham density-functional theory (DFT) and discuss different approximations to treat electron correlation effects. Under the final state rule, we consider two different treatments of the core-hole final state effects. In the core-hole pseudopotential (CHP) method~\cite{taillefumier2002x,gougoussis2009first,prendergast2006x}, we use a DFT-based approach and consider the static response to a core hole in the final state. In the linear response method~\cite{vinson2011bethe,ocean-3,puschnig2002optical,vorwerk2017addressing}, we use the many-body perturbation framework to describe the core electron excitation.

In the CHP method, the core hole on the absorber atom is treated explicitly by a core-hole pseudopotential, and the final state is solved self-consistently by allowing valence electrons to relax. The dipole contribution to Eq.~\ref{eq:goldenrule} can be approximated by
\begin{eqnarray} \label{eq:finalstate}
  \sigma_{CHP}(\omega)&=& 4\pi^2 \alpha\, \omega\, \sum_f \left | \bs{e} \cdot \langle \tilde{\psi}_f |\bs{r}| \psi_\alpha \rangle \right |^2  \delta(\tilde{\epsilon}_f-\epsilon_\alpha-\omega) \nonumber \\
     &=& -4\pi \alpha\, \omega\, \mathrm{Im}  \left [\bs{e}\cdot \langle \psi_\alpha | \bs{r} \, ( \omega - \tilde{H}_{DFT} + \mathrm{i}\eta ) ^{-1} \, \bs{r}| \psi_\alpha \rangle \cdot \bs{e} \right ],
\end{eqnarray}
where $|\psi_\alpha\rangle$ is the core-hole state before the excitation and $|\tilde{\psi}_f \rangle$ are empty states at the presence of the core hole, with $\epsilon_\alpha$ and $\tilde\epsilon_f$ the corresponding Kohn-Sham energy levels.

Alternatively, within many-body perturbation theory, neutral electronic excitations are described under the \emph{GW}-\textit{Bethe-Salpeter Equation} (BSE) framework, as a post-DFT treatment. Under the \emph{GW} approximation, the DFT exchange-correlation potential is replaced by the energy-dependent self-energy operator, which contains a screened exchange term and a Coulomb hole term. Except for a few cases~\cite{tang2022many,PhysRevB.86.195135,PhysRevB.94.035163}, the self-energy correction to empty states is often neglected in practical XANES calculations and existing studies primarily focus on correcting the core-hole energy level~\cite{golze2020accurate,doi:10.1021/acs.jctc.1c01180}. 
In this work, self-energy corrections to the DFT valence or conduction orbitals are not considered, and the correlation effects between the excited electron and core hole are described by the BSE. Specifically, $| \Psi_f \rangle $ in Eq.~\ref{eq:goldenrule} are approximated by the correlated electron-hole excitations $|S \rangle $ with excitation energies $\Omega_S$, which are the eigenvectors and eigenvalues of the BSE Hamiltonian~\cite{vinson2011bethe,puschnig2002optical,vorwerk2017addressing}, $\hat{H}^{BSE}$. This yields
\begin{equation}
     \sigma_{BSE}(\omega)=4\pi^2 \alpha\, \omega\, \sum_S \left | \bs{e}\cdot \langle S |\bs{r}| 0 \rangle \right |^2  \delta(\Omega_S-\omega),
\end{equation}
where $| 0 \rangle$ denotes the DFT ground state. 
For XANES, only transitions from the localized core-level orbitals are considered, and the transition matrix elements can be evaluated in real-space despite the periodic boundary conditions.
By introducing the single-particle velocity operator $\bs{v}=i\,[\hat{H}_{BSE},\bs{r}]$ and making use of the identity $\langle S |\bs{r}| 0 \rangle=-i\,\langle S |\bs{v}| 0 \rangle/\Omega_S$, one can prove that
\begin{equation}\label{eq:cs_bse}
     \sigma_{BSE}(\omega)=4\pi^2 \alpha / \omega\, \sum_S \left | \bs{e}\cdot \langle S |\bs{v}| 0 \rangle \right |^2  \delta(\Omega_S-\omega),
\end{equation}
which is proportional to the imaginary part of the macroscopic dielectric constant~\cite{rohlfing2000electron},
\begin{eqnarray}\label{eq:x-ray_bse}
     %\epsilon_2=16\pi / \omega^2\, \sum_S \left | \bs{\lambda}\cdot \langle S |\bs{v}| 0 \rangle \right |^2  \delta(\Omega_S-\omega).
     \epsilon_2 &\propto& 4\pi \sum_S \left | \bs{e}\cdot \langle S |\bs{v}| 0 \rangle \right |^2  \delta(\Omega_S-\omega) \nonumber \\
       &\propto& -4\pi \; \mathrm{Im} \sum_{\Phi, \Phi'} \left[ \bs{e}\cdot  \langle 0 |  \bs{v}^\dagger | \Phi \rangle \langle \Phi |  (\omega - \hat{H}_{BSE} + \mathrm{i} \eta )^{-1} | \Phi' \rangle \langle \Phi' | \bs{v} | 0 \rangle \cdot \bs{e} \right].
\end{eqnarray}
Here we have expanded $|S \rangle $ into the complete basis of single excitations $| \Phi_{c \alpha \mathbf{k}} \rangle = \hat{c}_{c \mathbf{k}}^\dagger \hat{c}_{\alpha \mathbf{k}} | 0 \rangle$ of a core hole $(\alpha)$ and excited electron $(c)$ with wavevector $\mathbf{k}$. Matrix elements of $\hat{H}_{BSE}$ are given by~\cite{vorwerk2017addressing}
\begin{equation}\label{eq:bse}
    H^{BSE}_{c \alpha \mathbf{k}, c' \alpha' \mathbf{k}'} = \Delta E_{c \alpha \mathbf{k}, c' \alpha' \mathbf{k}'} + 2V_{c \alpha \mathbf{k}, c' \alpha' \mathbf{k}'} + W_{c \alpha \mathbf{k}, c' \alpha' \mathbf{k}'},
\end{equation}
where $\Delta E_{c \alpha \mathbf{k}, c' \alpha' \mathbf{k}'}= \left( \epsilon_{c\mathbf{k}} - \epsilon_{\alpha}\right)\delta_{cc'}\delta_{\alpha \alpha'}\delta_{\mathbf{k}\mathbf{k}'}$ is the energy difference between the excited electron and the core hole. Because the core-hole state is localized in real space, the core-hole energy is dispersionless. Therefore we drop the momentum dependence in $\epsilon_{\alpha}$. The matrix elements of the bare electron-hole exchange and the screened direct interaction are given by
\begin{eqnarray}
    V_{c \alpha \mathbf{k}, c' \alpha' \mathbf{k}'} = \int d^3r d^3r' \psi^*_{\alpha\mathbf{k}}(\mathbf{r}) \psi_{c \mathbf{k}}(\mathbf{r}) v(\mathbf{r},\mathbf{r}') \psi^*_{c' \mathbf{k}'}(\mathbf{r}') \psi_{\alpha'\mathbf{k}'}(\mathbf{r}'), \label{eq:exc} \\
    W_{c \alpha \mathbf{k}, c' \alpha' \mathbf{k}'} = -\int d^3 r d^3r' \psi^*_{\alpha\mathbf{k}}(\mathbf{r})\psi_{\alpha'\mathbf{k}'}(\mathbf{r})w(\mathbf{r},\mathbf{r}') \psi_{c \mathbf{k}}(\mathbf{r}')\psi^*_{c' \mathbf{k}'}(\mathbf{r}'), \label{eq:direct}
\end{eqnarray}
where $v$ is the bare Coulomb kernel
%\begin{equation}\label{eq:direct}
%\end{equation}
and $w = \epsilon^{-1}v$ is the screened Coulomb kernel under the matrix notation, i.e., $w(\mathbf{r}, \mathbf{r'}) = \int d^3 \mathbf{r''}  \epsilon^{-1} (\mathbf{r}, \mathbf{r''}) v(\mathbf{r''}, \mathbf{r'})$.
We follow the standard implementation of the BSE method and take the static approximation in the direct term, dropping the frequency dependence in the dielectric tensor. 

Practically, it is often more convenient to evaluate the matrix elements in Eqs.~\ref{eq:exc} and~\ref{eq:direct} in reciprocal space, which leads to
\begin{align}
    V_{c \alpha \mathbf{k}, c' \alpha' \mathbf{k}'} &= \frac{1}{V} \sum_{\mathbf{G}} M^*_{\alpha c \mathbf{k}}(\mathbf{G},\mathbf{q}=0) v_{\mathbf{G}}(0) M_{\alpha' c' \mathbf{k}'}(\mathbf{G},\mathbf{q}=0), \label{eq:exc_imp} \\
    W_{c \alpha \mathbf{k}, c' \alpha' \mathbf{k}'} &= -\frac{1}{V} \sum_{\mathbf{G}\mathbf{G}'} M^*_{\alpha' \alpha}(\mathbf{G},\mathbf{q})w_{\mathbf{G},\mathbf{G}'}(\mathbf{q}) M_{c' c \mathbf{k}'}(\mathbf{G}', \mathbf{q})\delta_{\mathbf{q},\mathbf{k}-\mathbf{k}'}, \label{eq:direct_imp}
\end{align}
where $V$ is the volume of the crystal and $M_{ij\mathbf{k}}(\mathbf{G},\mathbf{q}) = \langle i \mathbf{k} | \mathrm{e}^{-\mathrm{i}(\mathbf{G}+\mathbf{q})\mathbf{r}}| j \mathbf{k}+\mathbf{q} \rangle$ with $\mathbf{G}$ being reciprocal lattice vectors.
%
%-----------------------------------------------------------------------------
\subsection{XAS Implementation in \ocean, \exciting, and \xs} \label{sec:implmentation}
%-----------------------------------------------------------------------------
In this work, we compare XAS spectra calculated from three popular \textit{ab initio} codes: \ocean~\cite{vinson2011bethe,ocean-3}, \exciting~\cite{gulans2014exciting}, and \xs~\cite{taillefumier2002x,gougoussis2009first}. Key aspects of the methodology and implementations of the three codes are summarized in Table~\ref{tab:code}.
While {\exciting} and {\ocean} compute the spectra within the BSE formalism, \xs{} employs the CHP formalism. Additionally, five important aspects in the numerical implementations are worth mentioning: 1) the boundary condition, 2) the treatment of the core state, 3) the basis set, 4) the size of simulation cell, and 5) the treatment of empty states.

All the three codes considered here employ periodic boundary conditions. Both \ocean{} and \xs{} are planewave pseudoptential codes. 
\xs{} is a module of the \qe{} code,\cite{giannozzi2009quantum,giannozzi2017advanced} while \ocean{} interfaces with \qe{} to generate the necessary DFT eigenstates. 
%. \ocean{} can be interfaced with both \qe{} and ABINIT~\cite{Gonze2020}. 
In the pseudopotential method, the bare Coulomb potential of the nuclei and the effects of tightly bound core electrons are replaced by a smooth effective potential only acting on valence electrons. Thus the core-hole wavefunction is not directly accessible in \ocean{} and \xs{}, and is obtained from a separate calculation on an isolated atom. \exciting, on the other hand, is a full-potential all-electron code that treats both core electrons and valence electrons explicitly. {\exciting} employs the augmented linearized plane-wave plus local orbital (LAPW+LO) basis set to account for both the strong variation of the wavefunction inside the core region and the smooth wavefunction in the interstitial region. As such, the core wavefunction is directly accessible in \exciting.

In {\xs},  due to the presence of the core hole at the absorbing atomic site, a large enough supercell is required to avoid spurious interaction between core-hole sites. In this work, we construct a supercell with the condition that the supercell size is no smaller than 9~\text{\AA} along each principle axis. Our own testing and prior work~\cite{yan2019ultrathin} indicate that this is sufficient to converge the XAS spectra relative to the supercell size. In the CHP simulations, the excited electron can be either neglected (i.e., a charged final state) or explicitly included in the system (i.e., a neutral final state). In our \xs{} calculations, the excited electron is treated as part of the system and placed at the bottom of the conduction band, which is known as the excited-electron and core-hole approach~\cite{prendergast2006x,PhysRevB.62.7901,PhysRevLett.67.2521}. 
{\xs} evaluates the resolvent in Eq.~\ref{eq:finalstate} using the Lanczos algorithm~\cite{lanczos1950iteration,lanczos1952solution} to avoid an explicit summation over final states. 
Because the Hamiltonian in Eq.~\ref{eq:finalstate} is based on DFT, only the density of valence electrons is required to construct the Hamiltonian, and therefore {\xs} calculations do not require empty bands explicitly.
%As a result, {\xs} calculations do not require the empty bands explicitly.

Both \exciting{} and \ocean{} require an explicit summation over the unoccupied states to construct $\hat{H}_{BSE}$ of the unit cell, and the number of unoccupied bands have to be converged carefully, because states with higher energies can contribute to the spectral weight at much lower excitation energies through the off-diagonal terms of $\hat{H}_{BSE}$. In \ocean{}, $(\omega - \hat{H}_{BSE} + \mathrm{i} \eta)^{-1}$ is calculated iteratively using the Lanczos method. This term in \exciting{} is computed by diagonalizing $\hat{H}_{BSE}$ explicitly. The \exciting{} code expands BSE matrix elements in reciprocal space following Eqs.~\ref{eq:exc_imp} and \ref{eq:direct_imp}, thus introducing an additional convergence with respect to the cut-off of $|\mathbf{G}+\mathbf{q}|$ for the summation over reciprocal lattice vectors $\mathbf{G}$, referred to as $|\mathbf{G}+\mathbf{q}|_{max}$. The random phase approximation (RPA) is used to compute the screened Coulomb interaction in Eq.~\ref{eq:direct_imp}, which involves another truncated summation over unoccupied states. The \ocean{} code calculates the core-hole screening in real space, using a hybrid RPA and model RPA approach~\cite{PhysRevB.103.245143}. This requires an estimation of $\epsilon_\infty$, the electronic contribution to the static dielectric constant, and this value is taken from the Materials Project~\cite{jain2013commentary,petousis2017high} for insulating systems or set to 10000 for metallic systems.
%and the $\mathbf{k}$-point mesh used to sample the Brillouin zone needs to be converged.

It should be noted that \ocean{} and \exciting{} output the imaginary part of the macroscopic dielectric constant $\epsilon_2$, while \xs{} outputs the absorption cross section $\sigma$. In all three codes it is assumed that the core-level response can be treated independently for each atom and x-ray edge. To have a direct cross-code spectral comparison, we convert the \xs{} output to $\epsilon_2$ using $\epsilon_2(\omega) = \sigma(\omega)/(\alpha \, \omega \, \Omega)$,
where $\Omega$ is the volume of the unit cell. All the cross-code spectral comparison is performed based on $\epsilon_2$.

\begin{table}[bt!]
% The best place to locate the table environment is directly after its first
% reference in text
\caption{\label{tab:code}
Summary of the main features of the three codes used in the XAS simulations.
}
\begin{ruledtabular}
\begin{tabular}{cccc}
 &
\textrm{\exciting}&
\textrm{\ocean}&
\textrm{\xs}
\\
\colrule
Method & Bethe-Salpeter Equation
 & Bethe-Salpeter Equation
 &  Core-hole Pseudopotential\\
Boundary condition & Periodic & Periodic &  Periodic\\
Treatment of core & Explicit, all-electron & Pseudopotential & Pseudopotential\\
Basis  & LAPW + LO & Planewave & Planewave\\
Simulation cell & Unit cell & Unit cell & Supercell\\
Explicit empty states & Yes & Yes & No \\
\end{tabular}
\end{ruledtabular}
\end{table}

\subsection{Edge Alignment}\label{sec:alignment}

The energy onset of an XAS edge encodes important information about the chemical environment around the absorbing atom. In particular, the edge shift can be correlated to chemical composition, as first observed by Bergengren in the study of phosphorus K-edge XAS~\cite{bergengren1920rontgenabsorption}. 
In any system with symmetrically inequivalent sites, including heterogeneous samples with mixtures of phases or stoichiometries, a proper accounting for core-level shifts is required to accurately compare simulated XAS to experiment. 
Based on the studies of K-edge and L-edge XAS~\cite{lindh1921kenntnis,stelling1930ztschr,kunzl1932linear},  Kunzl’s law~\cite{kunzl1932linear} states that the edge shifts are governed by the valency. The absorbing atom exhibits a positive (negative) edge shift when it is oxidized (reduced), which can be understood as the shielding effects of valence electrons on the ionic potential. In the case of most 3{\it d} transition metal K edges, the oxidation state of the metal site can be deduced from the edge shift~\cite{yamamoto2008assignment}. However, there are also exceptions, such as Cr, where the position of the main absorption alone is an ambiguous measure of the oxidation state~\cite{Farges2009,Tromp2007}.
In early studies, the \textit{effective charge} of the absorbing atom was determined empirically according to its chemical environment~\cite{suchet1965chemical}. 
%With the widely adopted \textit{ab initio} electronic structure methods, the edge shift, or core-level shift, can be calculated from first principles. 
The three codes we investigate here are all capable of calculating relative edge shifts, or core-level shifts, from first principles.  

\begin{figure}[bht!] 
\includegraphics[width=3.375 in]{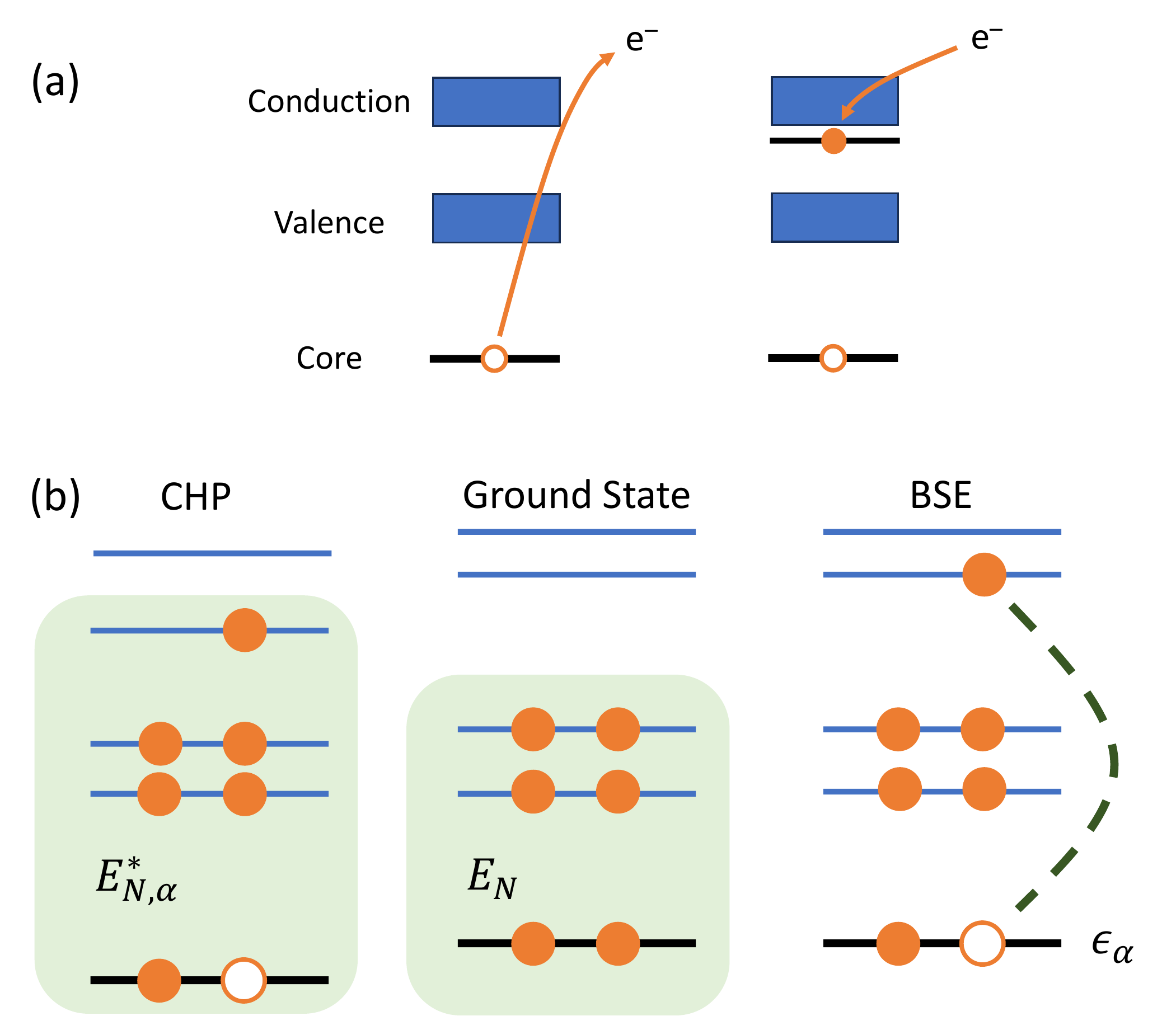}
\caption{\label{fig:edge_alignment_schematics} Physical pictures of excitation energies that determine the edge alignment in XAS simulations. a) Separation of the neutral excitation induced by XAS into a sequence of two processes. Left: core electron removal; Right: electron insertion at the presence of the core hole. b) Excitation energies including final state effects as implemented in CHP (left) and BSE (right) codes. The dashed line in the BSE diagram indicates the electron - core hole coupling terms.} 
\end{figure}

Several physical effects contribute to the XAS edge alignment. We can treat XAS as two sequential processes as shown in Fig.~\ref{fig:edge_alignment_schematics}a. In the core electron removal step, a core electron is removed from the system leaving a core hole behind in an $N-1$ electron system. This step corresponds to the physical process in x-ray photoemission spectroscopy (XPS). The energy needed to remove a core-level electron, i.e. the energy difference between the ground state $E_N$ and core-excited state $E_{N-1,\alpha}^*$, includes both the impact of the local potential on the core electron and the final state relaxation in the presence of the core hole. In the electron insertion step, an electron is put back to the conduction band. The electron--core-hole interaction affects the insertion energy. In particular, low energy excitations can involve the localization of the electron, binding to the core hole to form an exciton in insulators, with energies smaller than found in the independent particle approximation. These physical effects are treated under quite different approximations in the three codes. We first compare the methods for computing the electron removal energy, providing the basis for comparison to XPS shifts and probing the approximate treatment of the core-hole relaxation energy. Then, instead of considering the electron insertion process separately, we compare the methods based on the full excitation energy calculation.

The electron removal energy of the core level $\alpha$, in the quasi-particle picture, is given by its eigenenergy ($\epsilon_\alpha$) with the opposite sign. The leading contributions are the single particle term and electrostatics, which are well captured within DFT. In the case of \exciting{}, the DFT eigenvalue of the core level $\epsilon_\alpha^{\mathit{e}}$ is readily available. A complication arises for pseudopotential codes. In \ocean{}, the relative energy of the core level can be calculated by $\epsilon_\alpha^{\mathit{o}} = X^{\mathit{o}}_\alpha + \langle \psi_\alpha \vert V_{\mathrm{KS}}^{ps} \vert \psi_\alpha \rangle$, where $V_\mathrm{KS}^{ps}$ is the system-dependent total Kohn-Sham potential, $\psi_\alpha$ is the core-level wavefunction under the frozen core approximation, and $X_\alpha^o$ is a system-independent unknown that cancels out when comparing relative shifts. The DFT eigenenergies from \exciting{} or \ocean{} should give identical shifts up to the limit of the frozen-core approximation. The many-body exchange correlation effects capture the relaxation of the rest of the system in the presence of the core hole~\cite{doi:10.1021/acs.jctc.7b01192,doi:10.1021/acs.jctc.8b00458}. The state-of-the-art is to perform a self-energy ($\Sigma$) correction to $\epsilon_{\alpha}$ under the {\it GW} approximation. 
%In principle, $\Sigma_{\alpha}$ can be evaluated in \exciting{} (ref), but it is often neglected in practical studies due to the high computational cost. 
An approximate self-energy correction for the core level is implemented in \ocean{}. $\epsilon_\alpha^{\mathit{o}}$ is corrected by $\Sigma_{\alpha}=- \langle  \psi_\alpha \vert \sfrac{1}{2} W_c + V_{xc}\vert \psi_\alpha \rangle$, where $W_c$ is the static screened Coulomb potential of the core hole and $V_{xc}$ is the exchange-corelation potential~\cite{ocean-3,vxc}. Since this approximation uses the same screened potential needed for the BSE calculation, there is very little cost to include it. In \xs{}, the core electron removal energy, including initial and final state effects, is calculated using the total energy formulation within a $\Delta$SCF procedure, $\epsilon_\alpha^{\mathit{x}} = X^{\mathit{x}}_\alpha$ + $E^{*}_{N-1,\alpha} - E_N$, where $X^{\mathit{x}}_\alpha$ is an unknown constant due to the use of different pseudopotentials for the absorber atom in the ground state and core-hole excited state.  However, for every calculation of a given core level the same change in pseduopotentials can be used, making $X^{\mathit{x}}_\alpha$ system-independent. Consequently, $X^{\mathit{x}}_\alpha$ cancels out when comparing relative shifts. Note that both \xs{} and \ocean{} make use of the frozen-core approximation, since neither the core-hole pseudopotential nor $\psi_{\alpha}$, respectively, are allowed to change from system to system.

For the full calculation of the excitation energies, the CHP and BSE codes have very different implementations detailed in Fig.~\ref{fig:edge_alignment_schematics}b. In this study, we consider the excited core hole approximation in CHP, where the core electron is placed at the bottom of the conduction band. The electron--core-hole interaction is described through the self-consistently screened core-hole potential that takes into account the relaxation of valence electrons. Clearly, the Kohn-Sham energy levels and wavefunctions of the CHP final state are different from those of the ground state, as the attractive core-hole potential pulls the valence electrons down in energy. The relative alignment of the lowest excitation in \xs{} is carried out within the $\Delta$SCF procedure as $E^{*}_{N,\alpha} - E_N$, where $E^{*}_{N,\alpha}$ is the total energy of the CHP final state. In the two BSE codes (\ocean{} and \exciting{}), the final state effects are treated by the direct and exchange interactions of the electron-hole pairs, where the basis functions are the ground-state orbitals as indicated in Fig.~\ref{fig:edge_alignment_schematics}b. In a full {\it GW}-BSE implementation, the self-energy correction is applied to both the core level and to the conduction bands. However, this is often neglected in practical studies due to the high computational cost. As noted, in \ocean{} the self-energy correction for the core level is approximated by the static core-hole relaxation energy.

\section{Workflow}
In this study, we focus on the XANES region of the XAS spectra, which is more sensitive to the electronic structure than the extended x-ray absorption fine structure (EXAFS) region at the higher energy range. To ensure a rigorous XANES cross-code validation, we have developed a workflow that automates the generation of the input files with a set of carefully tested input parameters. From this set of parameters, we can calculate fully converged XANES spectra using different codes. Details of this workflow are described below and summarized in Fig.~\ref{fig:workflow}, which includes input parameters for ground state and spectral calculations. Based on this workflow, a Python package called \emph{Lightshow} has been developed to automate the XANES simulation input file generation for multiple codes~\cite{carbone2022lightshow}.

\subsection{Workflow for XANES spectra calculations}\label{sec:workflow}

\begin{figure}[htb] 
\includegraphics[width=6.0 in]{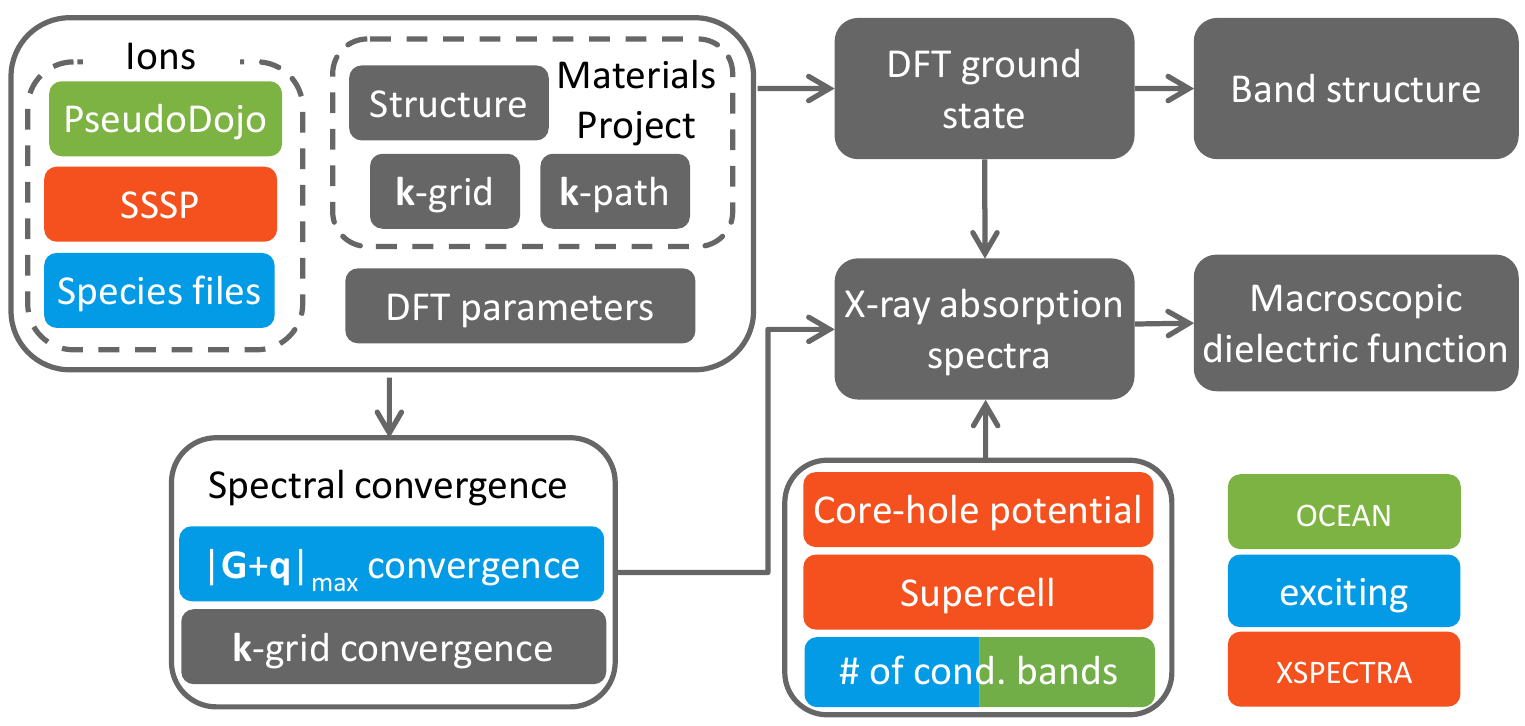}
\caption{\label{fig:workflow} Workflow of the XANES benchmark calculations. Computation steps common to all codes are shown in grey, \ocean{}-specific ones in green, \exciting{}-specific ones in blue, and \xs{}-specific ones in orange.} 
\end{figure}

Ground state input parameters (on the upper left of Fig.~\ref{fig:workflow}) contain three groups: parameters to treat core electrons, structural parameters, and DFT parameters. 

%\textit{Ionic potentials.} In \ocean{} and \xs{}, ionic potentials are represented by pseudopotentials taken from standardized pseudopotential libraries. Norm-conserving pseudopotentials used in \ocean{} calculations are taken from the PseudoDojo library~\cite{van2018pseudodojo}. Pseudopotentials in \xs{} are taken from the precision version of the Standard solid-state pseudopotentials (SSSP) library~\cite{lejaeghere2016reproducibility,prandini2018precision}, where Ti and O potentials are treated by the ultrasoft and the projector augmented wave~(PAW) schemes, respectively. In the \exciting{} code, the ionic configurations are stored in so-called \textit{species} files that contain all information on the atomic-sphere part of the LAPW+LO basis. The standard O \textit{species} file (i.e., for ground state calculations) does not include local orbitals, and the Ti \textit{species} file contains one \textit{s} and one \textit{p} local orbital, with reference energies at -62.238 and -39.209 eV, respectively. In order to obtain reliable trail energies in the wide range needed (about 40 eV above the CBM), we augment the Ti \textit{species} file with one \textit{d} orbital and one \textit{f} orbital both at a same reference energy of 30 eV, and the O \textit{species} file with one \textit{s} orbital and one \textit{p} orbital also at 30 eV. These additional local orbitals significantly improve the band structures between 25 to 40 eV above the CBM.

\textit{Core treatment.} In \ocean\ and \xs, the core electrons are not explicitly treated, and pseudopotentials are taken from stadardized pseudopotential libraries. Norm-conserving pseudopotentials used in \ocean{} calculations are taken from the PseudoDojo library~\cite{van2018pseudodojo}. Pseudopotentials in \xs{} are taken from the precision version of the Standard solid-state pseudopotentials (SSSP) library~\cite{lejaeghere2016reproducibility,prandini2018precision}, where Ti and O potentials are treated by the ultrasoft and the projector augmented wave~(PAW) schemes, respectively. In the \exciting\ code, the Kohn-Sham equation is solved for all electrons explicitly. The atomic basis functions, local orbitals and the respective trial energies are defined for each atomic species in the so-called \textit{species files}. In order to obtain accurate band structures, we augment the default LAPW basis for O with one \textit{s} and one \textit{p} orbital, both with reference energies of 30 eV. For Ti, we augment the default settings, which contains one \textit{s} orbital with a reference energy of $-$62.238 eV and one \textit{p} orbital with a reference energy of $-$39.209 eV, with one \textit{d} orbital and one \textit{f} orbital, both with reference energies of 30 eV. These local orbitals at high energies are crucial to improve the band structure in the range of 25~eV to 40 eV above the CBM.

\textit{Structural parameters.} Lattice parameters and atomic positions of the materials in the benchmark study are taken from the optimized structures obtained using the Perdew-Burke-Ernzerhof (PBE) functional~\cite{perdew1996generalized} reported in the Materials Project~\cite{jain2013commentary}. The set of $\mathbf{k}$-points required for a converged DFT ground state calculation and the high-symmetry $\mathbf{k}$-path for the visualization of the band structure are also taken from the Materials Project. From the ground state calculation, the workflow generates the band structure. 

\textit{DFT parameters.} We choose the PBE exchange-correlation functional in all calculations for its general applicability and the consistency with the PBE optimized crystal structures from the Materials Project. For \ocean{} and \xs{}, the plane-wave energy cutoffs for wavefunction and electron density employ the recommended values from the corresponding pseudopotential libraries. The total energy convergence threshold in the self-consistent field calculation is set to $10^{-10}$ Rydberg per atom. 
%Smearing is only for metals w/ ocean
A Gaussian smearing of 0.002 Rydberg is used in \xs{} and \exciting{} calculations. In \ocean{} calculations, a Gaussian smearing of 0.02~Rydberg is included only for metallic systems. This small difference in the choice of smearing between \xs{}/\exciting{} and \ocean{} does not affect our results, as supported by the comparison of the band structure from different codes in the \emph{Results} Section. %In addition, the ground state broadening parameters are more than three times smaller than the core-hole lifetime broadening parameter we used, so they do not affect the comparison of the XANES spectra.
%\todo{How about the default parameters used in \exciting{}? Exciting use default. rgkmax?} 
The DFT wavefunctions and orbital energies are the input of Eqs.~\ref{eq:finalstate} and~\ref{eq:x-ray_bse} for the subsequent XANES calculations using \ocean{}. \exciting{} adopts most of its default parameters for ground state properties. 

\textit{Spectral parameters.} %For all the three codes, the core-hole lifetime broadening applied to the Ti K-edge spectra was set to 0.89~eV half-width at half maximum. %~\cite{CAMPBELL20011}. 
For all the three codes, a Lorentzian broadening was applied to the Ti K-edge spectra with a half-width at half maximum of 0.89~eV. This value was chosen to give good resolution of the spectral features for measuring the similarity across different codes. For a quantitative comparison between simulation and experiment, multiple broadening mechanisms need to be considered, such as core-hole lifetime, the photo-electron lifetime in the final state and the resolution of the instrument~\cite{HEBERT200712}.

%For all three codes, a constant Lorentzian broadening of 0.89~eV half-width at half maximum was applied to the Ti K-edge spectra, to approximate the lifetime broadening in the system~\cite{CAMPBELL20011}.
%This approximates both the core-hole lifetime broadening as well as other sources of broadening, such as the photo-electron lifetime in the final state, are important for the comparison between theory and experiment. The electron lifetime broadening is typically included in the post-processing of the simulated spectrum in order to improve the quantitative agreement of the spectral shape with experiment. For the limited scope of comparison to experiment presented here, electron lifetime broadening is not included.

\xs{} calculations require a core-hole pseudopotential for the absorber atom. We use the same Ti core-hole PAW pseudotential as in the Ref.~\onlinecite{yan2019ultrathin}, which was optimized for the 4{\it p} scattering states generated using the \textit{atomic} module in the \qe{} code. %The core hole electronic configuration is obtained by by removing one electron from the 1{\it s} level of Ti element. This Ti core-hole PAW pseudotential can be found in the Supplementary Material~\cite{SI}. 
During the reconstruction of the all electron wavefunction in the \xs{} calculations, we set the PAW radius ($r\_{\textit{paw}}$) the same as the cutoff radius in the core-hole PAW pseudopotential. \xs{} uses the Lanczos iterative method to solve Eq.~\ref{eq:finalstate}. The maximum number of iterations is set to 5000, and the early exit threshold is set to $0.01$ for spectra $200$ iterations apart. 

Both \ocean{} and \exciting{} use electron-hole pairs as a basis to expand the core electron excitations, which involves an explicit summation over unoccupied bands. 
The number of empty bands ($n_c$) needed to span a given energy range can be estimated from the density of states of the non-interacting electron gas: $n_c = (\sqrt{2}/3\pi^2) (\Delta E)^{3/2} \Omega$, where $\Omega$ is the volume of the unit cell and $\Delta E$ is the desired energy range in Ha. In practice the energy range of a given $n_c$ is less than this formula estimates, and an input of $\Delta E \approx 50$~eV yielded a conduction band count sufficient to span approximately 40~eV in the systems we studied.
%The number of the empty bands~($n_c$) is system-dependent, and in practice we consider an energy range ($\Delta E$) of approximately $50$~eV above the conduction band minimum. The relation between $n_c$ and $\Delta E$ is roughly estimated using the energy dispersion of the non-interacting electron gas, i.e., $n_c = \sqrt{2}/3\pi^2 (\Delta E)^{3/2} \Omega$, where $\Omega$ is the volume of the unit cell.
%\begin{equation}
%N_b = 4/(3\pi^2) V  (\Delta E)^{3/2} \approx \frac{0.917 }{ \textrm{\AA}^3 \textrm{Ryd}^{3/2}} (\Delta E)^{3/2}V
%\end{equation}. 
While \exciting{} explicitly diagonalizes the BSE Hamiltonian, \ocean{} uses the Lanczos iterative method. The maximum number of iterations is set to $1000$, with an early exit criterion of $0.001$ for spectra $5$ iterations apart. The total number of empty bands for computing the RPA dielectric response function is determined by setting the energy range to $100$~eV in both \ocean{} and \exciting{}.

%\begin{eqnarray}
%E &= \frac{ \pi^2 }{2 L^2 } \left( n_x^2 + n_y^2 + n_z^2\right)^2 \nonumber  \\
%&\approx \frac{ \pi^2 }{2 V^{2/3}} n^2  \nonumber \\
%n[E] &= 2^{1/2}E^{1/2} V^{1/3} \pi^{-1} \nonumber  \\
%N &= \int_0^E n^2 dn \sin \phi d\phi d\theta = \frac{4 \pi}{3}n^3 \vert_{n=n[E]} \nonumber  \\
%N &= \frac{4}{3 \pi^2} (2E)^{3/2} V \nonumber \\
%N &= \frac{0.917 }{ \textrm{\AA}^3 \textrm{Ryd}^{3/2}} E^{3/2}V
%\end{eqnarray}
%

Finally, there are a set of spectral parameters that require careful convergence tests. Common to the three codes, the \emph{k}-grid used in the Brillouin zone (BZ) sampling needs to be converged, and generally XAS calculations require denser sampling of the reciprocal space than ground state calculations. In \exciting{} and \ocean{} calculations, we employ the same \emph{k}-grid shift of (0.125, 0.250, 0.375) in the reciprocal lattice vector coordinates. We use $\Gamma$-centered \emph{k}-grids to sample the BZ of the supercell in \xs{} calculations. Specific to the \exciting{} code, it requires the convergence with respect to the cutoff $|\mathbf{G}+\mathbf{q}|_{max}$ in Eqs.~\ref{eq:exc_imp} and~\ref{eq:direct_imp}. Details on the convergence study are discussed in the next section.

\subsection{Convergence of Spectral Parameters}

%Put in equation with explicit bands, k-points, and basis 
\subsubsection{Spectral similarity metric}
%Considering the computational time and accuracy in the calculated spectra, some important parameters should be adjusted properly to guarantee the results are accurate while the computational time is moderate. Although the purpose of this work is to compare the XANES spectra from different codes, to obtain the converged spectrum within each code is an important prerequisite. The spectrum can be regarded as converged if there are only negligible changes observed when a finer parameter is used in the calculations. 
In order to define the threshold for the convergence of spectral parameters, we introduce a spectral similarity metric. This definition is not unique and there are several possible choices, such as the Cosine similarity, L$_2$-normalized Euclidean distance, the Pearson correlation coefficient, and the Spearman’s rank correlation score~\cite{spearman1904}. Without losing the generality, in this study we choose the Spearman’s rank correlation score ($r_{sp}$) as our spectral similarity metric, and we have verified that the overall trend in the convergence behavior is similar when using the Cosine similarity or the Pearson correlation coefficient. Given two XANES spectra on the same grid, $\mathbf{\mu_1}$ and $\mathbf{\mu_2}$, $r_{sp}$ is defined as the Pearson correlation coefficient of the corresponding rank variables $R(\mathbf{\mu_1})$ and $R(\mathbf{\mu_2})$, 
\begin{equation} \label{eq:spearman}
  r_{sp} = \frac{\mathrm{cov}(R(\mathbf{\mu_1}),R(\mathbf{\mu_2}))}{\sigma_R(\mathbf{\mu_1})\sigma_R(\mathbf{\mu_2})},
\end{equation}
where the rank function returns the rank of a given value when compared with the rest of the values in a list, and $\mathrm{cov}$ and $\sigma$ denote the covariance and standard deviation of $R(\mathbf{\mu_1})$ and $R(\mathbf{\mu_2})$, respectively. If there are no repeated data points, a perfect Spearman’s rank correlation yields $+1$ or $-1$ when the two spectra are perfectly monotonically correlated or anti-correlated, whether linear or not. In this sense, the Spearman’s rank correlation score is more sensitive to the spectral shape than the Pearson correlation coefficient that focuses on the linear correlation. Further, we define the spectral similarity as 
\begin{equation} \label{eq:logspearman}
  s = \log_{10}(1-r_{sp}).
\end{equation} By visual inspection, we consider a spectrum to be converged with respect to the reference, when $r_{sp}\geq 0.999$ or equivalently $s\leq -3$. Under this threshold, two XANES spectra show negligible differences (see Fig.~S1b in the Supplementary Material~\cite{SI}). At an even better similarity score of  $s\leq -4$, two spectra are on top of each other as shown in Fig.~S1c.

\subsubsection{K-point grid resolution metric}
%The k-point mesh is the common parameter within the three codes used in this work that needs to be converged. Theoretically, using a $n_1 \times n_2 \times n_3$ k-point mesh in the unit cell is equivalent to using a Gamma point in $n_1 \times n_2 \times n_3$ supercell. In this work, to gauge the k-point mesh, we use the radius of the largest sphere that can fit into this $n_1 \times n_2 \times n_3$ supercell. We define this largest radius as effective radius, which is a quantity in the real space.
Because all the three codes use periodic boundary conditions, the implicit integral in Eqs.~\ref{eq:finalstate} and~\ref{eq:x-ray_bse} is evaluated as a spatial integral within the unit cell over crystal momentum within the first BZ. In practice, the integral over the BZ is carried out as a finite, discrete sum over {\it k} points on a regular mesh. It is important that this sampling is performed on a \emph{k}-grid with fine enough resolution, such that the numerical integration over a finite number of {\it k} points yields converged values. Although this convergence test can be performed trivially on a single system according to the size of the \emph{k}-mesh ($n_1$, $n_2$, $n_3$), a system-independent metric needs to be introduced, because the resolution of the \emph{k}-grid depends on both the mesh size and the size/shape of the BZ. For this purpose, we recognize that an $n_1 \times n_2 \times n_3$ \emph{k}-point mesh of the unit cell is equivalent to a $\Gamma$-point sampling of an $n_1 \times n_2 \times n_3$ supercell. The {\it k}-grid resolution can be quantified by the effective crystal size ($R_\mathrm{eff} $) defined as the shortest spacing between parallel faces of this supercell,
%\begin{equation}
%R_\mathrm{eff} = 2\pi \mathrm{min}_i \Big\{  \frac{n_i}{\vert b_i \vert} \Big\} = \mathrm{min}_i %\big\{ n_i \vert a_i \vert \sin[\theta_i] \big\},
%\end{equation}
%\todo{Is the equation correct?}
%\begin{eqnarray}
%b_1&=&2\pi \frac{a_2\times a_3}{\vert a_1\cdot a_2\times a_3 \vert} \\
%R_\mathrm{eff} &=& 2\pi \mathrm{min}_i \Big\{  \frac{n_i}{\vert b_i \vert} \Big\} = \mathrm{min}_i %\big\{ n_i \times a_i \times \cos\theta_i \big\}, \theta_i \text{is the angle between } a_i \text{ and } a_j\times a_k
%\end{eqnarray}
\begin{eqnarray}
R_\mathrm{eff} &=& 2\pi \, \mathrm{min}_i \Big\{  \frac{n_i}{\vert b_i \vert} \Big\}, \\
\vert b_i \vert &=&2\pi \epsilon_{ijk} \left| \frac{a_j\times a_k}{  a_i\cdot ( a_j\times a_k) } \right|, %\dlx{i\ne \{j,k\}, j\ne k,}
\end{eqnarray}
%where $|a_i|$ and $|b_i|$ are the lengths of the lattice and reciprocal lattice vectors, respectively, and $\theta_i$ is the angle between the two lattice vectors $a_{j,k\ne i}$. 
where $a$ and $b$ are the lattice vectors and reciprocal lattice vectors, respectively, and $\epsilon_{ijk}$ is the Levi-Civita symbol. Note that the magnitude of $b_i$ depends inversely on the magnitude of $a_i$ and directly on the angle between $a_i$ and $a_j\times a_k$. %the lattice vectors $a_j$ and $a_k$. 
%Although non-regular k-point meshes are capable of comparable sampling with a smaller total number of k-points\cite{XX}, but all three codes in this work adopt regular sampling.

%There are some other code dependent parameters that need to be tested. Since XSpectra employs the explicit core-hole method, in which the absorption atom is treated with a core-hole pseudo-potential while the other atoms are treated with the neutral pseudo-potentials. In this case, the absorption atom acts like a dopant in the system. To avoid a the dopant-dopant interaction, a relative large supercell is needed in the calculations. According to our tests, a supercell with lattice constants along each direction larger than 9~\AA{} can guarantee a converged spectra. In the following calculations, we adopt to this threshold for the supercell size. 

\subsection{Ti-O-10 Dataset}

In this work, we choose a dataset consisting of ten representative titanium oxide (Ti$_x$O$_y$) compounds referred to as the Ti-O-10 dataset, such that they cover a reasonably wide range of the materials space characterized by: insulator (5) and metal (5), oxidation state, coordination number, as well as the number of symmetrically nonequivalent absorber sites. This list includes common polymorphs observed experimentally: rutile (mp-2657), anatase (mp-390), and brookite (mp-1840). Details of the material attributes are listed in Table ~\ref{tab:materials}. 

\begin{table}[!htb] 
\caption{Key attributes of the systems in the Ti-O-10 dataset: Materials Project id (mpid), chemical formula, space group, band gap ($E_g$), oxidation state (OS), coordination number ($n_{coord}$), and number of nonequivalent sites ($n_{site}$).}
\begin{ruledtabular}
\label{tab:materials}
\begin{tabular}{ccccccc}
\textrm{mpid}&
\textrm{Formula}&
\textrm{Space group}&
\textrm{$E_g$ (eV)} \footnote{Band gap values were extracted from the Materials Project~\cite{jain2013commentary}, which were calculated with the PBE functional.}&
\textrm{OS}&
\textrm{$n_{coord}$}&
\textrm{$n_{site}$}
\\
\colrule
mp-10734  & Ti$_4$O$_5$  & I4/m               & 0.0   & 2.5+ & 6    & 1   \\
mp-1203   & TiO          & C2/m               & 0.0   & 2+   & 4, 5 & 3   \\
mp-1215   & Ti$_2$O      & P$\overline{3}$m1  & 0.0   & 1+   & 3    & 1   \\
mp-1840   & TiO$_2$ (brookite)      & Pbca               & 2.29  & 4+   & 6    & 1   \\
mp-2657   & TiO$_2$ (rutile)     & P4$_2$/mnm         & 1.77  & 4+   & 6    & 1   \\
mp-2664   & TiO          & Fm$\overline{3}$m  & 0.0   & 2+   & 6    & 1   \\
mp-390    & TiO$_2$ (anatase)      & I4$_1$/amd         & 2.06  & 4+   & 6    & 1   \\
mp-430    & TiO$_2$      & P2$_1$/c           & 2.23  & 4+   & 7    & 1   \\
mp-458    & Ti$_2$O$_3$  & R$\overline{3}$c   & 0.0   & 3+   & 6    & 1   \\
mvc-11115 & TiO$_2$      & R3m                & 2.46  & 4+   & 4, 6 & 2   \\   
\end{tabular}
\end{ruledtabular}
\end{table}\section{Results}

\subsection{Band Structure}

As the first step of the cross-code validation, we compare the band structure obtained from the three codes in order to assess the quality of pseudopotentials used for \ocean{} and \xs{} as well as the settings of the \textit{species} files for \exciting{}. Band gaps of the insulators calculated from the three codes are very close to each other, with \ocean{} / \xs{} and \exciting{} within $30$~meV as shown in Table~\ref{tab:bandgap}. This difference is the same as that within the pseudopotential codes. Note that these band gap values are calculated from a rough \emph{k}-grid used in the ground state calculations, as the main purpose is to compare different codes and not to get accurate band gaps. Therefore small differences are expected between values reported in Table~\ref{tab:bandgap} and those from the Materials Project (Table \ref{tab:materials}). Our results suggest that the error introduced by the pseudopotentials is negligible for the band gap.

\begin{table}[!htbp] 
\caption{Comparison of the DFT-PBE band gap (in eV) of the insulators from three codes.}
\begin{ruledtabular}
\label{tab:bandgap}
\begin{tabular}{cccc}
\textrm{mpid}&
\textrm{\xs{}}&
\textrm{\ocean{}}&
\textrm{\exciting{}}\\
\colrule
brookite (mp-1840)              & 2.35 & 2.37 & 2.36 \\
%\hline
rutile (mp-2657)              & 1.82 & 1.84 & 1.82 \\
%\hline
anatase (mp-390)               & 2.12 & 2.14 & 2.12 \\
%\hline
TiO$_2$ (mp-430)               & 2.29 & 2.32 & 2.29 \\
%\hline    
TiO$_2$ (mvc-11115)            & 2.50 & 2.51 & 2.50 \\
\end{tabular}
\end{ruledtabular}
\end{table}

\begin{table}[!htbp]
\caption{Comparison of the DFT energy levels. The first, second and third columns in each section correspond to energy differences (in meV) between \exciting{} and \ocean{} (E-O), \xs{} and \ocean{} (X-O), and \exciting{} and \xs{} (E-X), respectively. The largest RMSD and Max D for each energy range are highlighted in bold. }
\begin{ruledtabular}
\label{tab:groundstate}
\begin{tabular}{c|c|c|c|c|c}
\textrm{mpid}&
\textrm{comparison}&
\textrm{Val.\ band}&
\textrm{CBM + 10 eV}&
\textrm{CBM + 20 eV}&
\textrm{CBM + 30 eV}
\\
\colrule
& & X-O \quad E-O \quad  E-X & X-O \quad E-O \quad  E-X & X-O \quad E-O \quad  E-X & X-O \quad E-O \quad  E-X 
\\
%JTV is just trying to see how different separators look in the table
mp-390      & RMSD  & 16 \quad  18 \quad     2   
                    & \textbf{17} \quad  \textbf{17} \quad     2   
                    & 13 \quad  13 \quad     5 
                    & 14 \quad  13 \quad     7    \\
anatase     & Max D & 37 \quad  40 \quad     5    
                    & 32 \quad  \textbf{33} \quad     5   
                    & 32 \quad  33 \quad    12   
                    & 36 \quad  44 \quad    32   \\
\hline
mp-2657     & RMSD  & 16 \quad  18  \quad     3    
                    & \textbf{17} \quad  \textbf{17}  \quad     2   
                    & 14 \quad  14  \quad     5   
                    & 16 \quad  14  \quad     7   \\
rutile            & Max D & 38 \quad  40  \quad     5    
                    & 31 \quad  \textbf{33}  \quad     6   
                    & 31 \quad  38  \quad    16   
                    & 40 \quad  57  \quad    59   \\
\hline
mp-1840     & RMSD  & 16 \quad  \textbf{19}  \quad     4    
                    & 16 \quad  10  \quad     15   
                    & 12 \quad  15  \quad     \textbf{18}   
                    & 14 \quad  12  \quad     18    \\
brookite            & Max D & 37 \quad  42  \quad     8    
                    & 29 \quad  28  \quad     26   
                    & 29 \quad  37  \quad     27   
                    & 31 \quad  37  \quad     27   \\
\hline
mp-430      & RMSD  & 17 \quad  \textbf{19}  \quad      2    
                    & 13 \quad  12  \quad      3   
                    & 11 \quad  10  \quad      5   
                    & 14 \quad  11  \quad      6     \\
TiO$_2$            & Max D & 40 \quad  \textbf{43}  \quad      4    
                    & 22 \quad  22  \quad      6   
                    & 23 \quad  22 \quad      11   
                    & 34 \quad  37 \quad      16    \\
\hline
mvc-11115   & RMSD  &  15 \quad 16  \quad       2   
                    &  10 \quad  8  \quad       3  
                    &   8 \quad  8  \quad       5  
                    &  13 \quad 11  \quad       5    \\
TiO$_2$            & Max D &  38 \quad 37  \quad       5   
                    &  18 \quad 15  \quad       6 
                    &  18 \quad 29  \quad      12  
                    &  43 \quad 60  \quad      19   \\    
\hline
mp-10734    & RMSD  &  13 \quad  13  \quad        2   
                    &  5  \quad   8  \quad        4  
                    &  11 \quad   9  \quad        6  
                    &  22 \quad  16  \quad        9    \\
Ti$_4$O$_5$            & Max D &  34 \quad  35  \quad        5   
                    &  15 \quad  15  \quad        7  
                    &  36 \quad  30  \quad       20  
                    &  77 \quad  99  \quad       48   \\                                          
\hline
mp-1203     & RMSD  &  13 \quad   13 \quad      2   
                    &  4  \quad    6 \quad      4  
                    &  8  \quad    7 \quad      5  
                    &  19 \quad   15 \quad      6    \\
TiO            & Max D &  36 \quad   36 \quad      5   
                    &  13 \quad   14 \quad      7  
                    &  35 \quad   28 \quad     17  
                    &  69 \quad   90 \quad     44   \\  
\hline
mp-1215     & RMSD  &   10 \quad   10  \quad      2   
                    &   5  \quad    9  \quad      4  
                    &   9  \quad    8  \quad      9 
                    &   19 \quad   11  \quad     17    \\
Ti$_2$O            & Max D &   37 \quad   35  \quad      7   
                    &   13 \quad   14  \quad      8  
                    &   35 \quad   29  \quad     34  
                    &   70 \quad   89  \quad     77   \\  
\hline
mp-2664     & RMSD  &  13 \quad   12  \quad      2   
                    &  4  \quad    6  \quad      4  
                    &  10 \quad    8  \quad     10 
                    &  23 \quad   15  \quad     \textbf{25}   \\
TiO            & Max D &  36 \quad   35  \quad      6   
                    &  15 \quad   17  \quad      9  
                    &  \textbf{42} \quad   37 \quad      39  
                    &  82 \quad  124 \quad     \textbf{130}   \\    
\hline
mp-458      & RMSD  &  14 \quad   15  \quad       2   
                    &  5  \quad    7  \quad       3 
                    &  7  \quad    7  \quad       4 
                    &  16 \quad   11  \quad       9    \\
Ti$_2$O$_3$            & Max D &  34 \quad   35  \quad       4   
                    &  11 \quad   16  \quad       6  
                    &  26 \quad   23  \quad      20  
                    &  53 \quad   54  \quad      51   \\  
\end{tabular}
\end{ruledtabular}
\end{table}

\begin{figure}[htbp] 
\includegraphics[angle=270,width=\textwidth]{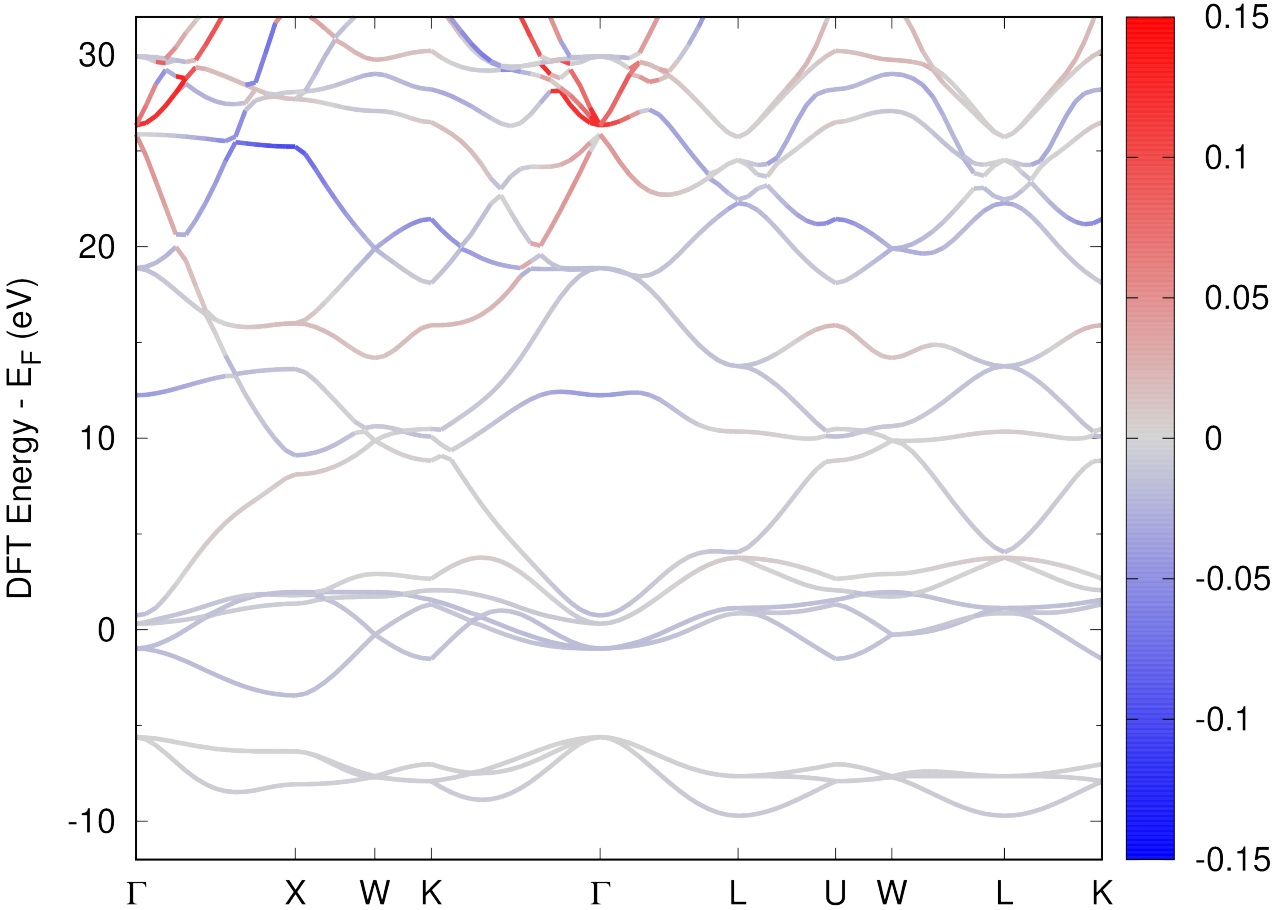}
\caption{\label{fig:bandstructure} 
%\todo{we need to change this figure for mp-2664} 
Band structure of mp-2664 calculated with the \exciting{} code. The colors signify the energy difference (in eV) between \exciting{} and \qe{} (with the PseudoDojo  pseudopotentials). The energy zero of \exciting{} is set to the Fermi level, the \qe{} alignment used in \ocean{} was determined by minimizing the RMSD over the occupied states down to $-20$~eV (a cutoff that neglects the O 2{\it s} states and the semi-core Ti 3{\it s} and 3{\it p} states).} 
\end{figure}

Next we perform a more comprehensive comparison of the DFT energy levels on the \emph{k}-grids of the ground state as shown in Table~\ref{tab:groundstate}. Since the absolute energy eigenvalues from different codes cannot be compared directly, prior to performing the comparison, the energy levels from different codes are aligned by minimizing the root mean squared deviation (RMSD) of the valence bands (see Appendix~\ref{app:rmsd}). The differences of the energy eigenvalues are characterized by RMSD and the maximum deviation (Max D). 
We collect the statistics for each phase and each code in four energy regions: the entire valence band and the conduction band spanning from the CBM to 10, 20, and 30~eV above the CBM.
For the entire valence band, comparing \ocean{} and \xs{}, across the phases the largest RMSD and Max D are 17 meV and 40 meV, respectively. The differences between \ocean{} / \xs{} and \exciting{} are similar, with the largest RMSD of 19 meV and Max D of 43 meV.  For most systems, \xs{} agrees better with \exciting{} (with the largest Max D of 8 meV) than \ocean{}, which is likely due to the optimized pseudopotentials in SSSP. The good agreement among the three codes extends to more than 20 eV above CBM. Within CBM+10 eV, the agreement between \ocean{} / \xs{} and \exciting{} is comparable to that for the valence band, with the largest RMSD of 17~meV and max~D of 33~meV. Even at CBM+20 eV, the same good agreement still holds with  the largest RMSD of 18 meV and Max D of 42 meV. At even higher energies, the differences increase more rapidly because it becomes more difficult to obtain accurate band structure for scattering states at very high energies.
In practice, it requires additional projectors in the construction of the pseudopotential or additional local basis functions in the LAPW+LO formalism optimized for high energy scattering states, respectively. Within CBM to CBM+30~eV, the RMSD increases only slightly to 25~meV, as it is averaged over the 30 eV energy range. However, the largest Max~D as a more senitive quantity grows up to 130 meV in the metallic system, mp-2664. To provide a direct visualization of the energy deviations between different codes, we plot the band structure of mp-2664 as obtained from \exciting{} and highlight the difference to that obtained using {\sc Quantum ESPRESSO} with the PseudoDojo pseudopotentials as used in \ocean{} in Fig.~\ref{fig:bandstructure}. All other materials have better agreement in the band structure comparison. We found that the large differences mainly come from the energy range between CBM+25 and CBM+30 eV along $\Gamma-X$, $\Gamma-L$ and $\Gamma-K$ lines. The band structure comparison shown in Table~\ref{tab:groundstate} and Fig.~\ref{fig:bandstructure} gives us the baseline error due to different treatment of the core electrons, i.e., choice of different pseudopotentials and pseudopotential versus all-electron, when comparing XANES spectra from the three codes. Further optimizing pseudopotentials or the LAPW+LO basis functions might improve the agreement at around CBM+30~eV, but this is non-trivial and beyond the scope of the current work. 

\subsection{$|\mathbf{G}+\mathbf{q}|_{\textrm{max}}$ Convergence}
\begin{figure}[hbtp] 
\includegraphics[width=\textwidth]{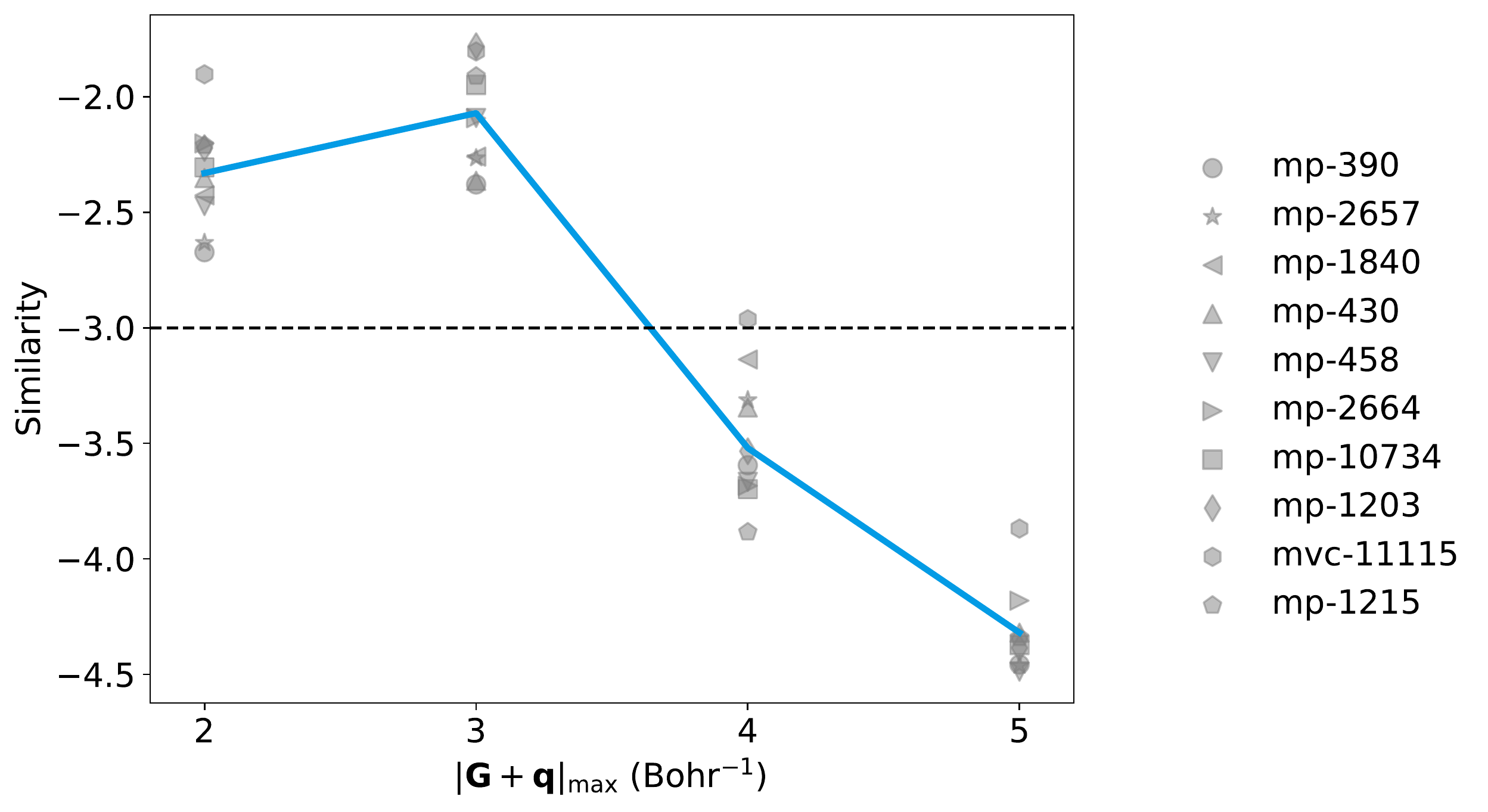}
\caption{\label{fig:gqmax} Convergence of the \exciting{} spectra with respect to $|\mathbf{G}+\mathbf{q}|_{\textrm{max}}$. The blue solid line indicates the average similarity at each $|\mathbf{G}+\mathbf{q}|_{\textrm{max}}$ value. } 
\end{figure}

The \exciting{} calculations require convergence of $|\mathbf{G}+\mathbf{q}|_{\textrm{max}}$ for the summation over reciprocal lattice vectors in Eqs.~\ref{eq:exc_imp} and \ref{eq:direct_imp}, which is controlled by the input parameter, GQMAX, in units of Bohr$^{-1}$. To study its convergence behavior, $|\mathbf{G}+\mathbf{q}|_{\textrm{max}}$ is varied  from 2 to 6 in steps of 1, except for mp-1840, where the highest $|\mathbf{G}+\mathbf{q}|_{\textrm{max}}$ value is set to 5 due to the computational cost. For each calculation, we compare the spectrum with the reference obtained with the highest $|\mathbf{G}+\mathbf{q}|_{\textrm{max}}$ and determine the threshold for convergence when the similarity with the reference drops below $-3$.  %We use the Spearman's rank correlation coefficient to characterize the spectral similarity and consider $\log_{10}(1-r_{sp})\leq -3$ as numerically converged. 
For systems with multiple sites (mp-1203 and mvc-11115), only the convergence for the first site is shown in Fig.~\ref{fig:gqmax} as the convergence behavior is similar for different sites. For most materials, converged spectra can be obtained with $|\mathbf{G}+\mathbf{q}|_{\textrm{max}}=4$, except for mvc-11115 where a higher value of  $|\mathbf{G}+\mathbf{q}|_{\textrm{max}}=4.2$ is required. In the following calculations, $|\mathbf{G}+\mathbf{q}|_{\textrm{max}}$ is set to 4.2 for mvc-11115, while for the rest of the materials a value of 4 is adopted.

\subsection{\emph{K}-grid Convergence}
\begin{figure}[htbp] 
\includegraphics[width=\textwidth]{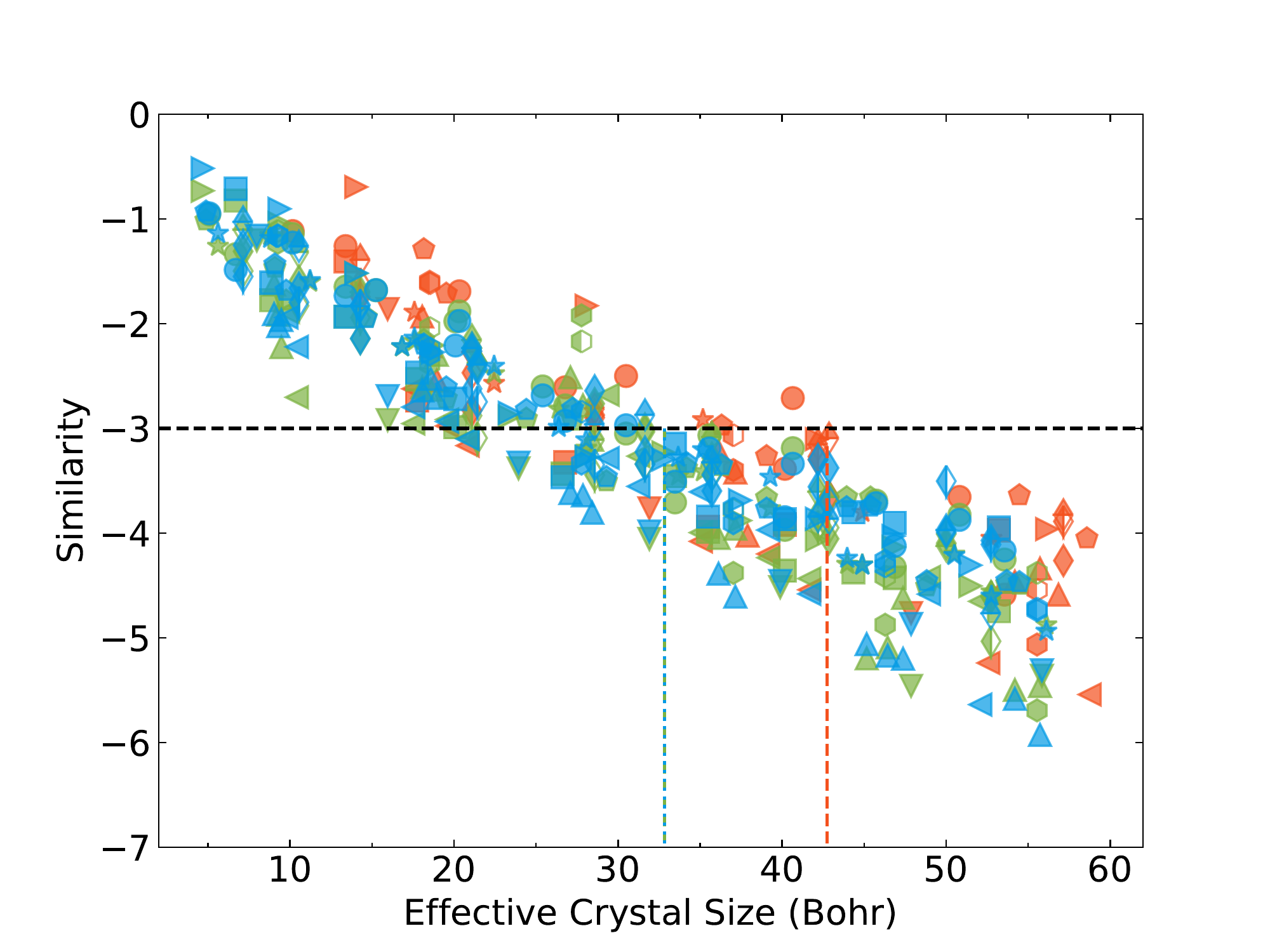}
\caption{\label{fig:convergence} Multi-code \emph{k}-grid convergence as a function of the effective crystal size (see text). The spectra obtained from different codes are represented by different colors: green (\ocean{}), orange (\xs{}) and blue (\exciting{}). The symbols designating the materials are the same as in Fig.~\ref{fig:gqmax}. Partially filled symbols with the same shape represent different absorption sites of the material. Vertical dashed lines denote the effective crystal size beyond which the similarity criterion is fully satisfied for each code. %Different materials are denoted by different markers: circle for mp-390, star for mp-2657, left triangle for mp-1840, up triangle for mp-430, down triangle for mp-458, right triangle for mp-2664, square for mp-10734, diamond for mp-1203, hexagon for mvc-11115 and pentagon for mp-1215. 
%\todo{Add labels of the materials to the figure.}
} 
\end{figure}

%Since the three codes used in this benchmark have different methodologies, it is expected that the convergence behavior would behave differently. As aforementioned, we use the Spearman's rank correlation coefficient ($S$) to characterize the convergence between two spectra. Since the $S$ between two spectra is always positive and close to 1, we further introduce the quantity $\mathrm{log}(1-S)$ to help determine the convergence of the spectra. If $\mathrm{log}(1-S)$ is -1, it means $S$ is around 0.9. The absolute value of the quantity $\mathrm{log}(1-S)$ can be regarded as count the number of nines after decimal. The more nines, the more converged spectra we have achieved.
We perform the \emph{k}-grid convergence study on the Ti-O-10 dataset against $R_\mathrm{eff}$ in a wide range up to about 60 Bohr.
For each material, we choose an over-converged dense \emph{k}-mesh as the reference, corresponding to  $R_\mathrm{eff}$ values between $55$ to $65$ Bohr. For a given $R_\mathrm{eff}$, there exist multiple choices of the \emph{k}-grid $\{n_i\}$, and we choose the smallest one ensuring  similar spacings between reciprocal lattice vectors. The calculated spectra on the coarse \emph{k}-grids are compared to the reference spectra, and we choose $s \leq -3$ as the threshold for convergence (see Fig.~S1). As shown in Fig.~\ref{fig:convergence}, overall the slopes for the three codes are similar. For each code, we perform a linear regression for data from each site and determine the intersection between the linear fit with the horizontal line of $s=-3$. The largest $R_\mathrm{eff}$ value at the intersection among all the absorption sites gives the convergence threshold of $R_\mathrm{eff}$, as indicated by green (\ocean{}), orange (\xs{}) and blue (\exciting{}) dashed lines in Fig.~\ref{fig:convergence}.  Out of the three, \ocean{} and \exciting{} converge at a similar pace ($R_\mathrm{eff} > 32.8$ Bohr), while \xs{} converges more slowly ($R_\mathrm{eff} > 42.7$ Bohr). The different \emph{k}-grid convergence behavior is likely due to the major difference in the methodology --- \ocean{}/\exciting{} perform BSE calculations on the unit cell and \xs{} performs CHP calculations on the supercell containing a core hole.

%, except for mvc-11115, where a larger $R_\mathrm{eff}$ is required due to an outlier at $R_\mathrm{eff}=27.8$ Bohr near the convergence threshold. A close examination of the k-grid convergence of mvc-11115 shows that the spectra are in good agreement with the reference, when $R_\mathrm{eff}\geq 18.5$ Bohr, and the small discrepancy mainly comes from the slightly larger peak intensity at $10.5$ eV (see Fig.~\ref{fig:mvc11115}a). The similarity at $R_\mathrm{eff}=27.8$ Bohr ($s=-2.93$) is slightly worse than that at $R_\mathrm{eff}=18.5$ Bohr ($s=-3.14$). A linear regression of the mvc11115 k-grid convergence data suggests the converged k-grid at around $R_\mathrm{eff}=21.1$ Bohr (see Fig.~\ref{fig:mvc11115}b). \xs{} and \exciting{} converge much slower, which requires $R_\mathrm{eff} > 43$ Bohr and $46$ Bohr, respectively.
%\todo{these two, mp-1215 and mp-2664, have the largest Max D for groundstate comparison, any correlation?}

\subsection{Cross-code Spectral Comparison} \label{sec:spectral_comparison}
For the cross-code comparison, one option is to use spectra computed at the converged \emph{k}-grid based on $R_\mathrm{eff}$ (see above section). However, any similarity measure better than $-3$ is not guaranteed to be meaningful. For this reason, we choose to compare the reference spectra obtained from the densest \emph{k}-grids used in the three codes. In a pairwise comparison, we first align two spectra truncated at 35 eV above the onset (see Section~\ref{sec:alignment}) and then calculate the similarity of the aligned truncated spectra.
%Since the three codes employ different methods, there exists differences of the energy scale for the spectra plots. Before calculating the similarity metrics (Spearman's rank correlation coefficient for example in this work), we first aligned the spectra by minimizing the Cosine similarity. 
The explicit number of bands needs to be provided in \ocean{} and \exciting{} calculations to yield a large enough energy range of about 35 eV. In this work, we estimate the required number of conduction bands based on the homogeneous electron gas model (see Section~\ref{sec:workflow}). The model turns out to work quite well for most of the systems, except for mp-2664, which has the smallest unit cell out of the 10 materials. For mp-2664, we manually increase the number of bands to ensure a sufficiently large energy range (see Appendix~\ref{app:bands}). In Fig.~\ref{fig:spectra}, we compare the spectra from the three codes for two representative systems: mp-2657 (rutile TiO$_2$, insulator) and mp-458 (Ti$_2$O$_3$, metal). We present the results on both the initial state rule (independent-particle approximation) and final state rule (including electron-hole interactions). Under the initial state rule, the core-hole effects are not considered, and the spectra reflect only the band structure effects. In practice, we switch off the exchange and direct coupling terms in BSE in \ocean{} and \exciting{}, and use the neutral Ti pseudopotential instead of the core-hole pseudopotential in \xs{}. Under the final state rule, the spectra correspond to the BSE or CHP Hamiltonian described in Section~\ref{sec:xas}. The full spectral similarity measure of the Ti-O-10 dataset (including all 3 code pairs) is summarized in Table~\ref{tab:spearman}.

The calculations for all three codes were carried out to converge the spectra up to 35 eV above the edge. As seen in Fig.~\ref{fig:spectra}, deviations start to appear above 35 eV as the impact of the finite number of empty states starts to emerge for \ocean{} and \exciting{}. The features in the high energy tail where the spectral intensity vanishes are very close in \ocean{} and \exciting{}, which is consistent with the good agreement of the band structure shown in Table~\ref{tab:groundstate}. On the other hand, \xs{} spectra extend into a much larger energy range because of the use of the Lanczos formalism that does not require empty states explicitly. At the independent particle level, the spectral shapes show nearly perfect agreement as can be seen in Figs.~\ref{fig:spectra}a and \ref{fig:spectra}c, respectively. The \xs{} \emph{vs.} \ocean{} comparison reveals high similarities of $s=-3.89$ for mp-2657 and $-$3.98  for mp-458, as expected within the pseudopotential codes. The \xs{} results overall show slightly stronger intensity than \ocean{} and \exciting{}. This is likely due to the use of different pseudopotentials, which can affect the shape of the conduction band wavefunctions as well as the transition matrix elements that depend on choice of the projectors. The convergence criteria based on Spearman’s rank correlation is more sensitive to spectral shape and less so to small differences in the absolute spectral intensity. The similarity of \exciting{} {\textit vs.} \ocean{} and \xs{} {\textit vs.} \exciting{} are slightly worse with $s=-3.17$ and $-$3.08 for mp-2657 and $-$2.71 and $-$2.59 for mp-458, respectively. Across the Ti-O-10 dataset, the average similarity at the independent particle level is $-$3.74 between \ocean{} and \xs{} and becomes slightly worse between \ocean{} / \xs{} and \exciting{} (better than $-$2.76) as shown in Table~\ref{tab:spearman}.

Overall, there is good agreement between the three codes at the interacting particle level, as shown in Figs.~\ref{fig:spectra}b and \ref{fig:spectra}d. We notice that better agreement is obtained from the two BSE codes and the treatment of core electrons (pseudopotential {\textit vs.} all electron) has a very small impact. As a result, noticeable differences emerge between \ocean{} / \exciting{} and \xs{}, especially regarding peak intensities. For example, near 20 eV in mp-2657, the first peak is higher than the second peak in the \ocean{} / \exciting{} spectra, while it is the opposite in the \xs{} spectrum. In addition, the shoulder peak in \ocean{} / \exciting{} near 10 eV is higher by about 22\% than that in \xs{}. As a result, \ocean{} and \exciting{} have high similarity values of $-$2.79 in mp-2657 and  $-$2.95 in mp-458, while $r_{sp}$ between \xs{} / \ocean{} and \exciting{} drops by an order of magnitude as compared to the independent particle level,  with similarity scores of $-$2.00 and $-$2.17 in mp-2657 and $-$2.79 and $-$2.71 in mp-458, respectively. From Table~\ref{tab:spearman}, one can see that at the interacting particle level the average similarity is $-$2.65 between \ocean{} and \exciting{}, which is slightly worse than the value of $-$2.93 at the independent particle level. This small additional deviation most likely results from the different numerical implementation of the BSE Hamiltonian, especially the dielectric screening (see Section~\ref{sec:implmentation}). The average similarity is about $-$2.02 between \xs{} and \ocean{} / \exciting{}. We do not find any substantial difference in the overall agreement between insulators and metals.

\begin{figure}[htbp] 
\includegraphics[width=\textwidth]{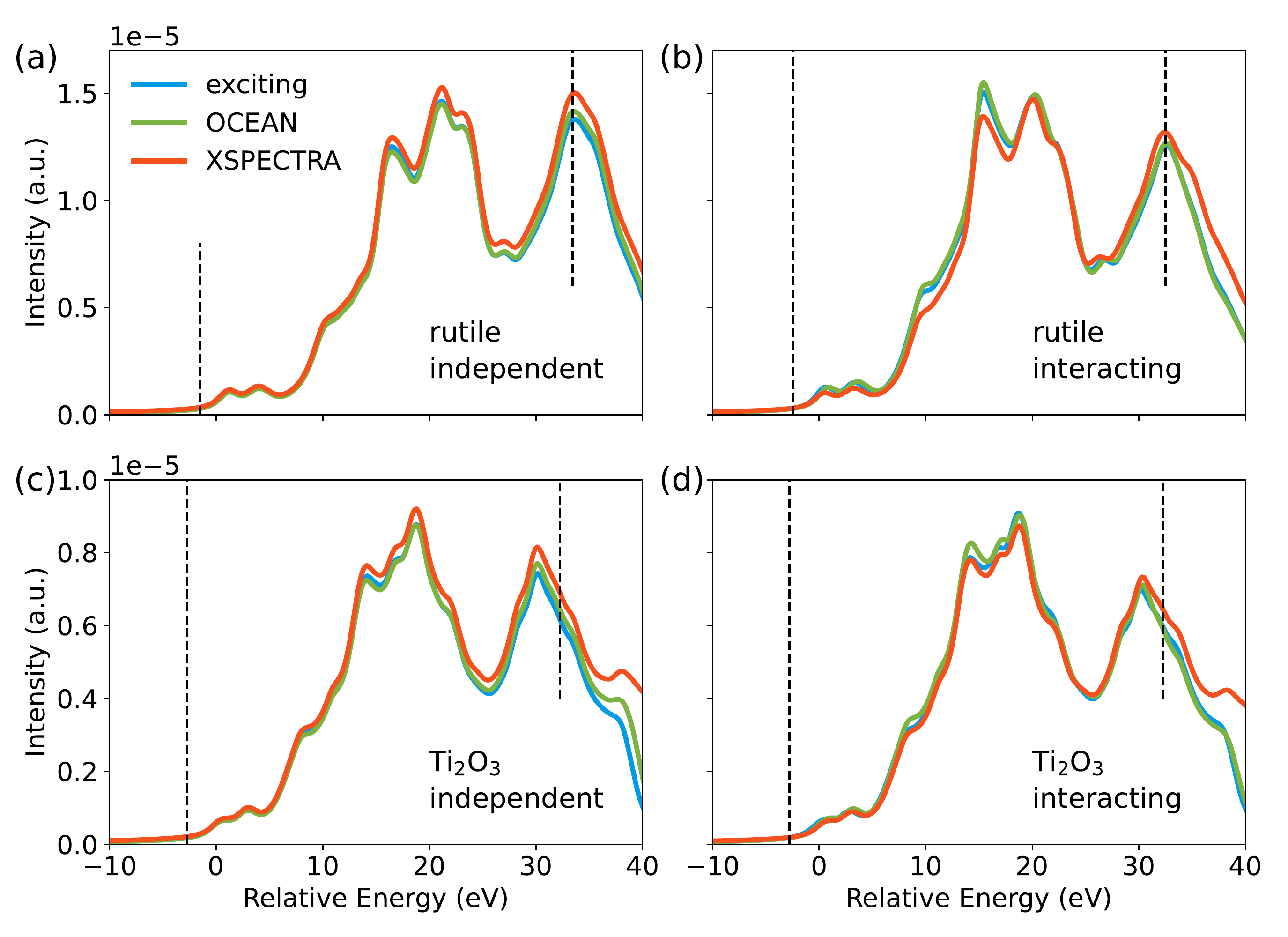}
\caption{\label{fig:spectra} Comparison of the spectra of rutile (mp-2657 ) and Ti$_2$O$_3$ (mp-458) from all three codes at both the independent and interacting particle level. The first and second vertical dashed lines indicate the absorption onset and 35 eV above it, where a quantitative comparison was performed. The energy is relative to the excitation to the CBM.}
\end{figure} 

\begin{table}[!htbp]
\caption{Spectral similarity of the Ti-O-10 dataset among the three codes at both the independent and interacting particle level. Values in the different columns refer to the comparison between \xs{} and \ocean{}, \exciting{} and \ocean{}, \exciting{} and \xs{}, respectively.}
\begin{ruledtabular}
\label{tab:spearman}
\begin{tabular}{cccc}
\textrm{mpid}& 
\textrm{\exciting{}  vs. \ocean{}}&
\textrm{\xs{}  vs. \ocean{}}&
\textrm{\xs{}  vs. \exciting{}}\\
      & independent \quad interacting & independent \quad interacting & independent \quad interacting \\
\colrule

mp-390    & -2.80 \quad -2.33 & -3.72 \quad -1.49 & -2.69 \quad -1.63  \\
mp-2657   & -3.17 \quad -2.79 & -3.89 \quad -2.00 & -3.08 \quad -2.17  \\
mp-1840   & -3.25 \quad -2.73 & -3.93 \quad -2.02 & -3.05  \quad -2.26  \\
mp-430    & -3.53 \quad -2.56 & -3.82 \quad -1.91 & -3.30  \quad -2.09  \\
mvc-11115:0 & -3.49 \quad -1.73 & -4.08 \quad -1.49 & -3.23  \quad -1.16  \\
mvc-11115:1 & -2.99 \quad -2.29 & -3.78 \quad -1.55 & -2.75  \quad -1.43  \\
mp-1203:0 & -2.82 \quad -2.87 & -3.63 \quad -2.08 & -2.67  \quad -2.16  \\
mp-1203:2 & -2.32 \quad -2.85 & -3.40 \quad -2.30 & -2.17  \quad -2.32  \\
mp-1203:4 & -2.78 \quad -2.83 & -3.20 \quad -2.01 & -2.48  \quad -1.92  \\
mp-10734  & -2.85 \quad -2.49 & -3.75 \quad -1.91 & -2.71  \quad -2.10  \\
mp-1215   & -2.72 \quad -2.96 & -3.66 \quad -2.40 & -2.62  \quad -2.16  \\
mp-2664   & -2.64 \quad -3.01 & -3.83 \quad -2.31 & -2.53  \quad -2.19  \\
mp-458    & -2.71 \quad -2.95 & -3.98 \quad -2.79 & -2.59  \quad -2.71  \\
\colrule
average & -2.93 \quad \textbf{-2.65} & \textbf{-3.74} \quad -2.02 & -2.76 \quad -2.02
\end{tabular}
\end{ruledtabular}
\end{table}

\if{0}
\subsection{Edge Alignment}
In Table~\ref{tab:cls}, we compare the edge alignment of the three codes. Because the three spectra for the same material have quite different energy ranges after the edge alignment, in the pair-wise comparison, we shift the first spectrum to align with the second one by maximizing the Cosine similarity between the two spectra as described in Section ~\ref{sec:alignment}. For example, in the ``\ocean{} {\it vs.} \exciting{}" column in Table~\ref{tab:cls}, the \exciting{} spectra are used as the reference. Then for each column the relative shifts ($\Delta \Delta \epsilon_\alpha$) are reported after subtracting the mean of the column (see Section ~\ref{sec:alignment}). A small standard deviation indicates that two codes yield similar edge alignment. As expected, the closest agreement is found between \exciting{} and \ocean{} without the adiabatic screening correction (the $-\frac{1}{2}W_c$ term), as both codes compute $\epsilon_\alpha$ at the ground state. 
Comparing the alignment of the independent particle spectra between \ocean{} and \exciting{} we find the standard deviation to be 0.10~eV (see Table~S1 in the Supplemental Material~\cite{SI}). We attribute this to the use of the frozen-core approximation in \ocean{}. Including interactions increases the discrepancy between \ocean{} and \exciting{} to 0.19~eV, most probably due to differences in the numerical implementation of the BSE between these two codes. 
%This comparison shows that the frozen core hole approximation is responsible for a small, but non-negligible error with a standard deviation of 0.19~eV. For energy alignment less than 0.19~eV, both \xs{} and \ocean{} may require an additional correction to account for the frozen core. 
After including the $-\frac{1}{2}W_c$ term in \ocean{}, the standard deviation between \exciting{} and \ocean{} increases to 0.91~eV, which highlights the significant contribution of the adiabatic screening effect of the core hole in the edge alignment. 

A similar magnitude in the standard deviation among the codes, ranging from $0.92$~eV to $1.56$ eV, is found in the other columns. Such degree of difference is expected, since the three chemical shifts are obtained with very different approximations: \exciting{} and \ocean{} are under the initial-state rule, \xs{}  includes final-state effects through a $\Delta$-SCF procedure. Without benchmarking against a set of reliable experimental results, it is not possible to state which, if any, of the three codes gives the best estimates for the core-level shifts. This is beyond the scope of the current work and could be subject of future research.

\begin{table}[!htb]
\caption{Difference in the relative core-level shifts in eV between code pairs in the interacting particle spectra. Comparisons with \ocean{} are done with (w/) and without (w/o) including the $-\frac{1}{2}W_c$ term to the core hole initial state (see text).}
\begin{ruledtabular}
\label{tab:cls}
\begin{tabular}{lccc}
\textrm{mpid:site}& 
\ocean{} {\it vs.} \exciting{}&
\ocean{} {\it vs.} \xs{}&
\xs{} {\it vs.} \exciting{}
\\
           & w/ $-\frac{1}{2}W_c$ \quad w/o $-\frac{1}{2}W_c$ & w/ $-\frac{1}{2}W_c$ \quad w/o $-\frac{1}{2}W_c$  \\
\colrule

mp-390    & $-$0.95 \quad\quad $-$0.11 & $-1.49$ \quad\quad  $-0.65$ & \;\;\;0.53  \\
mp-2657   & $-0.87$ \quad\quad $-$0.19 & $-1.73$ \quad\quad  $-1.05$ & \;\;\;0.86  \\
mp-1840   & $-0.76$ \quad\quad \;\;\;0.19 & $-1.68$ \quad\quad $-0.72$  & \;\;\;0.92  \\
mp-430    & $-0.56$ \quad\quad \;\;\;0.27 & $-1.47$ \quad\quad  $-0.63$  & \;\;\;0.87  \\
mvc-11115:0 & $-1.18$ \quad\quad $-$0.35 & $-1.30$ \quad\quad $-0.47$ &  \;\;\;0.23  \\
mvc-11115:1 & $-1.22$ \quad\quad $-$0.30 & $-1.27$ \quad\quad $-0.35$ &  \;\;\;0.06  \\
mp-1203:0 & \;\;\;0.75 \quad\quad \;\;\;0.09 & \;\;\;1.62 \quad\quad  \;\;\;0.96  & $-$0.87  \\
mp-1203:2 & \;\;\;0.81 \quad\quad \;\;\;0.08 & \;\;\;1.22 \quad\quad  \;\;\;0.49  & $-$0.39  \\
mp-1203:4 & \;\;\;0.57 \quad\quad $-$0.08 & \;\;\;2.27 \quad\quad  \;\;\;1.61  & $-$1.73  \\
mp-10734  & \;\;\;0.84 \quad\quad \;\;\;0.13 & \;\;\;0.35 \quad\quad  $-0.36$  & \;\;\;0.48  \\
mp-1215   & \;\;\;0.64 \quad\quad \;\;\;0.03 & \;\;\;2.34 \quad\quad  \;\;\;1.73  &   $-1.73$ \\
mp-2664   & \;\;\;1.10 \quad\quad \;\;\;0.16 & \;\;\;1.02 \quad\quad  \;\;\;0.07  & \;\;\;0.07  \\
mp-458    & \;\;\;0.82 \quad\quad \;\;\;0.07 & \;\;\;0.13 \quad\quad  $-0.62$  & \;\;\;0.69  \\
\hline
Standard deviation & \;\;\;{\bf 0.91} \quad\quad \;\;\;{\bf 0.19} & \;\;\;{\bf 1.56} \quad\quad \;\;\;{\bf 0.92} &  \;\;\;{\bf 0.93} \\
\end{tabular}
\end{ruledtabular}
\end{table}
\fi

 %\subsection{Comparison of the final-state effects between the linear response method and the core-hole potential method}
\subsection{Comparison of the final-state effects between the BSE and CHP methods}
\label{sec:final-state}

We observe that for the Ti-O-10 dataset, the results from the two BSE codes, {\sc ocean} and {\texttt exciting}, exhibit more spectral weight near the edge than those from the CHP code, {\sc xspectra} (See Fig.~\ref{fig:spectra} and Fig.~S2 \cite{SI}).  This trend is more pronounced in the insulating systems than in the metallic systems. This is indicative, but not proof, of a stronger core-hole potential, or, equivalently, weaker screening of the core hole in the BSE approach. 
%In order to understand the origin of the discrepancy between the CHP and BSE calculations, we have investigated several different methods for determining the induced density response of the valence and semi-core electrons in response to the core hole. 
To better understand the discrepancies between the CHP and BSE calculations, we examine the differences in the two theoretical approaches.
Since {\sc ocean} and {\sc exciting} results are nearly identical to each other, we will only use {\sc ocean} for this analysis.
 
%As we will show, when BSE calculations with {\sc ocean} use a screened core-hole potential that is similar to the effective screened core-hole potential implicit in {\sc xspects}, the two codes produce spectra in strong agreement with each other.

%We found that using a similar induced density, and hence a similar screened core-hole potential, in {\sc ocean} as what is used in {\sc xspectra} that the two codes produce spectra in strong agreement with each other. 

%Before addressing the calculation of the induced density response, we outline other factors between a CHP and BSE calculation that could give rise to differences in the final spectra. 
The differences between BSE and the final-state rule at the presence of a core hole have been discussed previously by Rehr, Soininen, and Shirley \cite{Rehr_2005}, and three main differences can be identified. 
Both the occupied and unoccupied electron states change when a core hole is created \cite{RevModPhys.62.929}. In CHP calculations, the valence band manifold is relaxed in presence of the core hole, while the BSE calculation uses the eigenstates of ground-state DFT. 
%First, the valence band manifold is different in the CHP calculation as it is relaxed in the presence of the core hole, while the BSE calculation uses the eigenstates of ground-state DFT. 
Although this difference can modify the weights of the transition matrix elements, as shown by Roychoudhury and Prendergast \cite{PhysRevB.107.035146}, this effect is expected to be small for the particular materials and edge studied here. The empty or mostly empty Ti {\it d}-bands sit several eV below the main-edge transitions into the Ti 4{\it p} states that also hybridize with neighboring oxygen orbitals. Additionally, Liang {\it et al.} investigated the role of secondary electron-hole excitations in a $\Delta$SCF context, and found that they played only a small role in the O K-edge XAS of TiO$_2$ (in contrast to later transition metal oxides) \cite{PhysRevLett.118.096402,PhysRevB.97.205127}. Second, the BSE Hamiltonian includes an exchange interaction between the excited electron -- core-hole pairs that is absent in CHP. The importance of this term can be investigated by selectively turning it off. 
Third, the screened core-hole potential is different between the CHP and BSE methods, owing to differences in the calculation of the induced density response to the core hole. Using \ocean{} we investigate several different approximations for calculating the induced density response.

%The difference between final-state rule (FSR) and BSE has been discussed previously by Rehr, Shhirley and Soininen [J J Rehr et al 2005 Phys. Scr. 2005 207].

\subsubsection{Induced density in response to the core-hole potential}
%Both the CHP and BSE approaches use a static approximation for the screening of the core-hole potential. The relevant energy is not the x-ray energy, but the energy that is exchanged through the direct interaction \cite{PhysRevB.29.5718,PhysRevB.67.115120}.
In response to the creation of a core hole, the electrons relax, screening the core-hole potential. 
In the CHP approach the core-hole potential and valence screening are all included in the self-consistent DFT calculation. In the BSE approach, electrons interact with the core hole through the screened potential $W$, which can be written using the dielectric response $\epsilon$ or in terms of an induced density $\rho_\mathrm{ind}$ and induced potential $V_\mathrm{ind}$ \cite{RevModPhys.74.601},
\begin{align}
W({\mathbf r}) &= \epsilon^{-1}({\mathbf r},{\mathbf r}') V_{\mathrm{ext}}({\mathbf r}') = V_{\mathrm{ext}}({\mathbf r}) + V_{\mathrm{ind}}({\mathbf r})\\
\rho_\mathrm{ind}({\mathbf r}) &= \chi({\mathbf r},{\mathbf r}') V_{\mathrm{ext}}({\mathbf r}') \\
V_\mathrm{ind}({\mathbf r}) &= v({\mathbf r},{\mathbf r}') \rho_\mathrm{ind}({\mathbf r}') % \\
%W &= ( 1 + v \chi ) v_{\textrm{ext}} \\
%\chi &= (1 - \chi^0 v )^{-1} \chi^0
\end{align}
where $v$ is the Coulomb operator and $\chi$ is the reducible polarizability, and implicit summation is assumed.
Note that both the CHP and BSE approaches use a static approximation for the screening of the core-hole potential \cite{PhysRevB.29.5718,PhysRevB.67.115120}.

Below we consider three methods to construct $V_\mathrm{ind}$. In the first method, (as is typical for a BSE calculation), the reducible polarizability is calculated within the RPA, 
\begin{align}
%\chi^0(1,2) &= -i G(1,2)G(2,1) \\
\chi^{\mathrm{RPA}} &= (1 - \chi^0 v )^{-1} \chi^0
\end{align}
where $\chi^0$ is the irreducible polarizability. In the second method, the many-body effects can be included through the exchange-correlation kernel, $f_{xc}$, 
\begin{equation}
\chi = [1 - \chi^0 (f_{xc} + v) ]^{-1} \chi^0
\end{equation}
where $V_{xc}$ is the exchange-correlation potential and $f_{xc} = \delta V_{xc} / \delta \rho$~\cite{PhysRevB.34.5390}. Under the linear response framework, normally the static approximation is also used, i.e., an adiabatic $f_{xc}$. Within {\sc ocean}, $f_{xc}$ can be included within the adiabatic local-density approximation (ALDA) \cite{PhysRevB.103.245143}. 

A third method for determining the screened potential is based on  the CHP method, where a self-consistent DFT calculation is carried out with a core hole and an excited electron at the bottom of the conduction band. 
%This screened core hole can also be used within a BSE code by noting that the induced density is the difference between two self-consistent DFT calculations, with (CH) and without the core hole (GS),
The induced density is calculated as the density difference between two self-consistent DFT calculations, with (CH) and without the core hole,
\begin{equation}
\rho_\mathrm{ind}^{\mathrm{CH}} = \rho^{\mathrm{CH}} - \rho^{\mathrm{GS}},
\end{equation}
where GS stands for ground state. This approach has been used previously for linear-response calculations by applying sufficiently small perturbing potentials~\cite{PhysRevB.31.5305,PhysRevB.29.7045}, but here we use the screening potential from a full core-hole. 

Both the linear response and self-consistent approaches entail approximations. In the case of linear response, the bare core-hole potential is not weak, and non-linearity in the screening response may be important. In the case of the self-consistent approach, the resulting screened potential is due to the electrons relaxing into the lowest energy configuration (in the presence of the core hole), a process that is not instantaneous. However, the relevant time scale for the screening is the inverse of the plasmon frequency ($\approx$ 10~eV to 20~eV) as compared to the core hole lifetime (inverse of the 1~eV broadening). These competing time scales apply equally to the linear response case. 
In some special cases, such as those with localized orbitals or defects, there may also be very slow screening processes that are incorrectly included in the self-consistent core-hole potential. However, this is not the case in the systems we are investigating here. 
%One can conceive of a system where the highest occupied orbital is a localized around a defect in the neighborhood of the absorbing atom where the time needed for the electron to hop over to the absorbing atom is longer than the core-hole lifetime, making a self-consistent calculation qualitatively incorrect, but this contrived example does not resemble the systems we are investigating here. 
The issue of screening time can be side-stepped entirely within a real-time formalism, such as real-time time-dependent DFT (RT-TDDFT) \cite{doi:10.1021/acs.chemrev.0c00223}, though x-ray implementations of RT-TDDFT tend to use non-periodic boundary conditions making comparisons with periodic boundary condition code calculations of extended systems difficult.

\subsubsection{Comparison of XANES spectra using different approximations}

\begin{figure}
\includegraphics[height=6.2in]{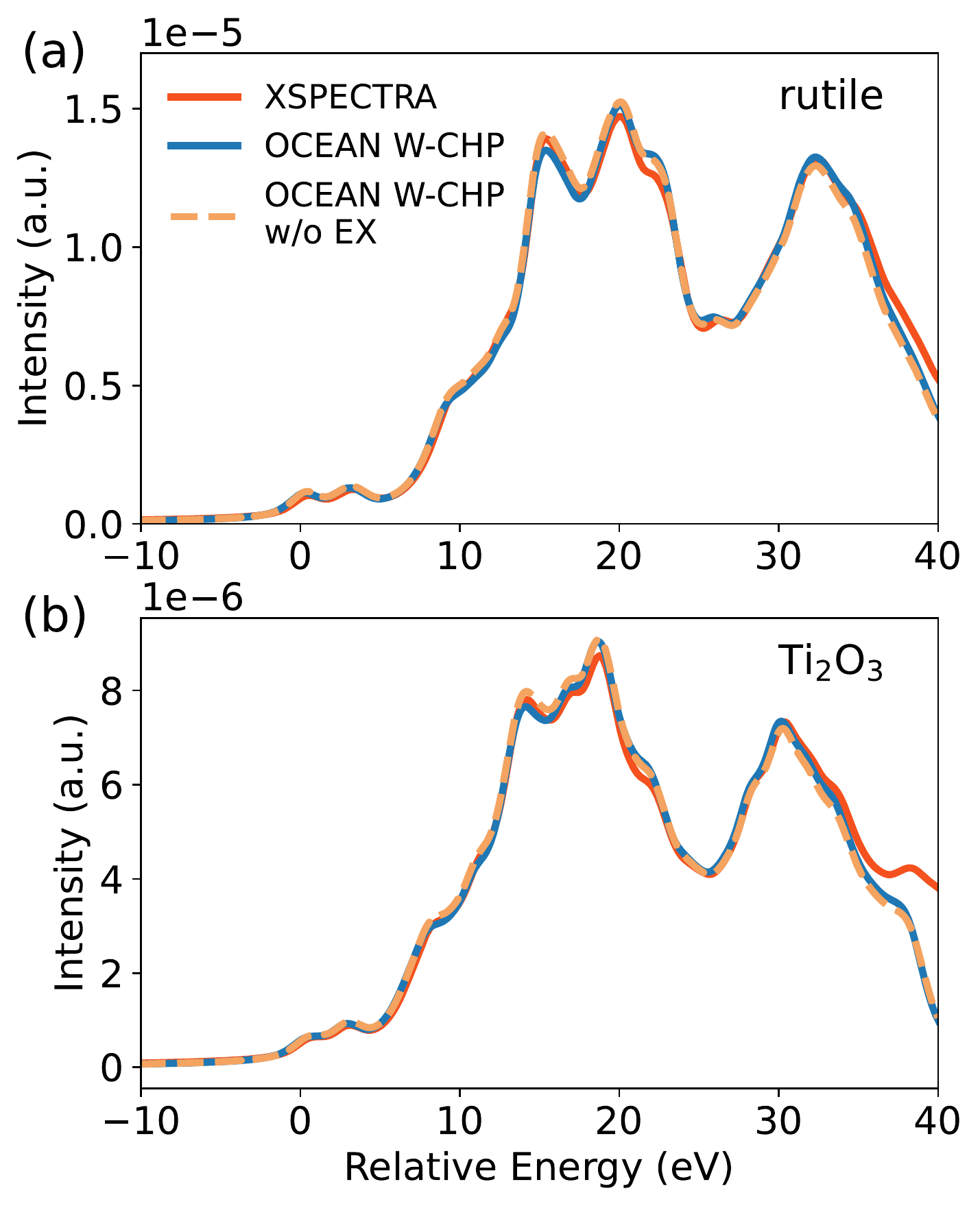}\\ % Here is how to import EPS art \\
\caption{Comparison of XAS spectra of TiO$_2$ (rutile, mp-2657) (a) and Ti$_2$O$_3$ (mp-458) (b) between {\sc xspectra} and {\sc ocean} using a screened core-hole based on the CHP method (W-CHP) with and without the exchange interaction (W-CHP w/o EX).
}
\label{rutile-exchange}
\end{figure}

\begin{figure}
  \includegraphics[height=4.5in]{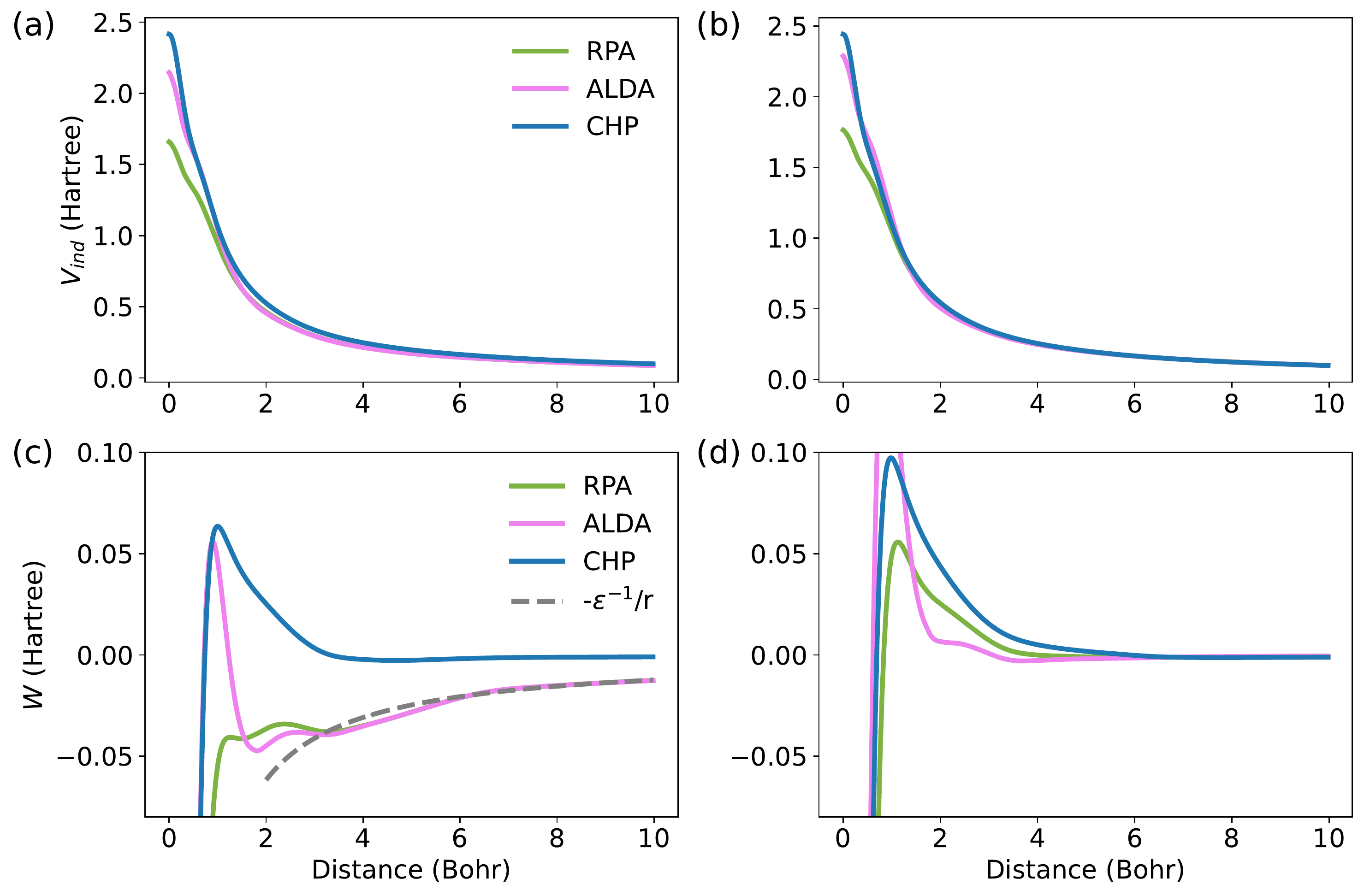}
\caption{Comparison of the induced potentials in response to a Ti 1{\it s} core hole in (a) rutile TiO$_2$ (mp-2657) and (b) Ti$_2$O$_3$ (mp-458) using three different methods for determining the induced density response. The RPA and CHP calculations of the induced potentials for the metallic Ti$_2$O$_3$ are in closer agreement than for TiO$_2$. The total screened core-hole potential is shown for (c) TiO$_2$ and (d) Ti$_2$O$_3$. The plots are zoomed in to show detail, with the full core-hole potential approaching 18~Hartree at the origin. For the metallic Ti$_2$O$_3$, all three approaches efficiently screen the core-hole potential, resulting in a total potential that is nearly zero at intermediate ranges. In the case of the insulating TiO$_2$, the two linear-response approaches both give the expected $-\epsilon_\infty^{-1}/r$ behavior at large distance. The grey dashed guide line uses the isotropic $\epsilon_\infty=8.1$ taken from the Materials Project \cite{jain2013commentary}.}
\label{induced-potentials}
\end{figure}

We first investigate the effect of the exchange interaction using a metal (Ti$_2$O$_3$, mp-458) and an insulator (rutile, mp-2657). {\sc ocean} calculations using the CHP-derived screened core-hole potential (W-CHP) are shown in Fig.~\ref{rutile-exchange} both with and without the exchange (W-CHP w/o EX). 
As the exchange term is repulsive, it slightly shifts spectral weight away from the edge, e.g., the small intensity decrease at the first peak in W-CHP as compared to W-CHP w/o EX, but it has little effect on the pre-edge features. While these are dipole transitions enabled due to hybridization between the titanium and neighboring oxygen atoms, the states are primarily 3{\it d} in nature, and the exchange terms between the 1{\it s} and 3{\it d} are small. This comparison highlights the limitation of a local or semi-local exchange potential in capturing the exchange interaction between a core hole and photoelectron. While small in the case of the K edge, the exchange term is vital for reproducing even qualitatively correct spectra for 3{\it d} transition metal L edges \cite{SHIRLEY2005}. The contribution of the exchange is associated with the local field effect of the dielectric response, which could play an important role in low-dimensional systems.

{\sc xspectra} and W-CHP w/o EX are the same level of theory, and, as we see in Fig.~\ref{rutile-exchange}, they produce nearly perfect agreement in rutile (mp-2657) and Ti$_2$O$_3$ (mp-458), except the small intensity differences at the second peak and the shoulder a couple of eV higher in energy. This indicates that, for this system, the unoccupied 4{\it p} orbitals that are probed in XAS are relatively unaffected by the self-consistent relaxation of the valence bands in the presence of the core hole. In the pre-edge region there is some evidence of these relaxation effects, where the intensities calculated by {\sc xspectra} might be suppressed by increased occupation of the on-site 3{\it d} orbitals due to the excited core electron.

\begin{figure}
\includegraphics[height=6.2in]{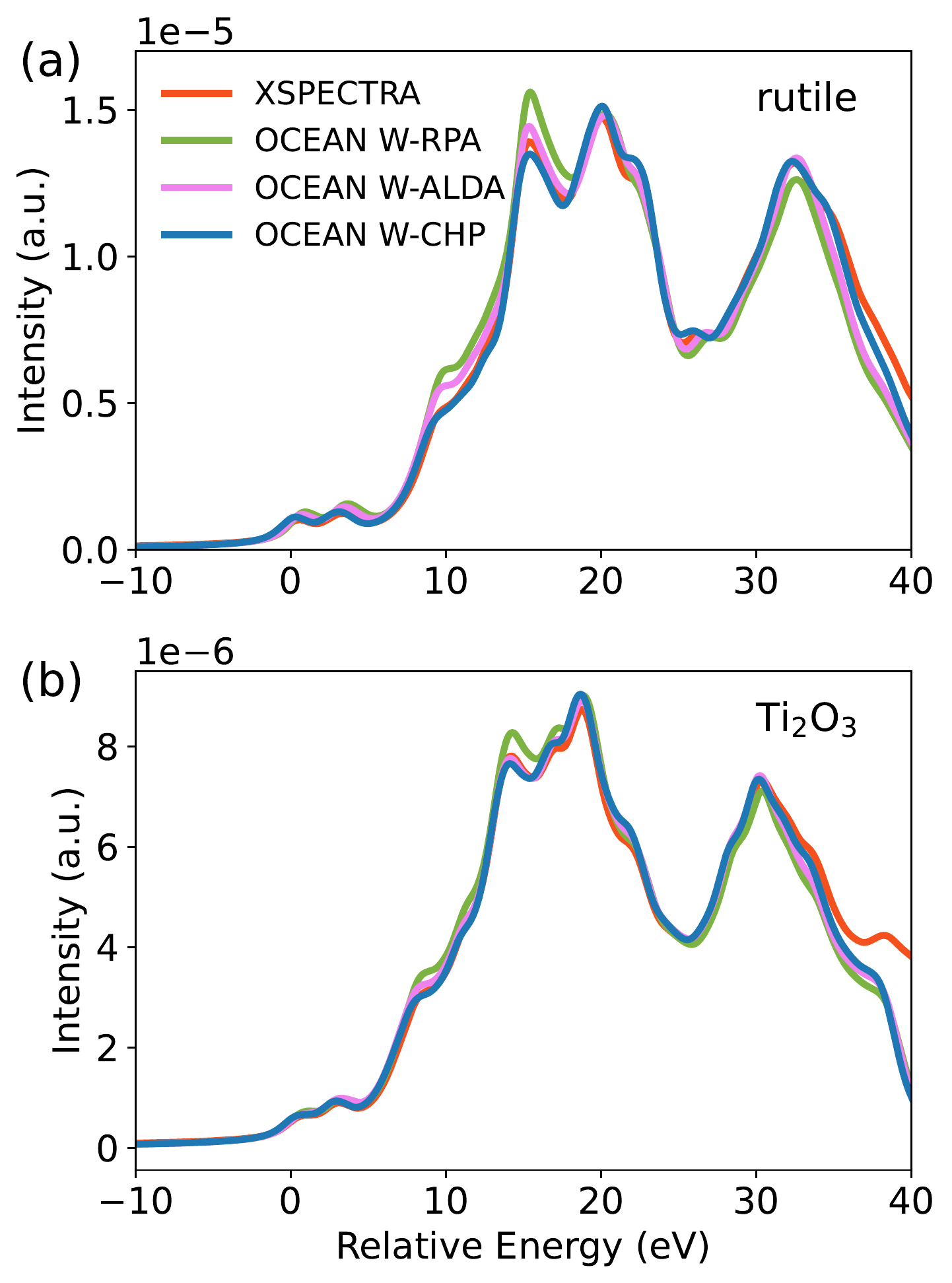}\\ % Here is how to import EPS art \\
\caption{ Comparison of the XAS of (a) rutile (mp-2657) and (b) Ti$_2$O$_3$ (mp-458) calculated using {\sc xspectra} and three different {\sc ocean} calculations under different approximations for the screened Coulomb interaction. The metallic Ti$_2$O$_3$ shows the same trends as the insulating rutile TiO$_2$, though the differences between the W-ALDA and W-CHP appear smaller.
}
\label{rutile-comp}
\end{figure}

Next we compare the three different methods for generating the screened core-hole potential $W$ introduced above: W-RPA, W-ALDA, and W-CHP.
%Next we compare {\sc xspectra} with {\sc ocean} using all three of the different methods for determining the screened core-hole $W$ identified above: W-RPA, W-ALDA, and W-CHP. Note, that these are all still BSE calculations, and that W-RPA is what was used for all previous {\sc ocean} calculations in this work. 
In Fig.~\ref{induced-potentials}a and \ref{induced-potentials}b, we show the three different induced potentials for rutile (mp-2657: insulator) and Ti$_2$O$_3$ (mp-458: metal). Both the CHP and ALDA induced potentials are noticeably stronger than the RPA one, leading to a reduction in the strength of the screened core-hole potential. This causes a reduction in strength of the resulting spectra near threshold. In the case of rutile TiO$_2$, visible discrepancies can be seen in the intermediate range up to around 6 Bohr, while in the metallic Ti$_2$O$_3$ (mp-458), the three approximations for the induced potential are all in agreement starting relatively close to the core hole, at approximately 1.5 Bohr. We further examine the trend in the total potential $W$ in Fig.~\ref{induced-potentials}c and \ref{induced-potentials}d. For the insulating TiO$_2$ case, the CHP approximation has a finite range, as the excited electron is retained in this approximation leading to local neutrality near the excited Ti site. In contrast, the BSE calculations consider the core hole charge explicitly, leading to the expected long range limit of $-\epsilon_{\infty}^{-1}/r$. %the long-range tail of the screened core-hole potential in the CHP approximation is notably weaker than the other two. This is because the CHP method entails perfect screening at a short distance of a few Ti-O bond lengths and $W$ can be understood as a local defect potential, while the linear response methods reproduce the response of a gapped system with macroscopic dielectric constant $-\epsilon_{\infty}^{-1}/r$ at the long range limit. 
In the metallic Ti$_2$O$_3$ (mp-458), as expected, all three total potentials are short-ranged.

In Fig.~\ref{rutile-comp}a we show a comparison of the polarization-averaged spectra for rutile TiO$_2$ under the dipole approximation, calculated with {\sc xspectra} as well as the three different approximations of $W$ within {\sc ocean}. Both of the modified {\sc ocean} calculations (W-ALDA and W-CHP) show better agreement with {\sc xspectra}, where the second peak has higher intensity as opposed to the previous W-RPA {\sc ocean} calculations (see Fig~6). Compared to W-RPA, the ALDA kernel has a non-negligible effect most noticeably at the first peak and the shoulder at $\approx 10$ eV, which shifts the spectral intensity to higher energy. The W-CHP spectrum exhibits an even lower intensity at the first peak and the shoulder than the W-ALDA spectrum, which can be attributed to the stronger screening potential under the self-consistent relaxation treatment at the presence of a full core hole than that from the linear response treatment as shown in Fig.~\ref{induced-potentials}a.  The same comparison is shown for the metallic Ti$_2$O$_3$ (mp-458) in Fig.~\ref{rutile-comp}b, and similar trends are observed but to a lesser extent, due to the smaller discrepancy in the screening potential in the metallic system as shown in Fig.~\ref{induced-potentials}b. 
%JTV -- For comparison, spectra in Fig.~\ref{rutile-comp} are shifted according to Section~\ref{sec:spectral_comparison} to achieve the best possible match. However, this procedure will lose the information on edge position. In Fig. S6, we plot {\sc ocean} spectra with different approximations in $W$ without shifting the spectra relative to each other. As we can see, W-CHP with and without exchange may underestimate the exciton binding energy as compared to W-RPA, as the pre-edge peaks and main peaks are moved to higher energy. A similar trend is found in Li K-edge XANES~\cite{olovsson2009all}. The W-ALDA spectra, on the other hand, shift the peaks significantly to lower energy.
Previous work showed that the exciton peak position differed between BSE and CHP calculations for the Li K edge of lithium halides \cite{olovsson2009all}. We do not see the same behavior here -- in Fig.~\ref{rutile-comp} peaks from different methods at the edge ($\approx15$~eV) and above the edge (30~eV) are very well aligned. This is due to the much weaker excitonic effects seen in the Ti K edge of titanium oxides, as evidenced by the similarity between the interacting and non-interacting spectra (Fig.~\ref{fig:spectra}) in contrast to the lithium halides (Fig.~2 of Ref.~\onlinecite{olovsson2009all}).

The W-CHP approximation yields a near-perfect screening of the core hole for rutile TiO$_2$ (Fig.~\ref{induced-potentials}(c)), due to the self-consistent density response as expected for an ionic system, where compensating charge resides closely around the absorbing titanium atom \cite{PhysRevB.79.235106}. However, as seen in Fig.~\ref{rutile-comp}(a), the calculations that use this potential ({\sc xspectra} and {\sc ocean} W-CHP) are largely in agreement with the calculations whose screened potential has the correct long-range form (W-RPA and W-ALDA). In addition, practical CHP calculations are also performed with a core hole whose occupancy (strength) can be varied between 1 (full core hole) and 1/2 (half core hole). The applicability of these approximations is system dependent, but the CHP method we use here enjoys success for both metallic and semi-conducting systems \cite{shirley2021core}. 

\subsection{Comparison with Experiment}
\label{expt_compare}
In this section, we compare the simulated spectra of rutile (mp-2657) and anatase (mp-390) to measurements from Carta \emph{et al.}~\cite{carta2015x}. 
%
%For first-principles approaches, like the three codes discussed in this work, the ideal is that only free parameters should be the structure (in other words, the choice of sample). The ground-state and x-ray excited-state electronic structure are functions of the structure, making any changes in the predicted spectrum an indirect result of changes to the input. 
%
%
While both the CHP and BSE methods attempt to simulate the full x-ray spectra, we 
note that there are several important effects that are not considered. These include the physical effects of the quasiparticle lifetime broadening, the effect of vibrational disorder, self-energy corrections, %secondary excitations, 
and satellite features. Below we summarize several limitations of this comparison. First, our calculations are carried out on perfectly ordered structures with the atoms fixed during the x-ray excitation. Including vibrational disorder has been shown to reveal symmetry-forbidden, but experimentally observed, ``dark states,'' suppress expected states~\cite{PhysRevB.90.205207,10.1063/1.4856835,PhysRevB.81.115125}, or change the intensities of pre-edge features~\cite{PhysRevB.81.115125,PhysRevB.98.014111}, while excited-state vibrational coupling can change the shape and energies of peaks~\cite{PhysRevB.98.214305}. A second large approximation is the use of orbitals from semi-local DFT as quasiparticle excitations of the x-ray excited state, neglecting many-body effects that could be incorporated with more accurate functionals~\cite{10.1063/5.0030493}, \emph{GW} self-energy corrections~\cite{PhysRevB.94.035163,doi:10.1073/pnas.2201258119}, the cumulant approximation~\cite{PhysRevB.101.245119}, and multi-configuration methods~\cite{Kariki2022}.
Lastly, the approaches we use here are limited to a single electron or single electron--core-hole pair approximation. It is well known that this approximation can fail for L$_{2,3}$ edges (2{\it p} $\rightarrow$ 3{\it d} transitions) where local $dd^*$ or ligand-to-metal charge transfer excitations play an important role~\cite{DEGROOT2021147061}. This is less of an issue at the K edge due the small quadrupole intensity and the reduced exchange interaction between the 3{\it d} electron and the 1{\it s} core hole as compared to a 2{\it p} hole~\cite{deGroot_2009}.  

%We discuss the remaining discrepancies between theory and experiment, and a systematic benchmark study against experiment is the subject of future work. 
Nevertheless, the comparison conducted here can provide the reader with a basic calibration for the degree to which the present methods capture the main features in experiment over the full XANES photon energy range. With this foundation, future investigations can incorporate further physical effects, building on prior work as well~\cite{PhysRevB.90.205207,10.1063/1.4856835,PhysRevB.81.115125,PhysRevB.81.115125,PhysRevB.98.014111,PhysRevB.98.214305,10.1063/5.0030493,doi:10.1021/acs.jctc.8b00458,PhysRevB.98.214305,PhysRevB.94.035163,doi:10.1073/pnas.2201258119,PhysRevB.101.245119,DEGROOT2021147061, deGroot_2009}. For example, the effects of vibrational disorder can be studied by statistically averaging atomic structures under a given temperature using the workflow developed in this study for each snapshot \cite{PhysRevB.90.205207}. A systematic benchmark study against experiment for a broader range of Titanium oxide polymorphs is the subject of future work.

%suppress expected states~\cite{PhysRevB.90.205207,10.1063/1.4856835,PhysRevB.81.115125}, or change the intensities of pre-edge features~\cite{PhysRevB.81.115125,PhysRevB.98.014111}, while excited-state vibrational coupling can change the shape and energies of peaks~\cite{PhysRevB.98.214305}. A second large approximation is the use of orbitals from semi-local DFT as quasiparticle excitations of the x-ray excited state, neglecting many-body effects that could be incorporated with more accurate functionals~\cite{PhysRevB.98.214305}, GW self-energy corrections~\cite{PhysRevB.94.035163,doi:10.1073/pnas.2201258119}, or the Cumulant approximation~\cite{PhysRevB.101.245119}.
%Lastly, the approaches we use here use a single electron or single electron--core-hole pair approximation. It is well known that this approximation can fail for L$_{2,3}$ edges (2{\it p} $\rightarrow$ 3{\it d} transitions) where local $dd^*$ or ligand-to-metal charge transfer excitations play an important role~\cite{DEGROOT2021147061}. This is less of an issue at the K edge due the small quadrupole intensity and the reduced exchange interaction between the 3{\it d} electron and the 1{\it s} core hole as compared to a 2{\it p} hole~\cite{deGroot_2009}.  

In this comparison, we present results including both dipole and quadrupole contributions from only {\sc ocean} and {\sc xspectra} because the results from the two BSE codes have been shown to be nearly identical. Since both rutile and anatase belong to the $4/mmm$ point group, the powder average of the quadrupole term is performed by a weighted average of three specially chosen photon orientations ($\{\mathbf{e}, \mathbf{q}\}$)~\cite{Brouder_1990}. The choice of photon orientations is given in Appendix~\ref{app:quadrupole}.
%($\mathbf{e}_i$, $\mathbf{q}_i$), where $i=1, 2, 3$~\cite{Brouder_1990}. Details of the derivation is given in Appendix~\ref{app:quadrupole}.

\begin{figure}
\includegraphics[height=3in]{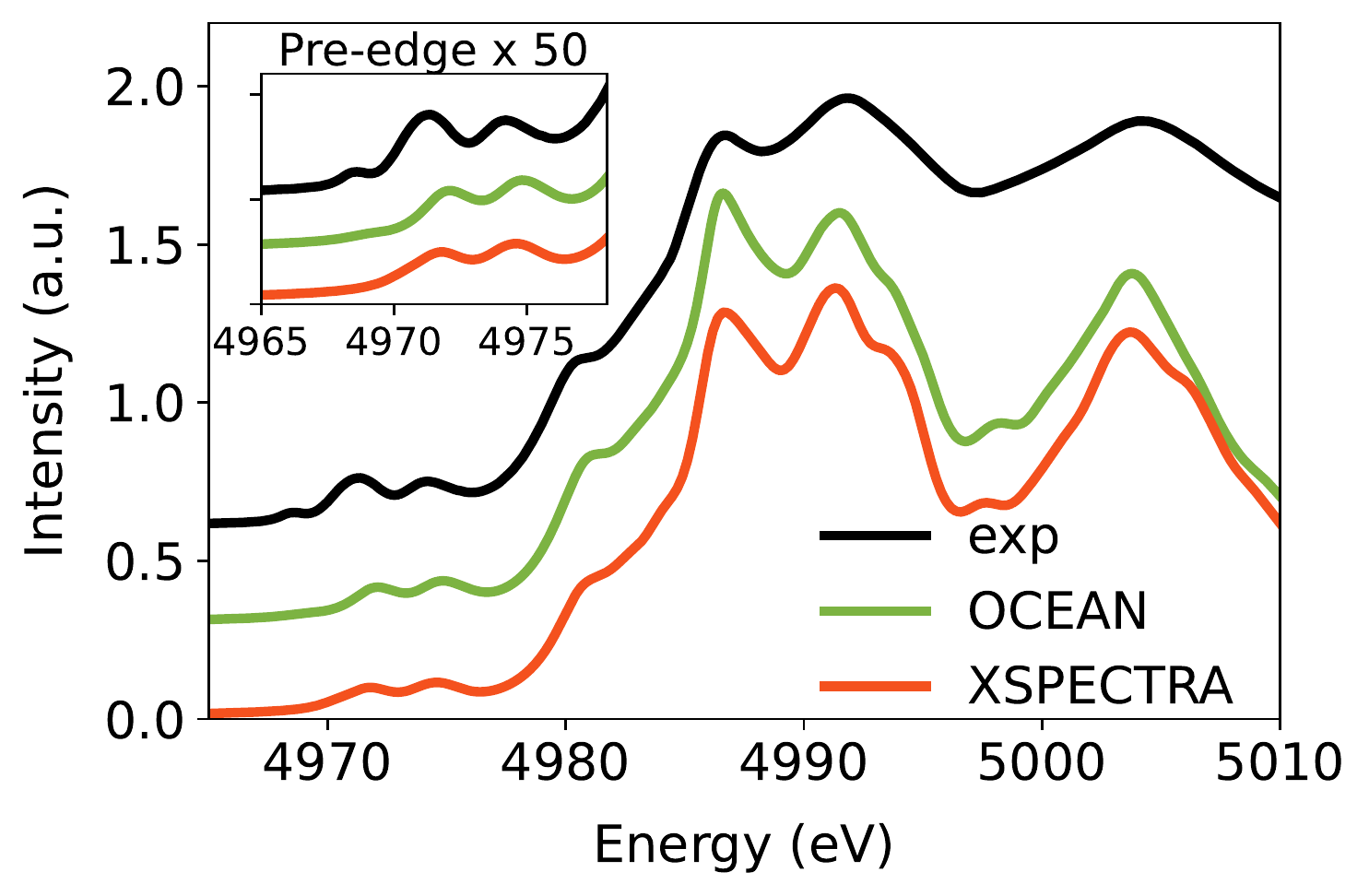}\\ % Here is how to import EPS art \\
\caption{Comparison of Ti K-edge XANES of rutile (mp-2657) between theory ({\sc ocean} and {\sc xspectra}) and experiment~\cite{carta2015x}. Inset shows the pre-edge magnified by fifty fold.}
\label{rutile-exp}
\end{figure}

\begin{figure}
\includegraphics[height=3in]{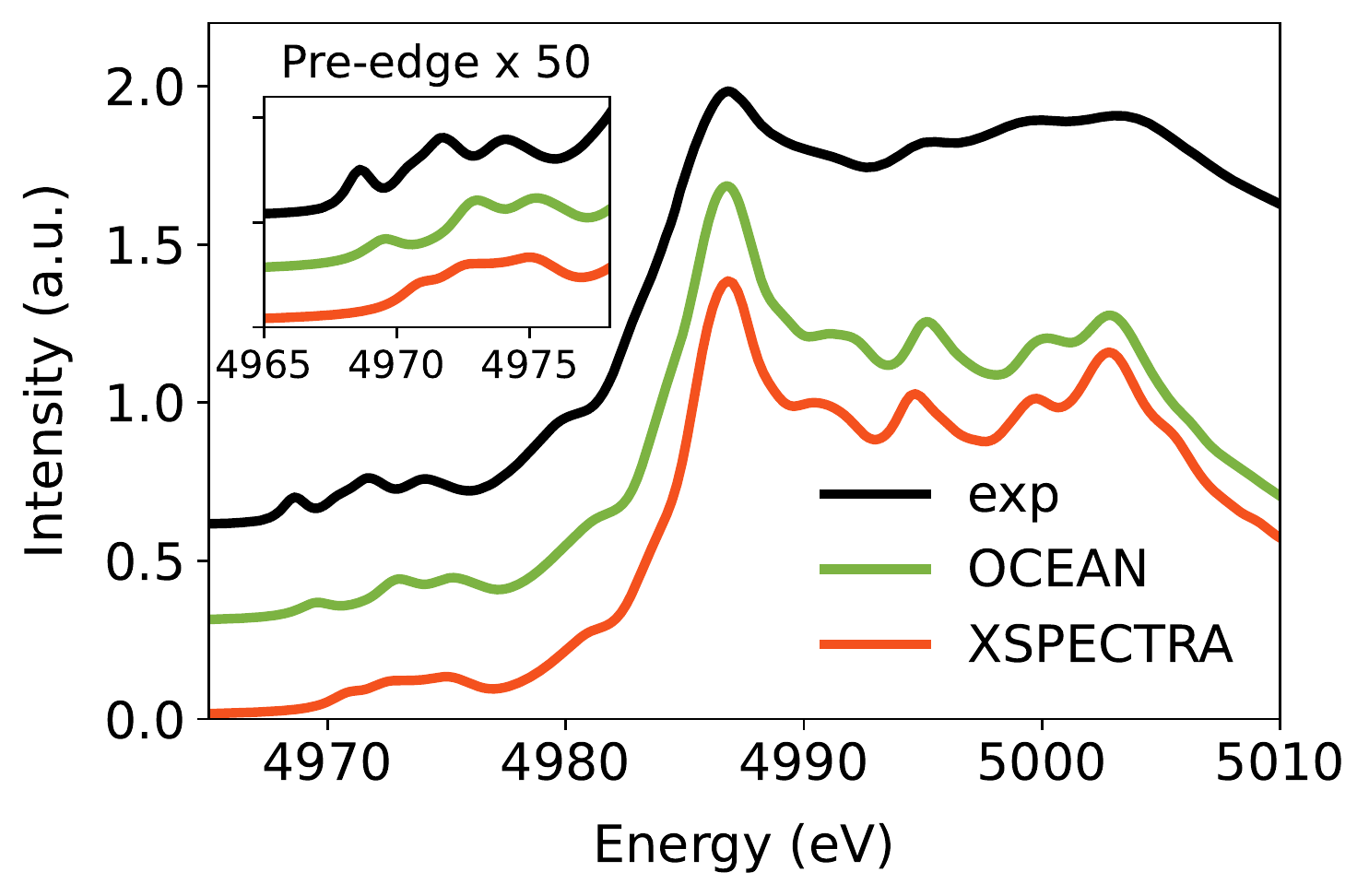}\\ % Here is how to import EPS art \\
\caption{ Comparison of Ti K-edge XANES of anatase (mp-390) between theory ({\sc ocean} and {\sc xspectra}) and experiment~\cite{carta2015x}. Inset shows the pre-edge magnified by fifty fold.}
\label{anatase-exp}
\end{figure}

The comparisons between theory and experiment for rutile and anatase TiO$_2$ are shown in Figs.~\ref{rutile-exp} and~\ref{anatase-exp}. The simulated spectra are aligned to the experiment at the first peak of the main edge near 4986 eV. In rutile, the experimental spectrum shows a shoulder peak around 4981 eV and three peaks in the main edge region at 4986.6~eV, 4991.8~eV, and 5004.0~eV. These main spectral features are well reproduced in both {\sc ocean} and {\sc xspectra}. The energy separation between the first two peaks are underestimated in the simulation by about 0.3~eV, likely due to the missing the self-energy correction at the PBE level of theory. For anatase, both simulation methods reproduce the main features, including the shoulder peak around 4980 eV, the pronounced main edge at 4986.8 eV and the three-peak feature between 4993~eV and 5005 eV. 

In the rutile experimental spectrum, the second peak has higher intensity than the first one. While {\sc xspectra} shows a similar pattern, it is the opposite in {\sc ocean}. This discrepancy between {\sc xspectra} and {\sc ocean} can be attributed to the different approximations for the screening potential.  As discussed in detail in the previous section, the screening potential in RPA is weaker than in the core-hole potential method, which results in stronger electron--core-hole coupling that redistributes oscillator strength towards lower energy in {\sc ocean} as compared to {\sc xspectra}. There is a similar trend in anatase, where the {\sc ocean} spectrum produces stronger low energy features than {\sc xspectra}, but to a lesser degree than rutile. However, it has been shown that when the shake-up satellites are considered as a post treatment of BSE by the convolution with the cumulant spectral function~\cite{kas2015real}, the overstated BSE intensity of the excitonically enhanced Ti $1s \rightarrow 4p$ transition in rutile is reduced and the resulting spectrum is in much better agreement with experiment~\cite{PhysRevB.101.245119}. Therefore one needs to be cautious about drawing quantitative conclusions when comparing theory to experiment on Ti K-edge XANES without considering the effects of satellites. Nevertheless, a systematic study of of the satellite effects is beyond the scope of the current work. 

The Ti K-edge XANES pre-edge features of rutile and anatase have been studied extensively~\cite{uozumi1992experimental,vedrinskii1998pre,cabaret1999pre,PhysRevLett.82.2398,danger2002quadrupolar,yamamoto2005core,woicik2007ferroelectric,wu2004quadrupolar,yamamoto2008assignment,de19902,PhysRevB.101.245119,vorwerk2017addressing,B926499J,PhysRevB.100.245207,Rossi:gb5100}, and the three pre-edge peaks have been assigned to 1) the dipole forbidden Ti $1s \rightarrow 3d$ $t_{2g}$ and 2) $1s \rightarrow 3d$ $e_{g}$ transitions under the crystal filed splitting and 3) the Ti $1s$ to nearest neighbor Ti $3d$ transitions mediated by the hybridization with ligand O $2p$ orbitals. The comparisons of the pre-edge features in rutile and anatase between theory and experiment are shown at the inset of Figs.~\ref{rutile-exp} and~\ref{anatase-exp}. Overall the pre-edge peaks are more pronounced in {\sc ocean} than {\sc xspectra}, as the stronger electron - core hole coupling in BSE sharpens the pre-edge features that have significant excitonic character. {\sc ocean} also shows better agreement with experiment in the pre-edge peak positions than {\sc xspectra}. A major discrepancy between theory and experiment is that the first pre-edge peak in rutile at 4968.6 eV is almost missing in both {\sc ocean} and {\sc xspectra}. The first pre-edge peak has a quadrupole nature, which is an order of magnitude smaller than the dipole contribution, as shown in Figs.~S4 and S5. The Ti octahedron is nearly ideal in rutile, but has a large bond angle distortion in anatase with two O-Ti-O bond angles at 154.67 degree in the mp-390 structure. As a result, the intensity of the Ti $1s \rightarrow 3d$ $t_{2g}$ excitation in rutile is negligible in simulated spectra based on the zero temperature structure. The temperature effects on XANES pre-edge features have been discussed in the literature~\cite{Durmeyer_2010,Tinte_2008,PhysRevB.76.214117,PhysRevB.98.014111}. In particular, Cockayne \emph{et al.} showed that the $t_{1u}^{(1)}$ phonon mode, which moves the Ti$^{4+}$ relative to its axial O neighbors, has a strong spectroscopic signature~\cite{PhysRevB.98.014111}. This centrosymmetry-breaking distortion can cause $p$ - $d$ mixing, which results in a finite pre-edge peak comparable to experiment~\cite{PhysRevB.98.014111}.

While important physical effects are visible in certain spectral features such as pre-edge peaks and relative peak intensities, the present methods represent the measured XANES spectra quite well over the 35 eV energy range. The predictive power of first-principles  XANES simulations in the whole spectral range is crucial to spectral analysis that employs fingerprinting and modern ML-based methods.

%\begin{itemize}
%\item Should include this Joly paper which uses FDMNES to look at the pre-edge including quadrupole terms.\cite{PhysRevLett.82.2398}
%\item SrTiO3 and BaTiO3 with and without the core hole and including quadrupole and dipole terms\cite{PhysRevB.71.245113}
%\item Calculations of the pre-edge of Ti K, including rutile and anatase using DFT approach \cite{B926499J}
%\item Experimental data for temp dependence of rutile pre-edge \cite{Durmeyer_2010}
%\item Anatase angle dependent measurement and FDMNES calculations with great results, but they fudged the 3d states to get the pre-edge correct. (Also the pre-edge strength is inverse with cluster size, and in my opinion they don't show proof of convergence) \cite{PhysRevB.100.245207}
%\item Same paper but for rutile \cite{Rossi:gb5100}
%\item Tinte and Shirley looked at the pre-edge of Ti K of SrTiO3 using a combination of BSE and a model for the vibrational coupling. Only looking at broadening of the pre-edge due to vibrations \cite{Tinte_2008}
%\item Direct discussion of distortions and how they cause the peak intensities in the pre-edge of rutile \cite{PhysRevB.76.214117}
%\item PbTiO3 paper that shows how the pre-edge of Ti K changes with temperature (BSE on top of AIMD) \cite{PhysRevB.98.014111}
%\item 
%\end{itemize}\subsection{Edge Alignment}
As discussed in Sec.~\ref{sec:alignment} all three codes make different approximations in calculating the core-level shifts, and none of the codes provides an absolute energy scale that can be compared to experiment without correction. All three attempt to provide correct relative energy shifts, using different levels of theory. 
Correct relative energy shifts means that a single parameter can be used to align the calculated Ti K-edge XAS from one code with experimental measurements, or, in our case, a single parameter can be used to align the calculations from two different codes. 
%This means that we could attempt to use a single parameter for each pair of codes to align all 13 spectra. 
However, we would not expect this single alignment to work equally well across all systems due to differences in theory and implementation between codes. In this section, we examine the differences to assess the key approximations that the three codes make.
%In this section we examine the differences in the relative shifts between pairs of codes to assess the key approximations they make.

We compare both the alignment of the XAS spectra and the 1{\it s} removal energies as explained in Sec.~\ref{sec:alignment}. 
For each pair of codes, system, and unique Ti site, we calculate the optimal relative alignmen of the x-ray spectra. 
The simulated spectra are truncated using an energy range of 35 eV from the onset (defined as the 2\% of the maximum intensity) to make sure that they cover similar energy ranges. 
We treat each spectrum as a vector. The similarity between two spectra ($\mathbf{\mu_1}$ and $\mathbf{\mu_2}$) are characterized by their Cosine similarity, $cos(\theta)=\frac{\mathbf{\mu_1} \cdot \mathbf{\mu_2}}{\norm{\mathbf{\mu_1}}\norm{\mathbf{\mu_2}}}$. We align
spectra from two codes using the optimal relative shift, %($\Delta \epsilon_\alpha^{\mathit{eo}}$, $\Delta \epsilon_\alpha^{\mathit{xo}}$ and $\Delta \epsilon_\alpha^{\mathit{xe}}$), 
which yields the highest Cosine similarity. 
We subtract out the average relative alignment -- the best fit single alignment parameter -- and report the system-by-system deviation. 
For the 1{\it s} removal energies, no fitting is necessary because we are directly comparing energies. However, as for the spectra, the use of pseudopotentials means that only the relative shifts are meaningful. 

First we compare the two BSE code \ocean{} and \exciting{} as shown in Table~\ref{tab:cls1} for (final state) spectra and 1{\it s} removal energies. For both, the two codes are using the same level of theory, except for different approximations in implementation. The 1{\it s} removal energies are calculated at the level of DFT, and they are given by the energy difference between the conduction band minimum (Fermi level for metals) and the energy of the 1{\it s} orbital. The two codes agree to $\pm0.13$~eV. 
We attribute this to the use of the frozen-core approximation in \ocean{}. As a check, we found that the agreement in shifts of the non-interacting XAS is similar at $\pm0.14$~eV (see Table~S1). Given the previously shown agreement in band structure (Table~\ref{tab:groundstate} and Fig.~\ref{fig:bandstructure}), the non-interacting XAS comparison is primarily a measure of the 1{\it s} energies.
%Given the previously shown agreement in band structure (Table~\ref{tab:groundstate} and Fig.~\ref{fig:bandstructure}), we attribute this to the use of the frozen core approximation in \ocean{}. The agreement of the non-interacting XAS is $\pm0.14$~eV (see Table~S1), indicating that the error mainly arises from the 1{\it s} energy level. 
In the case of the interacting XAS, the agreement in the relative shifts drops to $\pm0.20$~eV. Treating these errors as uncorrelated, the differences in the BSE implementation between \ocean{} and \exciting{} is responsible for approximately a $\pm0.15$~eV discrepancy in the core-level shifts. Like the spectral comparisons in Sec.~\ref{sec:spectral_comparison}, this shows that the two codes are in close agreement despite differences in basis sets and the treatment of core-level electrons.

\begin{table}[!htb]
\caption{The difference in the relative core-level shifts in eV between \ocean{} and \exciting{} for both 1{\it s} removal and XAS spectra. Both the frozen-core approximation in \ocean{} as well as the differences in the BSE implementation between the two codes are found to lead to small, but significant differences between the two codes. A negative number indicates that the \ocean{} spectrum is blue-shifted, relative to the \exciting{} one. $\sigma$ denotes the standard deviation of the differences.}
\begin{ruledtabular}
\label{tab:cls1}
\begin{tabular}{lcc}
%\textrm{mpid:site}& 
%\ocean{}  {\it v.s.} \exciting{} \\
\textrm{mpid:site}&  \ XAS & 1{\it s} removal 
\\
\colrule

mp-390     &$-0.17$  & $-$0.12\\
mp-2657   & $-0.23$ & $-0.15$   \\
mp-1840  & $\hphantom{-}0.14$  & $-0.14$  \\
mp-430    & $\hphantom{-}0.24$& $-0.12$  \\
mvc-11115:0 & $-0.33$ & $-0.09$ \\
mvc-11115:1 & $-0.34$  & $-0.16$ \\
mp-1203:0 & $\hphantom{-}0.13$ & \;\;\;0.10 \\
mp-1203:2  &  $\hphantom{-}0.11$ & $\hphantom{-}0.10$ \\
mp-1203:4 &  $-0.04$ & $-0.02$ \\
mp-10734  & $\hphantom{-}0.14$  & \;\;\;0.19 \\
mp-1215   & $\hphantom{-}0.10$ & \;\;\;0.11 \\
mp-2664   & $\hphantom{-}0.20$ & \;\;\;0.12  \\
mp-458    & $\hphantom{-}0.05$  & $\hphantom{-}0.18$ \\
\hline
$\sigma$ &  $\hphantom{-}${\bf 0.20} & \;\;\;{\bf 0.13}  \\
\end{tabular}
\end{ruledtabular}
\end{table}

Next, we turn to the comparison between CHP (\xs{}) and BSE (\ocean{}) codes. We start with the calculation of the 1{\it s} removal energy. This is a component of the shift seen in XAS (along with the energy cost of adding the core level electron to the conduction bands and the interaction energy of the exciton). 
%This is also an approximation to the shifts that are measured in x-ray photoemission spectroscopy (XPS). Here again, we are focused on the comparison between the two codes and their estimate of the change in the core level alignment between materials and sites. We therefore compare the changes in the calculated core level energy directly, while in XPS the measurements of extended systems must be taken with respect to the potential of an electrode (often the Fermi level of the material being studied).
%
In the case of \xs{}, the 1{\it s} removal energy is obtained from the $\Delta$SCF procedure explained in Sec.~\ref{sec:alignment}. The core-excited final state has a single Ti 1{\it s} core hole using the same core-hole pseudopotential as the spectral calculations. The core-excited state nominally has a net $+1$ charge, which, with periodic DFT, is neutralized by a uniform background charge. The effects of both the uniform background charge and the periodic images of the core hole fall off with supercell size.  We extrapolate the calculated core-level removal energies (at the minimum supercell size of 9~\hbox{\AA} and 13~\AA) to the infinite supercell limit. Note that the neutral excitation of the CHP method for x-ray spectra converges at smaller cell sizes, and as mentioned in Sec.~\ref{sec:implmentation}, a 9~\AA{} minimum lattice vector was found to be sufficient. In \ocean{}, the core-level removal energy is estimated as outlined in Sec.~\ref{sec:alignment} as the Hartree potential minus $\sfrac{1}{2}W_c$ to account for the relaxation of the system due to the removal of the core electron. 

%The comparison between the core-level removal energies between \xs{} and \ocean{} is shown in Table~\ref{tab:cls2} in the ``1{\it s} removal'' column without performing any spectral alignment. Again, the value of any individual difference is not meaningful due to the use of pseudopotentials, and we compare the relative shifts with respect to the average shift. 
From Table~\ref{tab:cls2}, we find that the two methods give reasonably similar 1{\it s} removal energies with a standard deviation of only $\pm0.19$~eV. There is a slight bimodal grouping of the results by metal (first six rows) and non-metal (last seven rows). For non-metals, the estimate of the 1{\it s} removal energy from \xs{} tends to be smaller than that of \ocean{} by 0.29~eV. 
This could indicate a sensitivity to the model dielectric that is used in \ocean{} to model the long-range response, different convergence behavior with respect to cell size of the $N-1$ core-hole calculation in \xs{}, or it could reflect differences between linear-response and self-consistent calculations of the density response.
%This parallels  our findings in Sec.~\ref{sec:final-state} where the self-consistent density response more efficiently screens the core-hole, lowering the energy cost of removing the core electron in \xs{}. 
%This could indicate a sensitivity to the model dielectric that is used in \ocean{} to model the long-range response, or it could be indicitative of the difference between linear-response and self-consistent calculations of the density response.

Finally, we compare the spectral shifts in XANES between \ocean{} and \xs{}, and we again find reasonable agreement between the codes and methods with a standard deviation of $\pm0.22$~eV. This would suggest that the contribution due to the different treatments of the final state effects is about $\pm 0.11$ eV, but we can not rule out some cancellation of errors.

\if{0}
\begin{table}[!htb]
\caption{Comparison of the relative core-level shifts in eV between \ocean{} and \xs{}. 
First, the Final State (w/o 1{\it s}) comparison reflects only the position of the conduction bands and the effect of the electron-hole interactions. Second, Final State (w/ $\Sigma_{1s}$) includes the previous effects and the relative shifts in the Ti 1{\it s} energies (including approximate self-energy corrections $\Sigma_{1s}$). Third, the \ocean{} alignment is the same, but \xs{} is shifted by the total energy differences between the neutral core-hole excited and the ground state. 
Finally, on the right, a comparison of the core-level removal energies for the Ti 1{\it s}. Unlike the comparison between BSE codes in Table~\ref{tab:cls1}, here the \ocean{} results include the $-\frac{1}{2}W_c$ correction to the core-hole energy (see text). The sign follows the same convention as in Table~\ref{tab:cls1}.%The sign is such that a negative number indicates that the \xs{} 1{\it s} is more strongly bound relative to the \ocean{} calculation for the given material, or, in the case of the XAS results, a negative number indicates that the \xs{} spectrum is red-shifted, relative to the \ocean{} one.
}
\begin{ruledtabular}
\label{tab:cls2}
\begin{tabular}{lccc|c}
%\textrm{mpid:site}&  &
%\ocean{}  {\it v.s.} \xs{} \\
\textrm{mpid:site}&  Final State (w/o 1{\it s}) & Final State (w/ $\Sigma_{1s}$) & Full & 1{\it s} removal 
\\
\colrule

mp-390    & $\hphantom{-}0.44$ & $\hphantom{-}0.24$ & $\hphantom{-}0.11$ & $\hphantom{-}0.21$\\
mp-2657   & $\hphantom{-}0.21$ & $\hphantom{-}0.07$ & $-0.11$ & $\hphantom{-}0.13$ \\
mp-1840   & $\hphantom{-}0.81$ & $\hphantom{-}0.53$ & $\hphantom{-}0.18$ & $\hphantom{-}0.29$ \\
mp-430    & $\hphantom{-}0.99$ & $\hphantom{-}0.71$ & $\hphantom{-}0.38$ & $\hphantom{-}0.27$\\
mvc-11115:0 & $\hphantom{-}0.68$ & $\hphantom{-}0.77$ & $\hphantom{-}0.33$ & $-0.09$\\
mvc-11115:1 & $\hphantom{-}0.82$ & $\hphantom{-}0.81$ & $\hphantom{-}0.23$ & $\hphantom{-}0.01$  \\
mp-1203:0 & $-0.53$ & $-0.52$ & $-0.14$ & $-0.01$ \\
mp-1203:2 & $-0.62$ & $-0.47$ & $\hphantom{-}0.00$ & $-0.14$ \\
mp-1203:4 & $-0.58$ & $-0.36$ & $ -0.21$ & $-0.22$ \\
mp-10734  & $-0.50$ &$-0.31$ & $-0.35$ & $-0.19$  \\
mp-1215   & $-0.56$ & $-0.44$ & $-0.06$ & $-0.12$ \\
mp-2664   & $-0.57$ & $-0.44$ & $-0.02$ & $-0.14$  \\
mp-458    & $-0.60$ &  $-0.59$ & $-0.35$ & $-0.00$  \\
\hline
$\sigma$ & $\hphantom{-}${\bf 0.66} & $\hphantom{-}${\bf 0.54} & $\hphantom{-}${\bf 0.23} & $\hphantom{-}${\bf 0.17}\\
\end{tabular}
\end{ruledtabular}
\end{table}
\else
\begin{table}[!htb]
\caption{The difference in the relative core-level shifts in eV between \ocean{} and \xs{} for both XAS and 1{\it s} removal.
Unlike the comparison between BSE codes in Table~\ref{tab:cls1}, here the \ocean{} results include the self-energy correction to the core-hole energy (see text). The sign follows the same convention as in Table~\ref{tab:cls1}. $\sigma$ denotes the standard deviation of the differences.
}
\begin{ruledtabular}
\label{tab:cls2}
\begin{tabular}{lcc}
%\textrm{mpid:site}&  &
%\ocean{}  {\it v.s.} \xs{} \\
\textrm{mpid:site}&  XAS & 1{\it s} removal 
\\
\colrule

mp-390    & $-0.06$ & $-0.25$\\
mp-2657   &  $\hphantom{-}0.14$ & $-0.17$ \\
mp-1840   & $-0.16$ & $-0.30$ \\
mp-430    &  $-0.36$ & $-0.30$\\
mvc-11115:0 & $-0.33$ & $\hphantom{-}0.08$\\
mvc-11115:1 &  $-0.23$ & $\hphantom{-}0.00$  \\
mp-1203:0 &  $\hphantom{-}0.11$ & $\hphantom{-}0.04$ \\
mp-1203:2 &  $-0.04$ & $\hphantom{-}0.18$ \\
mp-1203:4 &  $ \hphantom{-}0.17$ & $\hphantom{-}0.25$ \\
mp-10734  &  $\hphantom{-}0.31$ & $\hphantom{-}0.23$  \\
mp-1215   &  $\hphantom{-}0.09$ & $\hphantom{-}0.09$ \\
mp-2664   &  $\hphantom{-}0.02$ & $\hphantom{-}0.13$  \\
mp-458    & $\hphantom{-}0.34$ & $\hphantom{-}0.01$  \\
\hline
$\sigma$ & $\hphantom{-}${\bf 0.22} & $\hphantom{-}${\bf 0.19}\\
\end{tabular}
\end{ruledtabular}
\end{table}
\fi

\if{0}
Unlike \exciting{}, the CHP method in \xs{} includes an approximation to the relaxation of the system in response to the core hole. Therefore, for this comparison we include the adiabatic screening term, $-\frac{1}{2}W_c$, in \ocean{} to put it on the same theoretical footing. As seen in the right-hand column of Table~\ref{tab:cls2}, the overall agreement is poor. If we separate our systems and consider only the insulators, the first 5 materials (6 sites), the agreement is substantially better with a standard deviation of $\pm0.17$~eV, albeit for a limited sample size. (For comparison, a similar sub-sampling of 1{\it s} removal energies gives $\pm0.14$~eV.) The metals are slightly improved by this separation to $\pm0.56$~eV, but their agreement remains poor. This could be due to the partitioning between occupied and unoccupied states typically employed in BSE calculations. In \ocean{} (and \exciting{}), the valence-band states are taken from the ground-state calculation and are not relaxed in the presence of the core hole, whereas in the CHP method the occupied states are relaxed self-consistently in the presence of the core hole. 
\fi

\if{0}
In Table~\ref{tab:cls}, we compare the edge alignment of the three codes. Because the three spectra for the same material have quite different absolute energy scales (i.e., a consequence of pseudopotential and all electron calculations), in the pair-wise comparison, we shift the first spectrum to align with the second one by maximizing the Cosine similarity between the two spectra as described in Section ~\ref{sec:alignment}. For example, in the ``\ocean{} {\it vs.} \exciting{}" column in Table~\ref{tab:cls}, the \exciting{} spectra are used as the reference. Then for each column the relative shifts ($\Delta \Delta \epsilon_\alpha$) are reported after subtracting the mean of the column (see Section ~\ref{sec:alignment}). A small standard deviation indicates that two codes yield similar edge alignment. As expected, the closest agreement is found between \exciting{} and \ocean{} without the adiabatic screening correction (the $-\frac{1}{2}W_c$ term), as both codes compute $\epsilon_\alpha$ at the ground state. 
Comparing the alignment of the independent particle spectra between \ocean{} and \exciting{} we find the standard deviation to be 0.10~eV (see Table~S1 in the Supplemental Material~\cite{SI}). We attribute this to the use of the frozen-core approximation in \ocean{}. Including interactions increases the discrepancy between \ocean{} and \exciting{} to 0.19~eV, most probably due to differences in the numerical implementation of the BSE between these two codes. 
%This comparison shows that the frozen core hole approximation is responsible for a small, but non-negligible error with a standard deviation of 0.19~eV. For energy alignment less than 0.19~eV, both \xs{} and \ocean{} may require an additional correction to account for the frozen core. 
After including the $-\frac{1}{2}W_c$ term in \ocean{}, the standard deviation between \exciting{} and \ocean{} increases to 0.91~eV, which highlights the significant contribution of the adiabatic screening effect of the core hole in the edge alignment. 

A similar magnitude in the standard deviation among the codes, ranging from $0.92$~eV to $1.56$ eV, is found in the other columns. Such a degree of difference is expected, since the three chemical shifts are obtained with very different approximations: \exciting{} and \ocean{} are under the initial-state rule, \xs{}  includes final-state effects through a $\Delta$-SCF procedure. Without benchmarking against a set of reliable experimental results, it is not possible to state which, if any, of the three codes gives the best estimates for the core-level shifts. This is beyond the scope of the current work and could be the subject of future research.

\begin{table}[!htb]
\caption{Difference in the relative core-level shifts in eV between code pairs in the interacting particle spectra. Comparisons with \ocean{} are done with (w/) and without (w/o) including the $-\frac{1}{2}W_c$ term to the core hole initial state (see text).}
\begin{ruledtabular}
\label{tab:cls}
\begin{tabular}{lccc}
\textrm{mpid:site}& 
\ocean{} {\it vs.} \exciting{}&
\ocean{} {\it vs.} \xs{}&
\xs{} {\it vs.} \exciting{}
\\
           & w/ $-\frac{1}{2}W_c$ \quad w/o $-\frac{1}{2}W_c$ & w/ $-\frac{1}{2}W_c$ \quad w/o $-\frac{1}{2}W_c$  \\
\colrule

mp-390    & $-$0.95 \quad\quad $-$0.11 & $-1.49$ \quad\quad  $-0.65$ & \;\;\;0.53  \\
mp-2657   & $-0.87$ \quad\quad $-$0.19 & $-1.73$ \quad\quad  $-1.05$ & \;\;\;0.86  \\
mp-1840   & $-0.76$ \quad\quad \;\;\;0.19 & $-1.68$ \quad\quad $-0.72$  & \;\;\;0.92  \\
mp-430    & $-0.56$ \quad\quad \;\;\;0.27 & $-1.47$ \quad\quad  $-0.63$  & \;\;\;0.87  \\
mvc-11115:0 & $-1.18$ \quad\quad $-$0.35 & $-1.30$ \quad\quad $-0.47$ &  \;\;\;0.23  \\
mvc-11115:1 & $-1.22$ \quad\quad $-$0.30 & $-1.27$ \quad\quad $-0.35$ &  \;\;\;0.06  \\
mp-1203:0 & \;\;\;0.75 \quad\quad \;\;\;0.09 & \;\;\;1.62 \quad\quad  \;\;\;0.96  & $-$0.87  \\
mp-1203:2 & \;\;\;0.81 \quad\quad \;\;\;0.08 & \;\;\;1.22 \quad\quad  \;\;\;0.49  & $-$0.39  \\
mp-1203:4 & \;\;\;0.57 \quad\quad $-$0.08 & \;\;\;2.27 \quad\quad  \;\;\;1.61  & $-$1.73  \\
mp-10734  & \;\;\;0.84 \quad\quad \;\;\;0.13 & \;\;\;0.35 \quad\quad  $-0.36$  & \;\;\;0.48  \\
mp-1215   & \;\;\;0.64 \quad\quad \;\;\;0.03 & \;\;\;2.34 \quad\quad  \;\;\;1.73  &   $-1.73$ \\
mp-2664   & \;\;\;1.10 \quad\quad \;\;\;0.16 & \;\;\;1.02 \quad\quad  \;\;\;0.07  & \;\;\;0.07  \\
mp-458    & \;\;\;0.82 \quad\quad \;\;\;0.07 & \;\;\;0.13 \quad\quad  $-0.62$  & \;\;\;0.69  \\
\hline
$\sigma$ & \;\;\;{\bf 0.91} \quad\quad \;\;\;{\bf 0.19} & \;\;\;{\bf 1.56} \quad\quad \;\;\;{\bf 0.92} &  \;\;\;{\bf 0.93} \\
\end{tabular}
\end{ruledtabular}
\end{table}
\fi

\if{0}
Putting this here for safe keeping
\begin{table}[!htb]
\caption{The difference in the relative core-level shifts in eV between \ocean{} and \xs{}. On the left, a comparison of the core-level removal energies for the Ti 1{\it s}, and on the right the spectral alignment between the XAS calculations. Unlike the comparison between BSE codes in Table~\ref{tab:cls1}, here the \ocean{} results include the $-\frac{1}{2}W_c$ correction to the core-hole energy (see text).}
\begin{ruledtabular}
\label{tab:cls2}
\begin{tabular}{lc}
\textrm{mpid:site}& 
\ocean{}  {\it v.s.} \xs{} \\
&1{\it s} removal \;\quad XAS \;
\\
\colrule

mp-390    & $-$0.21  \quad\quad $-1.49$ \\
mp-2657   & $-0.12$  \quad\quad  $-1.73$ \\
mp-1840   & $-0.31$  \quad\quad $-1.68$ \\
mp-430    & $-0.27$ \quad\quad $-1.47$ \\
mvc-11115:0 & $\hphantom{-}0.07$ \quad\quad $-1.30$ \\
mvc-11115:1 & $-0.08$ \quad\quad $-1.27$  \\
mp-1203:0 & \;\;\;0.04 \quad\quad $\hphantom{-}1.62$ \\
mp-1203:2 & \;\;\;0.15 \quad\quad $\hphantom{-}1.22$ \\
mp-1203:4 & \;\;\;0.25 \quad\quad $\hphantom{-}2.27$ \\
mp-10734  & \;\;\;0.21 \quad\quad $\hphantom{-}0.35$  \\
mp-1215   & \;\;\;0.22 \quad\quad $\hphantom{-}2.34$ \\
mp-2664   & \;\;\;0.17  \quad\quad $\hphantom{-}1.02$  \\
mp-458    & $\hphantom{-}0.00$ \quad\quad $\hphantom{-}0.13$  \\
\hline
Standard deviation & \;\;\;{\bf 0.19} \quad\quad  $\hphantom{-}${\bf 1.56}\\
\end{tabular}
\end{ruledtabular}
\end{table}
\fi

\section{Conclusions}

We have developed an automated workflow to generate and standardize input files for simulating x-ray absorption spectra using three widely used codes: \ocean{}, \exciting{}, and \xs{}. 
By carefully converging the disparate settings for each of the three codes, the workflow ensures that the resulting spectra are free from numerical artifacts unrelated to the underlying theory or implementation details. We carried out quantitative comparisons between Ti K-edge XANES spectra from the three codes using ten representative titanium oxide compounds, referred to as the Ti-O-10 dataset.

We found that the two BSE-based codes, \exciting{} and \ocean{}, produce spectra measurably closer to each other than to the core-hole potential code \xs{}. We quantified the similarity between spectra using Spearman's rank correlation score.  We found that the average score between the two BSE codes is 0.998, while between either BSE code and \xs{} it is 0.990. While this difference appears small, it corresponds to visible differences in the spectra (see Fig.~\ref{fig:spectra}).  Nonetheless, the overall spectra are quite similar.  From these results, we draw several conclusions. First, despite significant implementation differences between the all-electron \exciting{} and pseudopotential-based \ocean{}, they produce nearly identical spectra, suggesting that both implementations are robust. Second, the BSE and CHP methods produce quite close agreement in main features and peak positions, though differences are noticeable, especially in the relative spectral intensities. Such discrepancies in spectral shape can be primarily attributed to the difference in the strength of the screened core-hole potential. Carrying out BSE calculations using the ALDA instead of RPA to screen the core hole greatly reduces the difference between the CHP and BSE results. 

We also conducted a theory (\ocean{} and \xs{}) to experiment comparison for rutile and anatase TiO$_2$, where the quandrupole contributions are included to give a more complete description of pre-edge features. Other effects, such as thermal disorder and satellite effects, although very important for a thorough comparison, are subject of future work. With these limitations, our main findings are the following. First, both \ocean{} and \xs{} reproduce quite well the main features of the experimental spectra in a wide energy range of about 35 eV, including the number of peaks, the overall spectral shape, and the positions of main and shoulder peaks. Second, \xs{} exhibits a seemingly better match to the relative intensity of the double-peak feature in rutile compared to \ocean{}. We found that the electron-core hole coupling is stronger in BSE under a weaker dielectric screening than CHP. As a result, the oscillator strength is redistributed more towards low energy in \ocean{} than \xs{}. This should be considered in the context of studies showing the impact of other physical effects. As shown by Woicik \emph{et al.}~\cite{PhysRevB.101.245119}, including the satellite effects using the cumulant method can modify the relative intensity of the two peaks and yield a much better agreement between \ocean{} and experiment. On the other hand, the cumulant correction is expected to worsen the agreement between \xs{} and experiment. Finally, overall \ocean{} pre-edge features are more pronounced and agree better with experiment than \xs{}. However, the Ti $1s \rightarrow 3d$ $t_{2g}$ peak in rutile is severely underestimated in both codes, due to the lack of the thermal disorder correction. %The current work is primarily limited to comparisons among simulated spectra. Future work should include detailed comparisons to measured spectra, 
%However, given the close agreement between all three methods on the Ti-O-10 set, such a comparison on the same materials is unlikely to definitively identify one code or method as superior to the others without 
This discussion underscores the importance of thoroughly accounting for additional effects such as vibrational disorder and satellites. On a material by material basis these have been shown to change positions and weights of spectral features to a degree comparable to the code- and method-dependent differences we have shown here.

Finally, we compared the relative edge alignment among the three codes. Between \ocean{} and \exciting{}, the standard deviation is 0.20 eV in the XAS spectra. This difference is partially due to the frozen core hole approximation used in \ocean{}, which gives arise a the standard deviation of 0.13 eV at the Kohn-Sham DFT level. Between \ocean{} and \xs{}, the standard deviation in the 1{\it s} removal energy is $0.19$ eV. The standard deviation in the XAS spectra is $0.22$ eV, though the two codes have quite different treatments for the self-energy correction and final state effects.

This study supports broad application of first-principles simulation for x-ray spectral analysis, in particular, modern data-driven methods that take into account the full spectral range of XANES. The automated workflow and the heuristics to achieve the numerical convergence provide a good standard for first-principles x-ray spectral simulations. Furthermore, the workflow can play a big role in developing simulated XAS spectral databases using high-throughput computing.

\section{Data Availability}
Data of the benchmark study, including input files, output files, and metadata~\cite{nomad-doi}, can be downloaded from the Novel Materials Discovery (NOMAD) Laboratory~\cite{draxl2019nomad} under the CC-BY-4.0 license.
\section{Acknowledgment}
We thank Dr.~Daniela Carta and Dr.~Themistoklis Prodromakis for providing us their measured Ti K-edge XANES spectra of rutile and anatase TiO$_2$. F.M., X.Q. and D.L. thank Eli Stavitski and Bruce Ravel for helpful discussion. This work is partially supported by the U.S. Department of Energy, Office of Science, Office of Basic Energy Sciences, under Award Numbers
FWP PS-030. The research used the theory and computation resources of the
Center for Functional Nanomaterials, which is a U.S. DOE Office of Science
Facility, and the Scientific Data and Computing Center, a component of the
Computational Science Initiative, at Brookhaven National Laboratory under Contract No. DE-SC0012704. This research also used resources of the National Energy Research Scientific Computing Center (NERSC), a U.S. Department of Energy Office of Science User Facility located at Lawrence Berkeley National Laboratory, operated under Contract No. DE-AC02-05CH11231 using NERSC award BES-ERCAP-20690, 18006 and 14811. This work received partial funding by the German Research Foundation (DFG) through the CRC 1404 (FONDA), Projektnummer 414984028, and the NFDI consortium FAIRmat – project 46019701. Certain software packages are identified to facilitate understanding. Such identification does not imply recommendations or endorsement by NIST.
%\begin{figure}[htb] 
%\includegraphics[width=6.0 in]{figures/convergence.png}
%\caption{\label{fig:convergence} Convergence of similarity metric with respect to the next k-mesh.} 
%\end{figure} 

%\begin{figure}[htb] 
%\includegraphics[width=6.0 in]{figures/convergence_last.png}
%\caption{\label{fig:convergence} Convergence of similarity metric with respect to the most converged k-mesh.} 
%\end{figure} 

\beginappendix
\section{Band structure RMSD comparison}\label{app:rmsd}
The comparisons for the ground-state band structure energies, given in Table~IV, are calculated using the regular {\it k}-point grids from the Materials Project. 
The RMSD is calculated in the standard way, but with pseudo-occupation numbers $f$ to limit the range of states, e.g., valence band, or limited energy ranges of the conduction band.
The RMSD between the energies $\epsilon$ from two codes $i$ and $j$ is given by, 
\begin{equation}
\mathrm{RMSD}[i,j] = \left[ \frac{\sum_{\mathbf{k}}w_\mathbf{k} \sum_n f_{n\mathbf{k}}^{ij}(\epsilon_{n\mathbf{k}}^i - \epsilon_{n\mathbf{k}}^j  )^2 }{\sum_\mathbf{k} w_\mathbf{k} \sum_n f_{n\mathbf{k}}^{ij} } \right]^{1/2}
\end{equation}
where the sum is over symmetry reduced {\it k}-points $\mathbf{k}$ with weights $w_\mathbf{k}$ and bands $n$. The 
occupation numbers are given by
\begin{align}
f_{n\mathbf{k}}^{ij} &= \sqrt{ f_{n\mathbf{k}}^{i} f_{n\mathbf{k}}^{j} } \\
f_{n\mathbf{k}}^{i}  &= \left[ e^{(\epsilon_{n\mathbf{k}}^i - \epsilon_L)/\gamma} + 1 \right]^{-1} \left[ e^{(\epsilon_H - \epsilon_{n\mathbf{k}}^i)/\gamma} + 1 \right]^{-1}
\end{align}
where the lower and upper bounds are given by $\epsilon_L$ and $\epsilon_H$, respectively, and a broadening $\gamma$ of 0.02~Rydberg is used for both. 

\section{Effect of number of bands}\label{app:bands}

\begin{figure}[htbp] 
\includegraphics[width=0.5\textwidth]{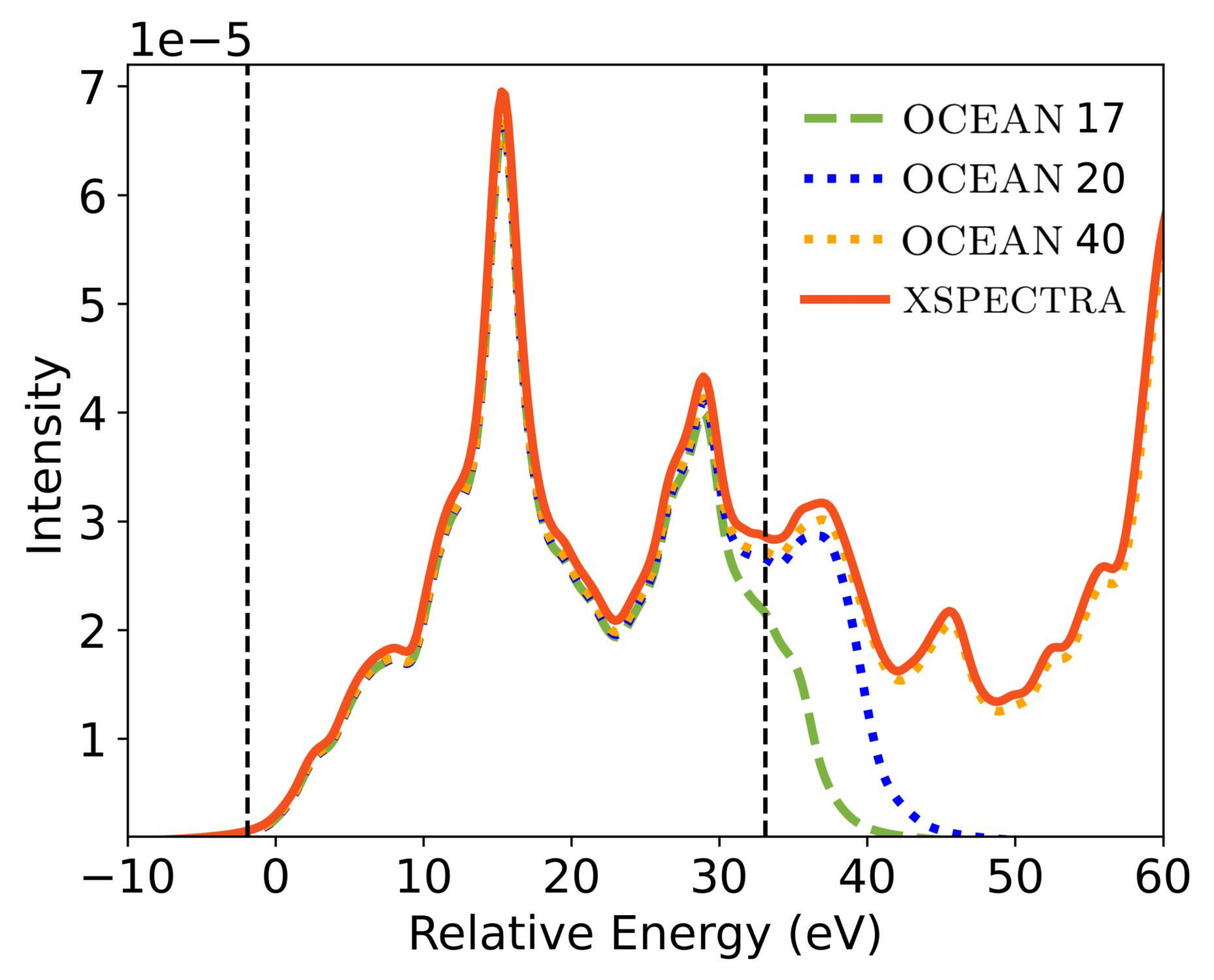}
\caption{\label{fig:numerofbands} The effect of number of empty bands on the XANES spectra of mp-2664 calculated at the independent particle level. \ocean{} 17 (20 or 40) corresponds to \ocean{} calculations with 17 (20 or 40) empty bands. The first dashed vertical line corresponds to our definition of the onset, and the second is 35 eV above onset.} 
\end{figure}

We carefully examined the energy range (35 eV from onset) used to compare the spectra. The estimation for the number of empty bands described in the main text based on the homogeneous electron gas model is not accurate enough for mp-2664, which is likely due to the small unit cell volume of mp-2664. In this case, comparing the effect of different number of bands, we found that by using 20 empty bands, the spectra converged within the energy range where the comparison of spectra are performed. Therefore, in this particular case, we use 20 empty bands instead of 17 from our homogeneous electron gas model.

\if{0}
\section{Bands to energies}

We start with a particle in a cube of volume $V$ 
\begin{equation}
    E_{abc} = \frac{1}{2}\frac{\pi^2}{V^{2/3}} (a^2+b^2+c^2)
\end{equation}
where $a,b,c$ are the integer indices along $\hat{x}, \hat{y}, \hat{z}$, respectively. 
Taking the continuous limit, the number of states N with energies between 0 and some cut-off $\Delta E$ is a sphere of radius $R$,
\begin{equation}
N = 4/3 \pi R^3
\end{equation}
where $R$ is defined by 
\begin{equation}
\Delta E = \frac{1}{2}\frac{\pi^2}{V^{2/3}} R^2 .
\end{equation}
Therefore 
\begin{equation}
    N = \frac{8\sqrt{2}}{3 \pi^2} E^{3/2} V \approx 0.382 E^{3/2} V .
\end{equation}
\fi

\section{Isotropic quadrupole transitions}
\label{app:quadrupole}

In Section~\ref{expt_compare} we compare calculations to measurements on powdered samples, requiring the calculaton of the isotropic spectra. 
For all three of the codes shown here, spectra are calculated for an explicit, cartesian polarization direction and, for quadrupole transitions, momentum vector. 
Both \ocean{} and {\sc xspectra} can be set to calculate only the quadrupole transition without the dipole operator. In the case of non-magnetic systems and linear polarization the dipole and quadrupole contributions do not interfere and can be summed separately. 
The dipole case is straightforward. Regardless of the symmetry of the system, an average of spectra calculated with three orthogonal polarization directions is equal to the isotropic average. 
The number of required calculations can be reduced to 1 or 2, depending on the symmetry of the crystal, but for simplicity we average over $\mathbf{e}= \{\hat{x}, \hat{y}, \hat{z}\}$.

The case of the quadrupole term $(\mathbf{e}\cdot\mathbf{r})(\mathbf{q}\cdot\mathbf{r})$ is significantly more complicated. 
Brouder detailed the angular dependence by writing the x-ray transitions in terms of spherical tensors.\cite{Brouder_1990} Following Brouder, we note that rutile and anatase both have a crystallographic point group 4/mmm. Their quadruple spectra therefore is given by Eq.~5.7 in Ref.~\onlinecite{Brouder_1990}
\begin{align}
\sigma(\mathbf{e},\mathbf{k}) &= \sigma^Q(0,0) + \sqrt{5/14} \left(3 \sin^2 \theta \sin^2 \psi - 1 \right) \sigma^Q(2,0) \nonumber \\
&+ 1/\sqrt{14}\left( 35 \sin^2\theta \cos^2\theta \cos^2 \psi + 5 \sin^2 \theta \sin^2\psi - 4 \right) \sigma^Q(4,0)  \nonumber \\
&+ \sqrt{5}\sin^2 \theta \left[ ( \cos^2\theta \cos^2\psi - \sin^2 \psi) \cos4\phi - 2 \cos \theta \sin \psi \cos \psi \sin 4\phi \right] \sigma^{Qr}(4,4)
\end{align}
where $\sigma^Q(a,b)$ are the elements of the tensor decomposition that are not zero by symmetry and $\sigma^Q(0,0)$ is the desired isotropic spectrum. 
The directions of the polarization and momentum are, in terms of the three angles, 
\begin{align}
\mathbf{e} &= (\sin\theta \cos \phi, \sin \theta \sin \phi, \cos \theta ) \nonumber \\
\mathbf{q} &= ( \cos \theta \cos\phi \cos \psi - \sin \phi \sin \psi, \nonumber \\
& \quad\;\; \cos \theta \sin\phi \cos \psi + \cos \phi \sin \psi, \nonumber \\
& \quad\;\; -\sin \theta  \cos \psi )\nonumber 
\end{align}
By a careful choice of polarization and momentum direction, three spectra can be combined to give only the isotropic response.
We choose:
\begin{align}
(\theta_1, \phi_1, \psi_1) &= (\acos[1/\sqrt{7}], -1/4 \atan[19/7\sqrt{11}], 1/2 \acos[2/9] ) \nonumber \\ 
(\theta_2, \phi_2, \psi_2) &= (-1/4 \acos[1/49], \pi / 8, 0 ) \nonumber \\ 
(\theta_3, \phi_3, \psi_3) &= (0, 0, 0) \nonumber 
\end{align}
giving
\begin{equation}
\sigma^Q(0,0) = \sigma^Q(\mathbf{e}_1, \mathbf{q}_1) - 1/15 \sigma^Q(\mathbf{e}_2, \mathbf{q}_2) + 1/15 \sigma^Q(\mathbf{e}_3, \mathbf{q}_3)
\end{equation}
which is added to the isotropic dipole spectra to give the total isotropic spectra including both dipole and quardupole terms.


\begin{thebibliography}{129}%
\makeatletter
\providecommand \@ifxundefined [1]{%
 \@ifx{#1\undefined}
}%
\providecommand \@ifnum [1]{%
 \ifnum #1\expandafter \@firstoftwo
 \else \expandafter \@secondoftwo
 \fi
}%
\providecommand \@ifx [1]{%
 \ifx #1\expandafter \@firstoftwo
 \else \expandafter \@secondoftwo
 \fi
}%
\providecommand \natexlab [1]{#1}%
\providecommand \enquote  [1]{``#1''}%
\providecommand \bibnamefont  [1]{#1}%
\providecommand \bibfnamefont [1]{#1}%
\providecommand \citenamefont [1]{#1}%
\providecommand \href@noop [0]{\@secondoftwo}%
\providecommand \href [0]{\begingroup \@sanitize@url \@href}%
\providecommand \@href[1]{\@@startlink{#1}\@@href}%
\providecommand \@@href[1]{\endgroup#1\@@endlink}%
\providecommand \@sanitize@url [0]{\catcode `\\12\catcode `\$12\catcode
  `\&12\catcode `\#12\catcode `\^12\catcode `\_12\catcode `\%12\relax}%
\providecommand \@@startlink[1]{}%
\providecommand \@@endlink[0]{}%
\providecommand \url  [0]{\begingroup\@sanitize@url \@url }%
\providecommand \@url [1]{\endgroup\@href {#1}{\urlprefix }}%
\providecommand \urlprefix  [0]{URL }%
\providecommand \Eprint [0]{\href }%
\providecommand \doibase [0]{http://dx.doi.org/}%
\providecommand \selectlanguage [0]{\@gobble}%
\providecommand \bibinfo  [0]{\@secondoftwo}%
\providecommand \bibfield  [0]{\@secondoftwo}%
\providecommand \translation [1]{[#1]}%
\providecommand \BibitemOpen [0]{}%
\providecommand \bibitemStop [0]{}%
\providecommand \bibitemNoStop [0]{.\EOS\space}%
\providecommand \EOS [0]{\spacefactor3000\relax}%
\providecommand \BibitemShut  [1]{\csname bibitem#1\endcsname}%
\let\auto@bib@innerbib\@empty
%</preamble>
\bibitem [{\citenamefont {Rehr}\ and\ \citenamefont
  {Albers}(2000)}]{rehr2000theoretical}%
  \BibitemOpen
  \bibfield  {author} {\bibinfo {author} {\bibfnamefont {John~J}\ \bibnamefont
  {Rehr}}\ and\ \bibinfo {author} {\bibfnamefont {Robert~C}\ \bibnamefont
  {Albers}},\ }\bibfield  {title} {\enquote {\bibinfo {title} {Theoretical
  approaches to x-ray absorption fine structure},}\ }\href {\doibase
  10.1103/RevModPhys.72.621} {\bibfield  {journal} {\bibinfo  {journal}
  {Reviews of Modern Physics}\ }\textbf {\bibinfo {volume} {72}},\ \bibinfo
  {pages} {621} (\bibinfo {year} {2000})}\BibitemShut {NoStop}%
\bibitem [{\citenamefont {Yamamoto}(2008)}]{yamamoto2008assignment}%
  \BibitemOpen
  \bibfield  {author} {\bibinfo {author} {\bibfnamefont {Takashi}\ \bibnamefont
  {Yamamoto}},\ }\bibfield  {title} {\enquote {\bibinfo {title} {Assignment of
  pre-edge peaks in k-edge x-ray absorption spectra of 3d transition metal
  compounds: Electric dipole or quadrupole?}}\ }\href {\doibase
  10.1002/xrs.1103} {\bibfield  {journal} {\bibinfo  {journal} {X-Ray
  Spectrometry: An International Journal}\ }\textbf {\bibinfo {volume} {37}},\
  \bibinfo {pages} {572--584} (\bibinfo {year} {2008})}\BibitemShut {NoStop}%
\bibitem [{\citenamefont {Saitoh}\ \emph {et~al.}(1995)\citenamefont {Saitoh},
  \citenamefont {Bocquet}, \citenamefont {Mizokawa}, \citenamefont {Namatame},
  \citenamefont {Fujimori}, \citenamefont {Abbate}, \citenamefont {Takeda},\
  and\ \citenamefont {Takano}}]{saitoh1995electronic}%
  \BibitemOpen
  \bibfield  {author} {\bibinfo {author} {\bibfnamefont {T}~\bibnamefont
  {Saitoh}}, \bibinfo {author} {\bibfnamefont {AE}~\bibnamefont {Bocquet}},
  \bibinfo {author} {\bibfnamefont {T}~\bibnamefont {Mizokawa}}, \bibinfo
  {author} {\bibfnamefont {H}~\bibnamefont {Namatame}}, \bibinfo {author}
  {\bibfnamefont {A}~\bibnamefont {Fujimori}}, \bibinfo {author} {\bibfnamefont
  {M}~\bibnamefont {Abbate}}, \bibinfo {author} {\bibfnamefont {Y}~\bibnamefont
  {Takeda}}, \ and\ \bibinfo {author} {\bibfnamefont {M}~\bibnamefont
  {Takano}},\ }\bibfield  {title} {\enquote {\bibinfo {title} {Electronic
  structure of la 1-x sr x mno 3 studied by photoemission and x-ray-absorption
  spectroscopy},}\ }\href {\doibase 10.1103/PhysRevB.51.13942} {\bibfield
  {journal} {\bibinfo  {journal} {Physical Review B}\ }\textbf {\bibinfo
  {volume} {51}},\ \bibinfo {pages} {13942} (\bibinfo {year}
  {1995})}\BibitemShut {NoStop}%
\bibitem [{\citenamefont {Prendergast}\ and\ \citenamefont
  {Galli}(2006)}]{prendergast2006x}%
  \BibitemOpen
  \bibfield  {author} {\bibinfo {author} {\bibfnamefont {David}\ \bibnamefont
  {Prendergast}}\ and\ \bibinfo {author} {\bibfnamefont {Giulia}\ \bibnamefont
  {Galli}},\ }\bibfield  {title} {\enquote {\bibinfo {title} {X-ray absorption
  spectra of water from first principles calculations},}\ }\href {\doibase
  10.1103/PhysRevLett.96.215502} {\bibfield  {journal} {\bibinfo  {journal}
  {Physical Review Letters}\ }\textbf {\bibinfo {volume} {96}},\ \bibinfo
  {pages} {215502} (\bibinfo {year} {2006})}\BibitemShut {NoStop}%
\bibitem [{\citenamefont {Zhang}\ \emph {et~al.}(2017)\citenamefont {Zhang},
  \citenamefont {Topsakal}, \citenamefont {Cama}, \citenamefont {Pelliccione},
  \citenamefont {Zhao}, \citenamefont {Ehrlich}, \citenamefont {Wu},
  \citenamefont {Zhu}, \citenamefont {Frenkel}, \citenamefont {Takeuchi} \emph
  {et~al.}}]{zhang2017multi}%
  \BibitemOpen
  \bibfield  {author} {\bibinfo {author} {\bibfnamefont {Wei}\ \bibnamefont
  {Zhang}}, \bibinfo {author} {\bibfnamefont {Mehmet}\ \bibnamefont
  {Topsakal}}, \bibinfo {author} {\bibfnamefont {Christina}\ \bibnamefont
  {Cama}}, \bibinfo {author} {\bibfnamefont {Christopher~J}\ \bibnamefont
  {Pelliccione}}, \bibinfo {author} {\bibfnamefont {Hu}~\bibnamefont {Zhao}},
  \bibinfo {author} {\bibfnamefont {Steven}\ \bibnamefont {Ehrlich}}, \bibinfo
  {author} {\bibfnamefont {Lijun}\ \bibnamefont {Wu}}, \bibinfo {author}
  {\bibfnamefont {Yimei}\ \bibnamefont {Zhu}}, \bibinfo {author} {\bibfnamefont
  {Anatoly~I}\ \bibnamefont {Frenkel}}, \bibinfo {author} {\bibfnamefont
  {Kenneth~J}\ \bibnamefont {Takeuchi}},  \emph {et~al.},\ }\bibfield  {title}
  {\enquote {\bibinfo {title} {Multi-stage structural transformations in
  zero-strain lithium titanate unveiled by in situ x-ray absorption
  fingerprints},}\ }\href {\doibase 10.1021/jacs.7b07628} {\bibfield  {journal}
  {\bibinfo  {journal} {Journal of the American Chemical Society}\ }\textbf
  {\bibinfo {volume} {139}},\ \bibinfo {pages} {16591--16603} (\bibinfo {year}
  {2017})}\BibitemShut {NoStop}%
\bibitem [{\citenamefont {Li}\ \emph {et~al.}(2015)\citenamefont {Li},
  \citenamefont {Zakharov}, \citenamefont {Zhao}, \citenamefont {Tappero},
  \citenamefont {Jung}, \citenamefont {Elsen}, \citenamefont {Baumann},
  \citenamefont {Nuzzo}, \citenamefont {Stach},\ and\ \citenamefont
  {Frenkel}}]{li2015complex}%
  \BibitemOpen
  \bibfield  {author} {\bibinfo {author} {\bibfnamefont {Y}~\bibnamefont {Li}},
  \bibinfo {author} {\bibfnamefont {D}~\bibnamefont {Zakharov}}, \bibinfo
  {author} {\bibfnamefont {S}~\bibnamefont {Zhao}}, \bibinfo {author}
  {\bibfnamefont {R}~\bibnamefont {Tappero}}, \bibinfo {author} {\bibfnamefont
  {U}~\bibnamefont {Jung}}, \bibinfo {author} {\bibfnamefont {A}~\bibnamefont
  {Elsen}}, \bibinfo {author} {\bibfnamefont {Ph}~\bibnamefont {Baumann}},
  \bibinfo {author} {\bibfnamefont {Ralph~G}\ \bibnamefont {Nuzzo}}, \bibinfo
  {author} {\bibfnamefont {EA}~\bibnamefont {Stach}}, \ and\ \bibinfo {author}
  {\bibfnamefont {AI}~\bibnamefont {Frenkel}},\ }\bibfield  {title} {\enquote
  {\bibinfo {title} {Complex structural dynamics of nanocatalysts revealed in
  operando conditions by correlated imaging and spectroscopy probes},}\ }\href
  {\doibase 10.1038/ncomms8583} {\bibfield  {journal} {\bibinfo  {journal}
  {Nature Communications}\ }\textbf {\bibinfo {volume} {6}},\ \bibinfo {pages}
  {1--6} (\bibinfo {year} {2015})}\BibitemShut {NoStop}%
\bibitem [{\citenamefont {Yano}\ and\ \citenamefont
  {Yachandra}(2009)}]{yano2009x}%
  \BibitemOpen
  \bibfield  {author} {\bibinfo {author} {\bibfnamefont {Junko}\ \bibnamefont
  {Yano}}\ and\ \bibinfo {author} {\bibfnamefont {Vittal~K}\ \bibnamefont
  {Yachandra}},\ }\bibfield  {title} {\enquote {\bibinfo {title} {X-ray
  absorption spectroscopy},}\ }\href@noop {} {\bibfield  {journal} {\bibinfo
  {journal} {Photosynthesis Research}\ }\textbf {\bibinfo {volume} {102}},\
  \bibinfo {pages} {241--254} (\bibinfo {year} {2009})}\BibitemShut {NoStop}%
\bibitem [{\citenamefont {De~Groot}\ and\ \citenamefont
  {Kotani}(2008)}]{de2008core}%
  \BibitemOpen
  \bibfield  {author} {\bibinfo {author} {\bibfnamefont {Frank}\ \bibnamefont
  {De~Groot}}\ and\ \bibinfo {author} {\bibfnamefont {Akio}\ \bibnamefont
  {Kotani}},\ }\href@noop {} {\emph {\bibinfo {title} {Core level spectroscopy
  of solids}}}\ (\bibinfo  {publisher} {CRC press},\ \bibinfo {year}
  {2008})\BibitemShut {NoStop}%
\bibitem [{\citenamefont {Haverkort}\ \emph {et~al.}(2012)\citenamefont
  {Haverkort}, \citenamefont {Zwierzycki},\ and\ \citenamefont
  {Andersen}}]{Haverkort2012}%
  \BibitemOpen
  \bibfield  {author} {\bibinfo {author} {\bibfnamefont {M.~W.}\ \bibnamefont
  {Haverkort}}, \bibinfo {author} {\bibfnamefont {M.}~\bibnamefont
  {Zwierzycki}}, \ and\ \bibinfo {author} {\bibfnamefont {O.~K.}\ \bibnamefont
  {Andersen}},\ }\bibfield  {title} {\enquote {\bibinfo {title} {Multiplet
  ligand-field theory using wannier orbitals},}\ }\href {\doibase
  10.1103/PhysRevB.85.165113} {\bibfield  {journal} {\bibinfo  {journal} {Phys.
  Rev. B}\ }\textbf {\bibinfo {volume} {85}},\ \bibinfo {pages} {165113}
  (\bibinfo {year} {2012})}\BibitemShut {NoStop}%
\bibitem [{\citenamefont {Liang}\ \emph {et~al.}(2017)\citenamefont {Liang},
  \citenamefont {Vinson}, \citenamefont {Pemmaraju}, \citenamefont {Drisdell},
  \citenamefont {Shirley},\ and\ \citenamefont
  {Prendergast}}]{PhysRevLett.118.096402}%
  \BibitemOpen
  \bibfield  {author} {\bibinfo {author} {\bibfnamefont {Yufeng}\ \bibnamefont
  {Liang}}, \bibinfo {author} {\bibfnamefont {John}\ \bibnamefont {Vinson}},
  \bibinfo {author} {\bibfnamefont {Sri}\ \bibnamefont {Pemmaraju}}, \bibinfo
  {author} {\bibfnamefont {Walter~S.}\ \bibnamefont {Drisdell}}, \bibinfo
  {author} {\bibfnamefont {Eric~L.}\ \bibnamefont {Shirley}}, \ and\ \bibinfo
  {author} {\bibfnamefont {David}\ \bibnamefont {Prendergast}},\ }\bibfield
  {title} {\enquote {\bibinfo {title} {Accurate x-ray spectral predictions: An
  advanced self-consistent-field approach inspired by many-body perturbation
  theory},}\ }\href {\doibase 10.1103/PhysRevLett.118.096402} {\bibfield
  {journal} {\bibinfo  {journal} {Phys. Rev. Lett.}\ }\textbf {\bibinfo
  {volume} {118}},\ \bibinfo {pages} {096402} (\bibinfo {year}
  {2017})}\BibitemShut {NoStop}%
\bibitem [{\citenamefont {Tang}\ \emph
  {et~al.}(2022{\natexlab{a}})\citenamefont {Tang}, \citenamefont {Li},
  \citenamefont {Zhang}, \citenamefont {Louie}, \citenamefont {Car},
  \citenamefont {Qiu},\ and\ \citenamefont {Wu}}]{tang2022many}%
  \BibitemOpen
  \bibfield  {author} {\bibinfo {author} {\bibfnamefont {Fujie}\ \bibnamefont
  {Tang}}, \bibinfo {author} {\bibfnamefont {Zhenglu}\ \bibnamefont {Li}},
  \bibinfo {author} {\bibfnamefont {Chunyi}\ \bibnamefont {Zhang}}, \bibinfo
  {author} {\bibfnamefont {Steven~G}\ \bibnamefont {Louie}}, \bibinfo {author}
  {\bibfnamefont {Roberto}\ \bibnamefont {Car}}, \bibinfo {author}
  {\bibfnamefont {Diana~Y}\ \bibnamefont {Qiu}}, \ and\ \bibinfo {author}
  {\bibfnamefont {Xifan}\ \bibnamefont {Wu}},\ }\bibfield  {title} {\enquote
  {\bibinfo {title} {Many-body effects in the x-ray absorption spectra of
  liquid water},}\ }\href@noop {} {\bibfield  {journal} {\bibinfo  {journal}
  {Proceedings of the National Academy of Sciences}\ }\textbf {\bibinfo
  {volume} {119}},\ \bibinfo {pages} {e2201258119} (\bibinfo {year}
  {2022}{\natexlab{a}})}\BibitemShut {NoStop}%
\bibitem [{\citenamefont {Chen}\ \emph
  {et~al.}(2021{\natexlab{a}})\citenamefont {Chen}, \citenamefont {Andrejevic},
  \citenamefont {Drucker}, \citenamefont {Nguyen}, \citenamefont {Xian},
  \citenamefont {Smidt}, \citenamefont {Wang}, \citenamefont {Ernstorfer},
  \citenamefont {Tennant}, \citenamefont {Chan},\ and\ \citenamefont
  {Li}}]{doi:10.1063/5.0049111}%
  \BibitemOpen
  \bibfield  {author} {\bibinfo {author} {\bibfnamefont {Zhantao}\ \bibnamefont
  {Chen}}, \bibinfo {author} {\bibfnamefont {Nina}\ \bibnamefont {Andrejevic}},
  \bibinfo {author} {\bibfnamefont {Nathan~C.}\ \bibnamefont {Drucker}},
  \bibinfo {author} {\bibfnamefont {Thanh}\ \bibnamefont {Nguyen}}, \bibinfo
  {author} {\bibfnamefont {R.~Patrick}\ \bibnamefont {Xian}}, \bibinfo {author}
  {\bibfnamefont {Tess}\ \bibnamefont {Smidt}}, \bibinfo {author}
  {\bibfnamefont {Yao}\ \bibnamefont {Wang}}, \bibinfo {author} {\bibfnamefont
  {Ralph}\ \bibnamefont {Ernstorfer}}, \bibinfo {author} {\bibfnamefont
  {D.~Alan}\ \bibnamefont {Tennant}}, \bibinfo {author} {\bibfnamefont {Maria}\
  \bibnamefont {Chan}}, \ and\ \bibinfo {author} {\bibfnamefont {Mingda}\
  \bibnamefont {Li}},\ }\bibfield  {title} {\enquote {\bibinfo {title} {Machine
  learning on neutron and x-ray scattering and spectroscopies},}\ }\href
  {\doibase 10.1063/5.0049111} {\bibfield  {journal} {\bibinfo  {journal}
  {Chemical Physics Reviews}\ }\textbf {\bibinfo {volume} {2}},\ \bibinfo
  {pages} {031301} (\bibinfo {year} {2021}{\natexlab{a}})}\BibitemShut
  {NoStop}%
\bibitem [{\citenamefont {Timoshenko}\ and\ \citenamefont
  {Frenkel}(2019)}]{timoshenko2019inverting}%
  \BibitemOpen
  \bibfield  {author} {\bibinfo {author} {\bibfnamefont {Janis}\ \bibnamefont
  {Timoshenko}}\ and\ \bibinfo {author} {\bibfnamefont {Anatoly~I}\
  \bibnamefont {Frenkel}},\ }\bibfield  {title} {\enquote {\bibinfo {title}
  {“inverting” x-ray absorption spectra of catalysts by machine learning in
  search for activity descriptors},}\ }\href {\doibase
  10.1021/acscatal.9b03599} {\bibfield  {journal} {\bibinfo  {journal} {Acs
  Catalysis}\ }\textbf {\bibinfo {volume} {9}},\ \bibinfo {pages}
  {10192--10211} (\bibinfo {year} {2019})}\BibitemShut {NoStop}%
\bibitem [{\citenamefont {Timoshenko}\ \emph {et~al.}(2017)\citenamefont
  {Timoshenko}, \citenamefont {Lu}, \citenamefont {Lin},\ and\ \citenamefont
  {Frenkel}}]{doi:10.1021/acs.jpclett.7b02364}%
  \BibitemOpen
  \bibfield  {author} {\bibinfo {author} {\bibfnamefont {Janis}\ \bibnamefont
  {Timoshenko}}, \bibinfo {author} {\bibfnamefont {Deyu}\ \bibnamefont {Lu}},
  \bibinfo {author} {\bibfnamefont {Yuewei}\ \bibnamefont {Lin}}, \ and\
  \bibinfo {author} {\bibfnamefont {Anatoly~I.}\ \bibnamefont {Frenkel}},\
  }\bibfield  {title} {\enquote {\bibinfo {title} {Supervised
  machine-learning-based determination of three-dimensional structure of
  metallic nanoparticles},}\ }\href {\doibase 10.1021/acs.jpclett.7b02364}
  {\bibfield  {journal} {\bibinfo  {journal} {The Journal of Physical Chemistry
  Letters}\ }\textbf {\bibinfo {volume} {8}},\ \bibinfo {pages} {5091--5098}
  (\bibinfo {year} {2017})},\ \bibinfo {note} {pMID: 28960990}\BibitemShut
  {NoStop}%
\bibitem [{\citenamefont {Marcella}\ \emph {et~al.}(2020)\citenamefont
  {Marcella}, \citenamefont {Liu}, \citenamefont {Timoshenko}, \citenamefont
  {Guan}, \citenamefont {Luneau}, \citenamefont {Shirman}, \citenamefont
  {Plonka}, \citenamefont {van~der Hoeven}, \citenamefont {Aizenberg},
  \citenamefont {Friend} \emph {et~al.}}]{marcella2020neural}%
  \BibitemOpen
  \bibfield  {author} {\bibinfo {author} {\bibfnamefont {Nicholas}\
  \bibnamefont {Marcella}}, \bibinfo {author} {\bibfnamefont {Yang}\
  \bibnamefont {Liu}}, \bibinfo {author} {\bibfnamefont {Janis}\ \bibnamefont
  {Timoshenko}}, \bibinfo {author} {\bibfnamefont {Erjia}\ \bibnamefont
  {Guan}}, \bibinfo {author} {\bibfnamefont {Mathilde}\ \bibnamefont {Luneau}},
  \bibinfo {author} {\bibfnamefont {Tanya}\ \bibnamefont {Shirman}}, \bibinfo
  {author} {\bibfnamefont {Anna~M}\ \bibnamefont {Plonka}}, \bibinfo {author}
  {\bibfnamefont {Jessi~ES}\ \bibnamefont {van~der Hoeven}}, \bibinfo {author}
  {\bibfnamefont {Joanna}\ \bibnamefont {Aizenberg}}, \bibinfo {author}
  {\bibfnamefont {Cynthia~M}\ \bibnamefont {Friend}},  \emph {et~al.},\
  }\bibfield  {title} {\enquote {\bibinfo {title} {Neural network assisted
  analysis of bimetallic nanocatalysts using x-ray absorption near edge
  structure spectroscopy},}\ }\href {\doibase 10.1039/D0CP02098B} {\bibfield
  {journal} {\bibinfo  {journal} {Physical Chemistry Chemical Physics}\
  }\textbf {\bibinfo {volume} {22}},\ \bibinfo {pages} {18902--18910} (\bibinfo
  {year} {2020})}\BibitemShut {NoStop}%
\bibitem [{\citenamefont {Liu}\ \emph {et~al.}(2021)\citenamefont {Liu},
  \citenamefont {Halder}, \citenamefont {Seifert}, \citenamefont {Marcella},
  \citenamefont {Vajda},\ and\ \citenamefont {Frenkel}}]{liu2021probing}%
  \BibitemOpen
  \bibfield  {author} {\bibinfo {author} {\bibfnamefont {Yang}\ \bibnamefont
  {Liu}}, \bibinfo {author} {\bibfnamefont {Avik}\ \bibnamefont {Halder}},
  \bibinfo {author} {\bibfnamefont {Soenke}\ \bibnamefont {Seifert}}, \bibinfo
  {author} {\bibfnamefont {Nicholas}\ \bibnamefont {Marcella}}, \bibinfo
  {author} {\bibfnamefont {Stefan}\ \bibnamefont {Vajda}}, \ and\ \bibinfo
  {author} {\bibfnamefont {Anatoly~I}\ \bibnamefont {Frenkel}},\ }\bibfield
  {title} {\enquote {\bibinfo {title} {Probing active sites in cu x pd y
  cluster catalysts by machine-learning-assisted x-ray absorption
  spectroscopy},}\ }\href {\doibase 10.1021/acsami.1c06714} {\bibfield
  {journal} {\bibinfo  {journal} {ACS Applied Materials \& Interfaces}\
  }\textbf {\bibinfo {volume} {13}},\ \bibinfo {pages} {53363--53374} (\bibinfo
  {year} {2021})}\BibitemShut {NoStop}%
\bibitem [{\citenamefont {Liu}\ \emph {et~al.}(2019)\citenamefont {Liu},
  \citenamefont {Marcella}, \citenamefont {Timoshenko}, \citenamefont {Halder},
  \citenamefont {Yang}, \citenamefont {Kolipaka}, \citenamefont {Pellin},
  \citenamefont {Seifert}, \citenamefont {Vajda}, \citenamefont {Liu} \emph
  {et~al.}}]{liu2019mapping}%
  \BibitemOpen
  \bibfield  {author} {\bibinfo {author} {\bibfnamefont {Yang}\ \bibnamefont
  {Liu}}, \bibinfo {author} {\bibfnamefont {Nicholas}\ \bibnamefont
  {Marcella}}, \bibinfo {author} {\bibfnamefont {Janis}\ \bibnamefont
  {Timoshenko}}, \bibinfo {author} {\bibfnamefont {Avik}\ \bibnamefont
  {Halder}}, \bibinfo {author} {\bibfnamefont {Bing}\ \bibnamefont {Yang}},
  \bibinfo {author} {\bibfnamefont {Lakshmi}\ \bibnamefont {Kolipaka}},
  \bibinfo {author} {\bibfnamefont {Michael~J}\ \bibnamefont {Pellin}},
  \bibinfo {author} {\bibfnamefont {Soenke}\ \bibnamefont {Seifert}}, \bibinfo
  {author} {\bibfnamefont {Stefan}\ \bibnamefont {Vajda}}, \bibinfo {author}
  {\bibfnamefont {Ping}\ \bibnamefont {Liu}},  \emph {et~al.},\ }\bibfield
  {title} {\enquote {\bibinfo {title} {Mapping xanes spectra on structural
  descriptors of copper oxide clusters using supervised machine learning},}\
  }\href {\doibase 10.1063/1.5126597} {\bibfield  {journal} {\bibinfo
  {journal} {The Journal of Chemical Physics}\ }\textbf {\bibinfo {volume}
  {151}},\ \bibinfo {pages} {164201} (\bibinfo {year} {2019})}\BibitemShut
  {NoStop}%
\bibitem [{\citenamefont {Guda}\ \emph {et~al.}(2021)\citenamefont {Guda},
  \citenamefont {Guda}, \citenamefont {Martini}, \citenamefont {Kravtsova},
  \citenamefont {Algasov}, \citenamefont {Bugaev}, \citenamefont {Kubrin},
  \citenamefont {Guda}, \citenamefont {{\v{S}}ot}, \citenamefont {van Bokhoven}
  \emph {et~al.}}]{guda2021understanding}%
  \BibitemOpen
  \bibfield  {author} {\bibinfo {author} {\bibfnamefont {Alexander~A}\
  \bibnamefont {Guda}}, \bibinfo {author} {\bibfnamefont {Sergey~A}\
  \bibnamefont {Guda}}, \bibinfo {author} {\bibfnamefont {Andrea}\ \bibnamefont
  {Martini}}, \bibinfo {author} {\bibfnamefont {AN}~\bibnamefont {Kravtsova}},
  \bibinfo {author} {\bibfnamefont {Alexander}\ \bibnamefont {Algasov}},
  \bibinfo {author} {\bibfnamefont {Aram}\ \bibnamefont {Bugaev}}, \bibinfo
  {author} {\bibfnamefont {SP}~\bibnamefont {Kubrin}}, \bibinfo {author}
  {\bibfnamefont {LV}~\bibnamefont {Guda}}, \bibinfo {author} {\bibfnamefont
  {P}~\bibnamefont {{\v{S}}ot}}, \bibinfo {author} {\bibfnamefont
  {JA}~\bibnamefont {van Bokhoven}},  \emph {et~al.},\ }\bibfield  {title}
  {\enquote {\bibinfo {title} {Understanding x-ray absorption spectra by means
  of descriptors and machine learning algorithms},}\ }\href {\doibase
  10.1038/s41524-021-00664-9} {\bibfield  {journal} {\bibinfo  {journal} {npj
  Computational Materials}\ }\textbf {\bibinfo {volume} {7}},\ \bibinfo {pages}
  {1--13} (\bibinfo {year} {2021})}\BibitemShut {NoStop}%
\bibitem [{\citenamefont {Li}\ \emph {et~al.}(2019)\citenamefont {Li},
  \citenamefont {Lu},\ and\ \citenamefont {Chan}}]{li2019deep}%
  \BibitemOpen
  \bibfield  {author} {\bibinfo {author} {\bibfnamefont {Liang}\ \bibnamefont
  {Li}}, \bibinfo {author} {\bibfnamefont {Mindren}\ \bibnamefont {Lu}}, \ and\
  \bibinfo {author} {\bibfnamefont {Maria~KY}\ \bibnamefont {Chan}},\
  }\bibfield  {title} {\enquote {\bibinfo {title} {A deep learning model for
  atomic structures prediction using x-ray absorption spectroscopic data},}\
  }\href {\doibase 10.48550/arXiv.1905.03928} {\bibfield  {journal} {\bibinfo
  {journal} {arXiv preprint arXiv:1905.03928}\ } (\bibinfo {year} {2019}),\
  10.48550/arXiv.1905.03928}\BibitemShut {NoStop}%
\bibitem [{\citenamefont {Aarva}\ \emph
  {et~al.}(2019{\natexlab{a}})\citenamefont {Aarva}, \citenamefont {Deringer},
  \citenamefont {Sainio}, \citenamefont {Laurila},\ and\ \citenamefont
  {Caro}}]{aarva2019understandingI}%
  \BibitemOpen
  \bibfield  {author} {\bibinfo {author} {\bibfnamefont {Anja}\ \bibnamefont
  {Aarva}}, \bibinfo {author} {\bibfnamefont {Volker~L}\ \bibnamefont
  {Deringer}}, \bibinfo {author} {\bibfnamefont {Sami}\ \bibnamefont {Sainio}},
  \bibinfo {author} {\bibfnamefont {Tomi}\ \bibnamefont {Laurila}}, \ and\
  \bibinfo {author} {\bibfnamefont {Miguel~A}\ \bibnamefont {Caro}},\
  }\bibfield  {title} {\enquote {\bibinfo {title} {Understanding x-ray
  spectroscopy of carbonaceous materials by combining experiments, density
  functional theory, and machine learning. part i: Fingerprint spectra},}\
  }\href {\doibase 10.1021/acs.chemmater.9b02049} {\bibfield  {journal}
  {\bibinfo  {journal} {Chemistry of Materials}\ }\textbf {\bibinfo {volume}
  {31}},\ \bibinfo {pages} {9243--9255} (\bibinfo {year}
  {2019}{\natexlab{a}})}\BibitemShut {NoStop}%
\bibitem [{\citenamefont {Aarva}\ \emph
  {et~al.}(2019{\natexlab{b}})\citenamefont {Aarva}, \citenamefont {Deringer},
  \citenamefont {Sainio}, \citenamefont {Laurila},\ and\ \citenamefont
  {Caro}}]{aarva2019understandingII}%
  \BibitemOpen
  \bibfield  {author} {\bibinfo {author} {\bibfnamefont {Anja}\ \bibnamefont
  {Aarva}}, \bibinfo {author} {\bibfnamefont {Volker~L.}\ \bibnamefont
  {Deringer}}, \bibinfo {author} {\bibfnamefont {Sami}\ \bibnamefont {Sainio}},
  \bibinfo {author} {\bibfnamefont {Tomi}\ \bibnamefont {Laurila}}, \ and\
  \bibinfo {author} {\bibfnamefont {Miguel~A.}\ \bibnamefont {Caro}},\
  }\bibfield  {title} {\enquote {\bibinfo {title} {Understanding x-ray
  spectroscopy of carbonaceous materials by combining experiments, density
  functional theory, and machine learning. part ii: Quantitative fitting of
  spectra},}\ }\href {\doibase 10.1021/acs.chemmater.9b02050} {\bibfield
  {journal} {\bibinfo  {journal} {Chemistry of Materials}\ }\textbf {\bibinfo
  {volume} {31}},\ \bibinfo {pages} {9256--9267} (\bibinfo {year}
  {2019}{\natexlab{b}})}\BibitemShut {NoStop}%
\bibitem [{\citenamefont {Routh}\ \emph {et~al.}(2021)\citenamefont {Routh},
  \citenamefont {Liu}, \citenamefont {Marcella}, \citenamefont {Kozinsky},\
  and\ \citenamefont {Frenkel}}]{routh2021latent}%
  \BibitemOpen
  \bibfield  {author} {\bibinfo {author} {\bibfnamefont {Prahlad~K}\
  \bibnamefont {Routh}}, \bibinfo {author} {\bibfnamefont {Yang}\ \bibnamefont
  {Liu}}, \bibinfo {author} {\bibfnamefont {Nicholas}\ \bibnamefont
  {Marcella}}, \bibinfo {author} {\bibfnamefont {Boris}\ \bibnamefont
  {Kozinsky}}, \ and\ \bibinfo {author} {\bibfnamefont {Anatoly~I}\
  \bibnamefont {Frenkel}},\ }\bibfield  {title} {\enquote {\bibinfo {title}
  {Latent representation learning for structural characterization of
  catalysts},}\ }\href {\doibase 10.1021/acs.jpclett.0c03792} {\bibfield
  {journal} {\bibinfo  {journal} {The Journal of Physical Chemistry Letters}\
  }\textbf {\bibinfo {volume} {12}},\ \bibinfo {pages} {2086--2094} (\bibinfo
  {year} {2021})}\BibitemShut {NoStop}%
\bibitem [{\citenamefont {Xiang}\ \emph {et~al.}(2022)\citenamefont {Xiang},
  \citenamefont {Huang}, \citenamefont {Li}, \citenamefont {Liu}, \citenamefont
  {Marcella}, \citenamefont {Routh}, \citenamefont {Li},\ and\ \citenamefont
  {Frenkel}}]{xiang2022solving}%
  \BibitemOpen
  \bibfield  {author} {\bibinfo {author} {\bibfnamefont {Shuting}\ \bibnamefont
  {Xiang}}, \bibinfo {author} {\bibfnamefont {Peipei}\ \bibnamefont {Huang}},
  \bibinfo {author} {\bibfnamefont {Junying}\ \bibnamefont {Li}}, \bibinfo
  {author} {\bibfnamefont {Yang}\ \bibnamefont {Liu}}, \bibinfo {author}
  {\bibfnamefont {Nicholas}\ \bibnamefont {Marcella}}, \bibinfo {author}
  {\bibfnamefont {Prahlad~K}\ \bibnamefont {Routh}}, \bibinfo {author}
  {\bibfnamefont {Gonghu}\ \bibnamefont {Li}}, \ and\ \bibinfo {author}
  {\bibfnamefont {Anatoly~I}\ \bibnamefont {Frenkel}},\ }\bibfield  {title}
  {\enquote {\bibinfo {title} {Solving the structure of “single-atom”
  catalysts using machine learning--assisted xanes analysis},}\ }\href
  {\doibase 10.1039/D1CP05513E} {\bibfield  {journal} {\bibinfo  {journal}
  {Physical Chemistry Chemical Physics}\ }\textbf {\bibinfo {volume} {24}},\
  \bibinfo {pages} {5116--5124} (\bibinfo {year} {2022})}\BibitemShut {NoStop}%
\bibitem [{\citenamefont {Trejo}\ \emph {et~al.}(2019)\citenamefont {Trejo},
  \citenamefont {Dadlani}, \citenamefont {De~La~Paz}, \citenamefont {Acharya},
  \citenamefont {Kravec}, \citenamefont {Nordlund}, \citenamefont {Sarangi},
  \citenamefont {Prinz}, \citenamefont {Torgersen},\ and\ \citenamefont
  {Dasgupta}}]{trejo2019elucidating}%
  \BibitemOpen
  \bibfield  {author} {\bibinfo {author} {\bibfnamefont {Orlando}\ \bibnamefont
  {Trejo}}, \bibinfo {author} {\bibfnamefont {Anup~L}\ \bibnamefont {Dadlani}},
  \bibinfo {author} {\bibfnamefont {Francisco}\ \bibnamefont {De~La~Paz}},
  \bibinfo {author} {\bibfnamefont {Shinjita}\ \bibnamefont {Acharya}},
  \bibinfo {author} {\bibfnamefont {Rob}\ \bibnamefont {Kravec}}, \bibinfo
  {author} {\bibfnamefont {Dennis}\ \bibnamefont {Nordlund}}, \bibinfo {author}
  {\bibfnamefont {Ritimukta}\ \bibnamefont {Sarangi}}, \bibinfo {author}
  {\bibfnamefont {Fritz~B}\ \bibnamefont {Prinz}}, \bibinfo {author}
  {\bibfnamefont {Jan}\ \bibnamefont {Torgersen}}, \ and\ \bibinfo {author}
  {\bibfnamefont {Neil~P}\ \bibnamefont {Dasgupta}},\ }\bibfield  {title}
  {\enquote {\bibinfo {title} {Elucidating the evolving atomic structure in
  atomic layer deposition reactions with in situ xanes and machine learning},}\
  }\href {\doibase 10.1021/acs.chemmater.9b03025} {\bibfield  {journal}
  {\bibinfo  {journal} {Chemistry of Materials}\ }\textbf {\bibinfo {volume}
  {31}},\ \bibinfo {pages} {8937--8947} (\bibinfo {year} {2019})}\BibitemShut
  {NoStop}%
\bibitem [{\citenamefont {Carbone}\ \emph {et~al.}(2020)\citenamefont
  {Carbone}, \citenamefont {Topsakal}, \citenamefont {Lu},\ and\ \citenamefont
  {Yoo}}]{carbone2020machine}%
  \BibitemOpen
  \bibfield  {author} {\bibinfo {author} {\bibfnamefont {Matthew~R}\
  \bibnamefont {Carbone}}, \bibinfo {author} {\bibfnamefont {Mehmet}\
  \bibnamefont {Topsakal}}, \bibinfo {author} {\bibfnamefont {Deyu}\
  \bibnamefont {Lu}}, \ and\ \bibinfo {author} {\bibfnamefont {Shinjae}\
  \bibnamefont {Yoo}},\ }\bibfield  {title} {\enquote {\bibinfo {title}
  {Machine-learning x-ray absorption spectra to quantitative accuracy},}\
  }\href {\doibase 10.1103/PhysRevLett.124.156401} {\bibfield  {journal}
  {\bibinfo  {journal} {Physical Review Letters}\ }\textbf {\bibinfo {volume}
  {124}},\ \bibinfo {pages} {156401} (\bibinfo {year} {2020})}\BibitemShut
  {NoStop}%
\bibitem [{\citenamefont {Tetef}\ \emph {et~al.}(2021)\citenamefont {Tetef},
  \citenamefont {Govind},\ and\ \citenamefont
  {Seidler}}]{tetef2021unsupervised}%
  \BibitemOpen
  \bibfield  {author} {\bibinfo {author} {\bibfnamefont {Samantha}\
  \bibnamefont {Tetef}}, \bibinfo {author} {\bibfnamefont {Niranjan}\
  \bibnamefont {Govind}}, \ and\ \bibinfo {author} {\bibfnamefont {Gerald~T}\
  \bibnamefont {Seidler}},\ }\bibfield  {title} {\enquote {\bibinfo {title}
  {Unsupervised machine learning for unbiased chemical classification in x-ray
  absorption spectroscopy and x-ray emission spectroscopy},}\ }\href {\doibase
  10.1039/D1CP02903G} {\bibfield  {journal} {\bibinfo  {journal} {Physical
  Chemistry Chemical Physics}\ }\textbf {\bibinfo {volume} {23}},\ \bibinfo
  {pages} {23586--23601} (\bibinfo {year} {2021})}\BibitemShut {NoStop}%
\bibitem [{\citenamefont {Carbone}\ \emph {et~al.}(2019)\citenamefont
  {Carbone}, \citenamefont {Yoo}, \citenamefont {Topsakal},\ and\ \citenamefont
  {Lu}}]{carbone2019classification}%
  \BibitemOpen
  \bibfield  {author} {\bibinfo {author} {\bibfnamefont {Matthew~R}\
  \bibnamefont {Carbone}}, \bibinfo {author} {\bibfnamefont {Shinjae}\
  \bibnamefont {Yoo}}, \bibinfo {author} {\bibfnamefont {Mehmet}\ \bibnamefont
  {Topsakal}}, \ and\ \bibinfo {author} {\bibfnamefont {Deyu}\ \bibnamefont
  {Lu}},\ }\bibfield  {title} {\enquote {\bibinfo {title} {Classification of
  local chemical environments from x-ray absorption spectra using supervised
  machine learning},}\ }\href {\doibase 10.1103/PhysRevMaterials.3.033604}
  {\bibfield  {journal} {\bibinfo  {journal} {Physical Review Materials}\
  }\textbf {\bibinfo {volume} {3}},\ \bibinfo {pages} {033604} (\bibinfo {year}
  {2019})}\BibitemShut {NoStop}%
\bibitem [{\citenamefont {Torrisi}\ \emph {et~al.}(2020)\citenamefont
  {Torrisi}, \citenamefont {Carbone}, \citenamefont {Rohr}, \citenamefont
  {Montoya}, \citenamefont {Ha}, \citenamefont {Yano}, \citenamefont {Suram},\
  and\ \citenamefont {Hung}}]{Torrisi2020}%
  \BibitemOpen
  \bibfield  {author} {\bibinfo {author} {\bibfnamefont {Steven~B.}\
  \bibnamefont {Torrisi}}, \bibinfo {author} {\bibfnamefont {Matthew~R.}\
  \bibnamefont {Carbone}}, \bibinfo {author} {\bibfnamefont {Brian~A.}\
  \bibnamefont {Rohr}}, \bibinfo {author} {\bibfnamefont {Joseph~H.}\
  \bibnamefont {Montoya}}, \bibinfo {author} {\bibfnamefont {Yang}\
  \bibnamefont {Ha}}, \bibinfo {author} {\bibfnamefont {Junko}\ \bibnamefont
  {Yano}}, \bibinfo {author} {\bibfnamefont {Santosh~K.}\ \bibnamefont
  {Suram}}, \ and\ \bibinfo {author} {\bibfnamefont {Linda}\ \bibnamefont
  {Hung}},\ }\bibfield  {title} {\enquote {\bibinfo {title} {Random forest
  machine learning models for interpretable x-ray absorption near-edge
  structure spectrum-property relationships},}\ }\href {\doibase
  10.1038/s41524-020-00376-6} {\bibfield  {journal} {\bibinfo  {journal} {npj
  Computational Materials}\ }\textbf {\bibinfo {volume} {6}},\ \bibinfo {pages}
  {109} (\bibinfo {year} {2020})}\BibitemShut {NoStop}%
\bibitem [{\citenamefont {Zheng}\ \emph {et~al.}(2020)\citenamefont {Zheng},
  \citenamefont {Chen}, \citenamefont {Chen},\ and\ \citenamefont
  {Ong}}]{zheng2020random}%
  \BibitemOpen
  \bibfield  {author} {\bibinfo {author} {\bibfnamefont {Chen}\ \bibnamefont
  {Zheng}}, \bibinfo {author} {\bibfnamefont {Chi}\ \bibnamefont {Chen}},
  \bibinfo {author} {\bibfnamefont {Yiming}\ \bibnamefont {Chen}}, \ and\
  \bibinfo {author} {\bibfnamefont {Shyue~Ping}\ \bibnamefont {Ong}},\
  }\bibfield  {title} {\enquote {\bibinfo {title} {Random forest models for
  accurate identification of coordination environments from x-ray absorption
  near-edge structure},}\ }\href {\doibase 10.1016/j.patter.2020.100013}
  {\bibfield  {journal} {\bibinfo  {journal} {Patterns}\ }\textbf {\bibinfo
  {volume} {1}},\ \bibinfo {pages} {100013} (\bibinfo {year}
  {2020})}\BibitemShut {NoStop}%
\bibitem [{\citenamefont {Rankine}\ \emph {et~al.}(2020)\citenamefont
  {Rankine}, \citenamefont {Madkhali},\ and\ \citenamefont
  {Penfold}}]{rankine2020deep}%
  \BibitemOpen
  \bibfield  {author} {\bibinfo {author} {\bibfnamefont {Conor~D}\ \bibnamefont
  {Rankine}}, \bibinfo {author} {\bibfnamefont {Marwah~MM}\ \bibnamefont
  {Madkhali}}, \ and\ \bibinfo {author} {\bibfnamefont {Thomas~J}\ \bibnamefont
  {Penfold}},\ }\bibfield  {title} {\enquote {\bibinfo {title} {A deep neural
  network for the rapid prediction of x-ray absorption spectra},}\ }\href
  {\doibase 10.1021/acs.jpca.0c03723} {\bibfield  {journal} {\bibinfo
  {journal} {The Journal of Physical Chemistry A}\ }\textbf {\bibinfo {volume}
  {124}},\ \bibinfo {pages} {4263--4270} (\bibinfo {year} {2020})}\BibitemShut
  {NoStop}%
\bibitem [{\citenamefont {Yan}\ \emph {et~al.}(2019)\citenamefont {Yan},
  \citenamefont {Topsakal}, \citenamefont {Selcuk}, \citenamefont {Lyons},
  \citenamefont {Zhang}, \citenamefont {Wu}, \citenamefont {Waluyo},
  \citenamefont {Stavitski}, \citenamefont {Attenkofer}, \citenamefont {Yoo}
  \emph {et~al.}}]{yan2019ultrathin}%
  \BibitemOpen
  \bibfield  {author} {\bibinfo {author} {\bibfnamefont {Danhua}\ \bibnamefont
  {Yan}}, \bibinfo {author} {\bibfnamefont {Mehmet}\ \bibnamefont {Topsakal}},
  \bibinfo {author} {\bibfnamefont {Sencer}\ \bibnamefont {Selcuk}}, \bibinfo
  {author} {\bibfnamefont {John~L}\ \bibnamefont {Lyons}}, \bibinfo {author}
  {\bibfnamefont {Wenrui}\ \bibnamefont {Zhang}}, \bibinfo {author}
  {\bibfnamefont {Qiyuan}\ \bibnamefont {Wu}}, \bibinfo {author} {\bibfnamefont
  {Iradwikanari}\ \bibnamefont {Waluyo}}, \bibinfo {author} {\bibfnamefont
  {Eli}\ \bibnamefont {Stavitski}}, \bibinfo {author} {\bibfnamefont {Klaus}\
  \bibnamefont {Attenkofer}}, \bibinfo {author} {\bibfnamefont {Shinjae}\
  \bibnamefont {Yoo}},  \emph {et~al.},\ }\bibfield  {title} {\enquote
  {\bibinfo {title} {Ultrathin amorphous titania on nanowires: Optimization of
  conformal growth and elucidation of atomic-scale motifs},}\ }\href {\doibase
  10.1021/acs.nanolett.8b04888} {\bibfield  {journal} {\bibinfo  {journal}
  {Nano letters}\ }\textbf {\bibinfo {volume} {19}},\ \bibinfo {pages}
  {3457--3463} (\bibinfo {year} {2019})}\BibitemShut {NoStop}%
\bibitem [{\citenamefont {Kuban}\ \emph
  {et~al.}(2022{\natexlab{a}})\citenamefont {Kuban}, \citenamefont {Rigamonti},
  \citenamefont {Scheidgen},\ and\ \citenamefont {Draxl}}]{kuban2022I}%
  \BibitemOpen
  \bibfield  {author} {\bibinfo {author} {\bibfnamefont {Martin}\ \bibnamefont
  {Kuban}}, \bibinfo {author} {\bibfnamefont {Santiago}\ \bibnamefont
  {Rigamonti}}, \bibinfo {author} {\bibfnamefont {Markus}\ \bibnamefont
  {Scheidgen}}, \ and\ \bibinfo {author} {\bibfnamefont {Claudia}\ \bibnamefont
  {Draxl}},\ }\bibfield  {title} {\enquote {\bibinfo {title} {Density-of-states
  similarity descriptor for unsupervised learning from materials data},}\
  }\href {\doibase 10.1038/s41597-022-01754-z} {\bibfield  {journal} {\bibinfo
  {journal} {Sci. Data}\ }\textbf {\bibinfo {volume} {9}},\ \bibinfo {pages}
  {646} (\bibinfo {year} {2022}{\natexlab{a}})}\BibitemShut {NoStop}%
\bibitem [{\citenamefont {Kuban}\ \emph
  {et~al.}(2022{\natexlab{b}})\citenamefont {Kuban}, \citenamefont {Gabaj},
  \citenamefont {Aggoune}, \citenamefont {Vona}, \citenamefont {Rigamonti},\
  and\ \citenamefont {Draxl}}]{kuban2022II}%
  \BibitemOpen
  \bibfield  {author} {\bibinfo {author} {\bibfnamefont {Martin}\ \bibnamefont
  {Kuban}}, \bibinfo {author} {\bibfnamefont {Šimon}\ \bibnamefont {Gabaj}},
  \bibinfo {author} {\bibfnamefont {Wahib}\ \bibnamefont {Aggoune}}, \bibinfo
  {author} {\bibfnamefont {Cecilia}\ \bibnamefont {Vona}}, \bibinfo {author}
  {\bibfnamefont {Santiago}\ \bibnamefont {Rigamonti}}, \ and\ \bibinfo
  {author} {\bibfnamefont {Claudia}\ \bibnamefont {Draxl}},\ }\bibfield
  {title} {\enquote {\bibinfo {title} {Similarity of materials and data-quality
  assessment by fingerprinting},}\ }\href {\doibase 10.1557/s43577-022-00339-w}
  {\bibfield  {journal} {\bibinfo  {journal} {MRS Bulletin}\ }\textbf {\bibinfo
  {volume} {47}},\ \bibinfo {pages} {1} (\bibinfo {year}
  {2022}{\natexlab{b}})}\BibitemShut {NoStop}%
\bibitem [{\citenamefont {Rankine}\ and\ \citenamefont
  {Penfold}(2022)}]{rankine2022accurate}%
  \BibitemOpen
  \bibfield  {author} {\bibinfo {author} {\bibfnamefont {Conor~Douglas}\
  \bibnamefont {Rankine}}\ and\ \bibinfo {author} {\bibfnamefont
  {TJ}~\bibnamefont {Penfold}},\ }\bibfield  {title} {\enquote {\bibinfo
  {title} {Accurate, affordable, and generalizable machine learning simulations
  of transition metal x-ray absorption spectra using the xanesnet deep neural
  network},}\ }\href {\doibase 10.1063/5.0087255} {\bibfield  {journal}
  {\bibinfo  {journal} {The Journal of Chemical Physics}\ }\textbf {\bibinfo
  {volume} {156}},\ \bibinfo {pages} {164102} (\bibinfo {year}
  {2022})}\BibitemShut {NoStop}%
\bibitem [{\citenamefont {Ghose}\ \emph {et~al.}(2023)\citenamefont {Ghose},
  \citenamefont {Segal}, \citenamefont {Meng}, \citenamefont {Liang},
  \citenamefont {Hybertsen}, \citenamefont {Qu}, \citenamefont {Stavitski},
  \citenamefont {Yoo}, \citenamefont {Lu},\ and\ \citenamefont
  {Carbone}}]{Ghose2023}%
  \BibitemOpen
  \bibfield  {author} {\bibinfo {author} {\bibfnamefont {Animesh}\ \bibnamefont
  {Ghose}}, \bibinfo {author} {\bibfnamefont {Mikhail}\ \bibnamefont {Segal}},
  \bibinfo {author} {\bibfnamefont {Fanchen}\ \bibnamefont {Meng}}, \bibinfo
  {author} {\bibfnamefont {Zhu}\ \bibnamefont {Liang}}, \bibinfo {author}
  {\bibfnamefont {Mark~S.}\ \bibnamefont {Hybertsen}}, \bibinfo {author}
  {\bibfnamefont {Xiaohui}\ \bibnamefont {Qu}}, \bibinfo {author}
  {\bibfnamefont {Eli}\ \bibnamefont {Stavitski}}, \bibinfo {author}
  {\bibfnamefont {Shinjae}\ \bibnamefont {Yoo}}, \bibinfo {author}
  {\bibfnamefont {Deyu}\ \bibnamefont {Lu}}, \ and\ \bibinfo {author}
  {\bibfnamefont {Matthew~R.}\ \bibnamefont {Carbone}},\ }\bibfield  {title}
  {\enquote {\bibinfo {title} {Uncertainty-aware predictions of molecular x-ray
  absorption spectra using neural network ensembles},}\ }\href {\doibase
  10.1103/PhysRevResearch.5.013180} {\bibfield  {journal} {\bibinfo  {journal}
  {Phys. Rev. Res.}\ }\textbf {\bibinfo {volume} {5}},\ \bibinfo {pages}
  {013180} (\bibinfo {year} {2023})}\BibitemShut {NoStop}%
\bibitem [{\citenamefont {Mathew}\ \emph {et~al.}(2018)\citenamefont {Mathew},
  \citenamefont {Zheng}, \citenamefont {Winston}, \citenamefont {Chen},
  \citenamefont {Dozier}, \citenamefont {Rehr}, \citenamefont {Ong},\ and\
  \citenamefont {Persson}}]{MP-Kedges}%
  \BibitemOpen
  \bibfield  {author} {\bibinfo {author} {\bibfnamefont {Kiran}\ \bibnamefont
  {Mathew}}, \bibinfo {author} {\bibfnamefont {Chen}\ \bibnamefont {Zheng}},
  \bibinfo {author} {\bibfnamefont {Donald}\ \bibnamefont {Winston}}, \bibinfo
  {author} {\bibfnamefont {Chi}\ \bibnamefont {Chen}}, \bibinfo {author}
  {\bibfnamefont {Alan}\ \bibnamefont {Dozier}}, \bibinfo {author}
  {\bibfnamefont {John~J.}\ \bibnamefont {Rehr}}, \bibinfo {author}
  {\bibfnamefont {Shyue~Ping}\ \bibnamefont {Ong}}, \ and\ \bibinfo {author}
  {\bibfnamefont {Kristin~A.}\ \bibnamefont {Persson}},\ }\bibfield  {title}
  {\enquote {\bibinfo {title} {High-throughput computational x-ray absorption
  spectroscopy},}\ }\href {\doibase 10.1038/sdata.2018.151} {\bibfield
  {journal} {\bibinfo  {journal} {Scientific Data}\ }\textbf {\bibinfo {volume}
  {5}},\ \bibinfo {pages} {180151} (\bibinfo {year} {2018})}\BibitemShut
  {NoStop}%
\bibitem [{\citenamefont {Chen}\ \emph
  {et~al.}(2021{\natexlab{b}})\citenamefont {Chen}, \citenamefont {Chen},
  \citenamefont {Zheng}, \citenamefont {Dwaraknath}, \citenamefont {Horton},
  \citenamefont {Cabana}, \citenamefont {Rehr}, \citenamefont {Vinson},
  \citenamefont {Dozier}, \citenamefont {Kas}, \citenamefont {Persson},\ and\
  \citenamefont {Ong}}]{MP-Ledges}%
  \BibitemOpen
  \bibfield  {author} {\bibinfo {author} {\bibfnamefont {Yiming}\ \bibnamefont
  {Chen}}, \bibinfo {author} {\bibfnamefont {Chi}\ \bibnamefont {Chen}},
  \bibinfo {author} {\bibfnamefont {Chen}\ \bibnamefont {Zheng}}, \bibinfo
  {author} {\bibfnamefont {Shyam}\ \bibnamefont {Dwaraknath}}, \bibinfo
  {author} {\bibfnamefont {Matthew~K.}\ \bibnamefont {Horton}}, \bibinfo
  {author} {\bibfnamefont {Jordi}\ \bibnamefont {Cabana}}, \bibinfo {author}
  {\bibfnamefont {John}\ \bibnamefont {Rehr}}, \bibinfo {author} {\bibfnamefont
  {John}\ \bibnamefont {Vinson}}, \bibinfo {author} {\bibfnamefont {Alan}\
  \bibnamefont {Dozier}}, \bibinfo {author} {\bibfnamefont {Joshua~J.}\
  \bibnamefont {Kas}}, \bibinfo {author} {\bibfnamefont {Kristin~A.}\
  \bibnamefont {Persson}}, \ and\ \bibinfo {author} {\bibfnamefont
  {Shyue~Ping}\ \bibnamefont {Ong}},\ }\bibfield  {title} {\enquote {\bibinfo
  {title} {Database of ab initio l-edge x-ray absorption near edge
  structure},}\ }\href {\doibase 10.1038/s41597-021-00936-5} {\bibfield
  {journal} {\bibinfo  {journal} {Scientific Data}\ }\textbf {\bibinfo {volume}
  {8}},\ \bibinfo {pages} {153} (\bibinfo {year}
  {2021}{\natexlab{b}})}\BibitemShut {NoStop}%
\bibitem [{\citenamefont {Shibata}\ \emph {et~al.}(2022)\citenamefont
  {Shibata}, \citenamefont {Kikumasa}, \citenamefont {Kiyohara},\ and\
  \citenamefont {Mizoguchi}}]{Shibata2022}%
  \BibitemOpen
  \bibfield  {author} {\bibinfo {author} {\bibfnamefont {Kiyou}\ \bibnamefont
  {Shibata}}, \bibinfo {author} {\bibfnamefont {Kakeru}\ \bibnamefont
  {Kikumasa}}, \bibinfo {author} {\bibfnamefont {Shin}\ \bibnamefont
  {Kiyohara}}, \ and\ \bibinfo {author} {\bibfnamefont {Teruyasu}\ \bibnamefont
  {Mizoguchi}},\ }\bibfield  {title} {\enquote {\bibinfo {title} {Simulated
  carbon k edge spectral database of organic molecules},}\ }\href {\doibase
  10.1038/s41597-022-01303-8} {\bibfield  {journal} {\bibinfo  {journal}
  {Scientific Data}\ }\textbf {\bibinfo {volume} {9}},\ \bibinfo {pages} {214}
  (\bibinfo {year} {2022})}\BibitemShut {NoStop}%
\bibitem [{\citenamefont {Shirley}\ \emph {et~al.}(2021)\citenamefont
  {Shirley}, \citenamefont {Pettersson},\ and\ \citenamefont
  {Prendergast}}]{shirley2021core}%
  \BibitemOpen
  \bibfield  {author} {\bibinfo {author} {\bibfnamefont {Eric~L}\ \bibnamefont
  {Shirley}}, \bibinfo {author} {\bibfnamefont {LGM}\ \bibnamefont
  {Pettersson}}, \ and\ \bibinfo {author} {\bibfnamefont {D}~\bibnamefont
  {Prendergast}},\ }\bibfield  {title} {\enquote {\bibinfo {title} {Core-hole
  potentials and related effects},}\ }in\ \href {\doibase
  10.1107/S1574870720011039} {\emph {\bibinfo {booktitle} {International Tables
  for Crystallography Volume I: X-ray absorption spectroscopy and related
  techniques}}},\ \bibinfo {editor} {edited by\ \bibinfo {editor}
  {\bibfnamefont {C.~T.}\ \bibnamefont {Chantler}}, \bibinfo {editor}
  {\bibfnamefont {F.}~\bibnamefont {Boscherini}}, \ and\ \bibinfo {editor}
  {\bibfnamefont {B.}~\bibnamefont {Bunker}}}\ (\bibinfo  {publisher} {Wiley
  Online Library},\ \bibinfo {year} {2021})\BibitemShut {NoStop}%
\bibitem [{\citenamefont {Draxl}\ and\ \citenamefont
  {Cocchi}(2021)}]{draxl2021}%
  \BibitemOpen
  \bibfield  {author} {\bibinfo {author} {\bibfnamefont {Claudia}\ \bibnamefont
  {Draxl}}\ and\ \bibinfo {author} {\bibfnamefont {Caterina}\ \bibnamefont
  {Cocchi}},\ }\bibfield  {title} {\enquote {\bibinfo {title} {exciting
  core-level spectroscopy},}\ }in\ \href {\doibase 10.1107/S1574870720011039}
  {\emph {\bibinfo {booktitle} {International Tables for Crystallography Volume
  I: X-ray absorption spectroscopy and related techniques}}},\ \bibinfo
  {editor} {edited by\ \bibinfo {editor} {\bibfnamefont {C.~T.}\ \bibnamefont
  {Chantler}}, \bibinfo {editor} {\bibfnamefont {F.}~\bibnamefont
  {Boscherini}}, \ and\ \bibinfo {editor} {\bibfnamefont {B.}~\bibnamefont
  {Bunker}}}\ (\bibinfo  {publisher} {Wiley Online Library},\ \bibinfo {year}
  {2021})\BibitemShut {NoStop}%
\bibitem [{\citenamefont {Rehr}\ \emph
  {et~al.}(2005{\natexlab{a}})\citenamefont {Rehr}, \citenamefont {Soininen},\
  and\ \citenamefont {Shirley}}]{rehr2005final}%
  \BibitemOpen
  \bibfield  {author} {\bibinfo {author} {\bibfnamefont {JJ}~\bibnamefont
  {Rehr}}, \bibinfo {author} {\bibfnamefont {JA}~\bibnamefont {Soininen}}, \
  and\ \bibinfo {author} {\bibfnamefont {Eric~L}\ \bibnamefont {Shirley}},\
  }\bibfield  {title} {\enquote {\bibinfo {title} {Final-state rule vs the
  bethe-salpeter equation for deep-core x-ray absorption spectra},}\ }\href
  {\doibase 10.1238/Physica.Topical.115a00207} {\bibfield  {journal} {\bibinfo
  {journal} {Physica Scripta}\ }\textbf {\bibinfo {volume} {2005}},\ \bibinfo
  {pages} {207} (\bibinfo {year} {2005}{\natexlab{a}})}\BibitemShut {NoStop}%
\bibitem [{\citenamefont {Fossard}\ \emph {et~al.}(2017)\citenamefont
  {Fossard}, \citenamefont {Hug}, \citenamefont {Gilmore}, \citenamefont {Kas},
  \citenamefont {Rehr}, \citenamefont {Vila},\ and\ \citenamefont
  {Shirley}}]{PhysRevB.95.115112}%
  \BibitemOpen
  \bibfield  {author} {\bibinfo {author} {\bibfnamefont {F.}~\bibnamefont
  {Fossard}}, \bibinfo {author} {\bibfnamefont {G.}~\bibnamefont {Hug}},
  \bibinfo {author} {\bibfnamefont {K.}~\bibnamefont {Gilmore}}, \bibinfo
  {author} {\bibfnamefont {J.~J.}\ \bibnamefont {Kas}}, \bibinfo {author}
  {\bibfnamefont {J.~J.}\ \bibnamefont {Rehr}}, \bibinfo {author}
  {\bibfnamefont {F.~D.}\ \bibnamefont {Vila}}, \ and\ \bibinfo {author}
  {\bibfnamefont {E.~L.}\ \bibnamefont {Shirley}},\ }\bibfield  {title}
  {\enquote {\bibinfo {title} {Quantitative first-principles calculations of
  valence and core excitation spectra of solid ${\mathrm{c}}_{60}$},}\ }\href
  {\doibase 10.1103/PhysRevB.95.115112} {\bibfield  {journal} {\bibinfo
  {journal} {Phys. Rev. B}\ }\textbf {\bibinfo {volume} {95}},\ \bibinfo
  {pages} {115112} (\bibinfo {year} {2017})}\BibitemShut {NoStop}%
\bibitem [{\citenamefont {Lejaeghere}\ \emph {et~al.}(2016)\citenamefont
  {Lejaeghere}, \citenamefont {Bihlmayer}, \citenamefont {Bj{\"o}rkman},
  \citenamefont {Blaha}, \citenamefont {Bl{\"u}gel}, \citenamefont {Blum},
  \citenamefont {Caliste}, \citenamefont {Castelli}, \citenamefont {Clark},
  \citenamefont {Dal~Corso} \emph {et~al.}}]{lejaeghere2016reproducibility}%
  \BibitemOpen
  \bibfield  {author} {\bibinfo {author} {\bibfnamefont {Kurt}\ \bibnamefont
  {Lejaeghere}}, \bibinfo {author} {\bibfnamefont {Gustav}\ \bibnamefont
  {Bihlmayer}}, \bibinfo {author} {\bibfnamefont {Torbj{\"o}rn}\ \bibnamefont
  {Bj{\"o}rkman}}, \bibinfo {author} {\bibfnamefont {Peter}\ \bibnamefont
  {Blaha}}, \bibinfo {author} {\bibfnamefont {Stefan}\ \bibnamefont
  {Bl{\"u}gel}}, \bibinfo {author} {\bibfnamefont {Volker}\ \bibnamefont
  {Blum}}, \bibinfo {author} {\bibfnamefont {Damien}\ \bibnamefont {Caliste}},
  \bibinfo {author} {\bibfnamefont {Ivano~E}\ \bibnamefont {Castelli}},
  \bibinfo {author} {\bibfnamefont {Stewart~J}\ \bibnamefont {Clark}}, \bibinfo
  {author} {\bibfnamefont {Andrea}\ \bibnamefont {Dal~Corso}},  \emph
  {et~al.},\ }\bibfield  {title} {\enquote {\bibinfo {title} {Reproducibility
  in density functional theory calculations of solids},}\ }\href {\doibase
  10.1126/science.aad3000} {\bibfield  {journal} {\bibinfo  {journal}
  {Science}\ }\textbf {\bibinfo {volume} {351}},\ \bibinfo {pages} {aad3000}
  (\bibinfo {year} {2016})}\BibitemShut {NoStop}%
\bibitem [{\citenamefont {van Setten}\ \emph {et~al.}(2015)\citenamefont {van
  Setten}, \citenamefont {Caruso}, \citenamefont {Sharifzadeh}, \citenamefont
  {Ren}, \citenamefont {Scheffler}, \citenamefont {Liu}, \citenamefont
  {Lischner}, \citenamefont {Lin}, \citenamefont {Deslippe}, \citenamefont
  {Louie} \emph {et~al.}}]{van2015gw}%
  \BibitemOpen
  \bibfield  {author} {\bibinfo {author} {\bibfnamefont {Michiel~J}\
  \bibnamefont {van Setten}}, \bibinfo {author} {\bibfnamefont {Fabio}\
  \bibnamefont {Caruso}}, \bibinfo {author} {\bibfnamefont {Sahar}\
  \bibnamefont {Sharifzadeh}}, \bibinfo {author} {\bibfnamefont {Xinguo}\
  \bibnamefont {Ren}}, \bibinfo {author} {\bibfnamefont {Matthias}\
  \bibnamefont {Scheffler}}, \bibinfo {author} {\bibfnamefont {Fang}\
  \bibnamefont {Liu}}, \bibinfo {author} {\bibfnamefont {Johannes}\
  \bibnamefont {Lischner}}, \bibinfo {author} {\bibfnamefont {Lin}\
  \bibnamefont {Lin}}, \bibinfo {author} {\bibfnamefont {Jack~R}\ \bibnamefont
  {Deslippe}}, \bibinfo {author} {\bibfnamefont {Steven~G}\ \bibnamefont
  {Louie}},  \emph {et~al.},\ }\bibfield  {title} {\enquote {\bibinfo {title}
  {Gw 100: Benchmarking g0w0 for molecular systems},}\ }\href {\doibase
  10.1021/acs.jctc.5b00453} {\bibfield  {journal} {\bibinfo  {journal} {Journal
  of Chemical Theory and Computation}\ }\textbf {\bibinfo {volume} {11}},\
  \bibinfo {pages} {5665--5687} (\bibinfo {year} {2015})}\BibitemShut {NoStop}%
\bibitem [{\citenamefont {Caruso}\ \emph {et~al.}(2016)\citenamefont {Caruso},
  \citenamefont {Dauth}, \citenamefont {Van~Setten},\ and\ \citenamefont
  {Rinke}}]{caruso2016benchmark}%
  \BibitemOpen
  \bibfield  {author} {\bibinfo {author} {\bibfnamefont {Fabio}\ \bibnamefont
  {Caruso}}, \bibinfo {author} {\bibfnamefont {Matthias}\ \bibnamefont
  {Dauth}}, \bibinfo {author} {\bibfnamefont {Michiel~J}\ \bibnamefont
  {Van~Setten}}, \ and\ \bibinfo {author} {\bibfnamefont {Patrick}\
  \bibnamefont {Rinke}},\ }\bibfield  {title} {\enquote {\bibinfo {title}
  {Benchmark of gw approaches for the gw 100 test set},}\ }\href {\doibase
  10.1021/acs.jctc.6b00774} {\bibfield  {journal} {\bibinfo  {journal} {Journal
  of Chemical Theory and Computation}\ }\textbf {\bibinfo {volume} {12}},\
  \bibinfo {pages} {5076--5087} (\bibinfo {year} {2016})}\BibitemShut {NoStop}%
\bibitem [{\citenamefont {Vinson}\ \emph {et~al.}(2014)\citenamefont {Vinson},
  \citenamefont {Jach}, \citenamefont {Elam},\ and\ \citenamefont
  {Denlinger}}]{PhysRevB.90.205207}%
  \BibitemOpen
  \bibfield  {author} {\bibinfo {author} {\bibfnamefont {John}\ \bibnamefont
  {Vinson}}, \bibinfo {author} {\bibfnamefont {Terrence}\ \bibnamefont {Jach}},
  \bibinfo {author} {\bibfnamefont {W.~T.}\ \bibnamefont {Elam}}, \ and\
  \bibinfo {author} {\bibfnamefont {J.~D.}\ \bibnamefont {Denlinger}},\
  }\bibfield  {title} {\enquote {\bibinfo {title} {Origins of extreme
  broadening mechanisms in near-edge x-ray spectra of nitrogen compounds},}\
  }\href {\doibase 10.1103/PhysRevB.90.205207} {\bibfield  {journal} {\bibinfo
  {journal} {Phys. Rev. B}\ }\textbf {\bibinfo {volume} {90}},\ \bibinfo
  {pages} {205207} (\bibinfo {year} {2014})}\BibitemShut {NoStop}%
\bibitem [{\citenamefont {Pascal}\ \emph {et~al.}(2014)\citenamefont {Pascal},
  \citenamefont {Boesenberg}, \citenamefont {Kostecki}, \citenamefont
  {Richardson}, \citenamefont {Weng}, \citenamefont {Sokaras}, \citenamefont
  {Nordlund}, \citenamefont {McDermott}, \citenamefont {Moewes}, \citenamefont
  {Cabana},\ and\ \citenamefont {Prendergast}}]{10.1063/1.4856835}%
  \BibitemOpen
  \bibfield  {author} {\bibinfo {author} {\bibfnamefont {Tod~A.}\ \bibnamefont
  {Pascal}}, \bibinfo {author} {\bibfnamefont {Ulrike}\ \bibnamefont
  {Boesenberg}}, \bibinfo {author} {\bibfnamefont {Robert}\ \bibnamefont
  {Kostecki}}, \bibinfo {author} {\bibfnamefont {Thomas~J.}\ \bibnamefont
  {Richardson}}, \bibinfo {author} {\bibfnamefont {Tsu-Chien}\ \bibnamefont
  {Weng}}, \bibinfo {author} {\bibfnamefont {Dimosthenis}\ \bibnamefont
  {Sokaras}}, \bibinfo {author} {\bibfnamefont {Dennis}\ \bibnamefont
  {Nordlund}}, \bibinfo {author} {\bibfnamefont {Eamon}\ \bibnamefont
  {McDermott}}, \bibinfo {author} {\bibfnamefont {Alexander}\ \bibnamefont
  {Moewes}}, \bibinfo {author} {\bibfnamefont {Jordi}\ \bibnamefont {Cabana}},
  \ and\ \bibinfo {author} {\bibfnamefont {David}\ \bibnamefont
  {Prendergast}},\ }\bibfield  {title} {\enquote {\bibinfo {title} {{Finite
  temperature effects on the X-ray absorption spectra of lithium compounds:
  First-principles interpretation of X-ray Raman measurements}},}\ }\href
  {\doibase 10.1063/1.4856835} {\bibfield  {journal} {\bibinfo  {journal} {The
  Journal of Chemical Physics}\ }\textbf {\bibinfo {volume} {140}} (\bibinfo
  {year} {2014}),\ 10.1063/1.4856835},\ \bibinfo {note} {034107}\BibitemShut
  {NoStop}%
\bibitem [{\citenamefont {Brouder}\ \emph {et~al.}(2010)\citenamefont
  {Brouder}, \citenamefont {Cabaret}, \citenamefont {Juhin},\ and\
  \citenamefont {Sainctavit}}]{PhysRevB.81.115125}%
  \BibitemOpen
  \bibfield  {author} {\bibinfo {author} {\bibfnamefont {Christian}\
  \bibnamefont {Brouder}}, \bibinfo {author} {\bibfnamefont {Delphine}\
  \bibnamefont {Cabaret}}, \bibinfo {author} {\bibfnamefont {Am\'elie}\
  \bibnamefont {Juhin}}, \ and\ \bibinfo {author} {\bibfnamefont {Philippe}\
  \bibnamefont {Sainctavit}},\ }\bibfield  {title} {\enquote {\bibinfo {title}
  {Effect of atomic vibrations on the x-ray absorption spectra at the $k$ edge
  of al in $\ensuremath{\alpha}{\text{-al}}_{2}{\text{o}}_{3}$ and of ti in
  ${\text{tio}}_{2}$ rutile},}\ }\href {\doibase 10.1103/PhysRevB.81.115125}
  {\bibfield  {journal} {\bibinfo  {journal} {Phys. Rev. B}\ }\textbf {\bibinfo
  {volume} {81}},\ \bibinfo {pages} {115125} (\bibinfo {year}
  {2010})}\BibitemShut {NoStop}%
\bibitem [{\citenamefont {Cockayne}\ \emph {et~al.}(2018)\citenamefont
  {Cockayne}, \citenamefont {Shirley}, \citenamefont {Ravel},\ and\
  \citenamefont {Woicik}}]{PhysRevB.98.014111}%
  \BibitemOpen
  \bibfield  {author} {\bibinfo {author} {\bibfnamefont {Eric}\ \bibnamefont
  {Cockayne}}, \bibinfo {author} {\bibfnamefont {Eric~L.}\ \bibnamefont
  {Shirley}}, \bibinfo {author} {\bibfnamefont {Bruce}\ \bibnamefont {Ravel}},
  \ and\ \bibinfo {author} {\bibfnamefont {Joseph~C.}\ \bibnamefont {Woicik}},\
  }\bibfield  {title} {\enquote {\bibinfo {title} {Local atomic geometry and ti
  $1s$ near-edge spectra in ${\mathrm{pbtio}}_{3}$ and
  ${\mathrm{srtio}}_{3}$},}\ }\href {\doibase 10.1103/PhysRevB.98.014111}
  {\bibfield  {journal} {\bibinfo  {journal} {Phys. Rev. B}\ }\textbf {\bibinfo
  {volume} {98}},\ \bibinfo {pages} {014111} (\bibinfo {year}
  {2018})}\BibitemShut {NoStop}%
\bibitem [{\citenamefont {Geondzhian}\ and\ \citenamefont
  {Gilmore}(2018)}]{PhysRevB.98.214305}%
  \BibitemOpen
  \bibfield  {author} {\bibinfo {author} {\bibfnamefont {Andrey}\ \bibnamefont
  {Geondzhian}}\ and\ \bibinfo {author} {\bibfnamefont {Keith}\ \bibnamefont
  {Gilmore}},\ }\bibfield  {title} {\enquote {\bibinfo {title} {Demonstration
  of resonant inelastic x-ray scattering as a probe of exciton-phonon
  coupling},}\ }\href {\doibase 10.1103/PhysRevB.98.214305} {\bibfield
  {journal} {\bibinfo  {journal} {Phys. Rev. B}\ }\textbf {\bibinfo {volume}
  {98}},\ \bibinfo {pages} {214305} (\bibinfo {year} {2018})}\BibitemShut
  {NoStop}%
\bibitem [{\citenamefont {Vinson}(2020)}]{10.1063/5.0030493}%
  \BibitemOpen
  \bibfield  {author} {\bibinfo {author} {\bibfnamefont {John}\ \bibnamefont
  {Vinson}},\ }\bibfield  {title} {\enquote {\bibinfo {title} {{Faster exact
  exchange in periodic systems using single-precision arithmetic}},}\ }\href
  {\doibase 10.1063/5.0030493} {\bibfield  {journal} {\bibinfo  {journal} {The
  Journal of Chemical Physics}\ }\textbf {\bibinfo {volume} {153}},\ \bibinfo
  {pages} {204106} (\bibinfo {year} {2020})}\BibitemShut {NoStop}%
\bibitem [{\citenamefont {Vinson}\ \emph {et~al.}(2016)\citenamefont {Vinson},
  \citenamefont {Jach}, \citenamefont {M\"uller}, \citenamefont
  {Unterumsberger},\ and\ \citenamefont {Beckhoff}}]{PhysRevB.94.035163}%
  \BibitemOpen
  \bibfield  {author} {\bibinfo {author} {\bibfnamefont {John}\ \bibnamefont
  {Vinson}}, \bibinfo {author} {\bibfnamefont {Terrence}\ \bibnamefont {Jach}},
  \bibinfo {author} {\bibfnamefont {Matthias}\ \bibnamefont {M\"uller}},
  \bibinfo {author} {\bibfnamefont {Rainer}\ \bibnamefont {Unterumsberger}}, \
  and\ \bibinfo {author} {\bibfnamefont {Burkhard}\ \bibnamefont {Beckhoff}},\
  }\bibfield  {title} {\enquote {\bibinfo {title} {Quasiparticle lifetime
  broadening in resonant x-ray scattering of
  ${\mathrm{nh}}_{4}{\mathrm{no}}_{3}$},}\ }\href {\doibase
  10.1103/PhysRevB.94.035163} {\bibfield  {journal} {\bibinfo  {journal} {Phys.
  Rev. B}\ }\textbf {\bibinfo {volume} {94}},\ \bibinfo {pages} {035163}
  (\bibinfo {year} {2016})}\BibitemShut {NoStop}%
\bibitem [{\citenamefont {Tang}\ \emph
  {et~al.}(2022{\natexlab{b}})\citenamefont {Tang}, \citenamefont {Li},
  \citenamefont {Zhang}, \citenamefont {Louie}, \citenamefont {Car},
  \citenamefont {Qiu},\ and\ \citenamefont {Wu}}]{doi:10.1073/pnas.2201258119}%
  \BibitemOpen
  \bibfield  {author} {\bibinfo {author} {\bibfnamefont {Fujie}\ \bibnamefont
  {Tang}}, \bibinfo {author} {\bibfnamefont {Zhenglu}\ \bibnamefont {Li}},
  \bibinfo {author} {\bibfnamefont {Chunyi}\ \bibnamefont {Zhang}}, \bibinfo
  {author} {\bibfnamefont {Steven~G.}\ \bibnamefont {Louie}}, \bibinfo {author}
  {\bibfnamefont {Roberto}\ \bibnamefont {Car}}, \bibinfo {author}
  {\bibfnamefont {Diana~Y.}\ \bibnamefont {Qiu}}, \ and\ \bibinfo {author}
  {\bibfnamefont {Xifan}\ \bibnamefont {Wu}},\ }\bibfield  {title} {\enquote
  {\bibinfo {title} {Many-body effects in the x-ray absorption spectra of
  liquid water},}\ }\href {\doibase 10.1073/pnas.2201258119} {\bibfield
  {journal} {\bibinfo  {journal} {Proceedings of the National Academy of
  Sciences}\ }\textbf {\bibinfo {volume} {119}},\ \bibinfo {pages}
  {e2201258119} (\bibinfo {year} {2022}{\natexlab{b}})}\BibitemShut {NoStop}%
\bibitem [{\citenamefont {Woicik}\ \emph {et~al.}(2020)\citenamefont {Woicik},
  \citenamefont {Weiland}, \citenamefont {Jaye}, \citenamefont {Fischer},
  \citenamefont {Rumaiz}, \citenamefont {Shirley}, \citenamefont {Kas},\ and\
  \citenamefont {Rehr}}]{PhysRevB.101.245119}%
  \BibitemOpen
  \bibfield  {author} {\bibinfo {author} {\bibfnamefont {J.~C.}\ \bibnamefont
  {Woicik}}, \bibinfo {author} {\bibfnamefont {C.}~\bibnamefont {Weiland}},
  \bibinfo {author} {\bibfnamefont {C.}~\bibnamefont {Jaye}}, \bibinfo {author}
  {\bibfnamefont {D.~A.}\ \bibnamefont {Fischer}}, \bibinfo {author}
  {\bibfnamefont {A.~K.}\ \bibnamefont {Rumaiz}}, \bibinfo {author}
  {\bibfnamefont {E.~L.}\ \bibnamefont {Shirley}}, \bibinfo {author}
  {\bibfnamefont {J.~J.}\ \bibnamefont {Kas}}, \ and\ \bibinfo {author}
  {\bibfnamefont {J.~J.}\ \bibnamefont {Rehr}},\ }\bibfield  {title} {\enquote
  {\bibinfo {title} {Charge-transfer satellites and chemical bonding in
  photoemission and x-ray absorption of $\mathrm{SrTi}{\mathrm{o}}_{3}$ and
  rutile $\mathrm{Ti}{\mathrm{o}}_{2}$: Experiment and first-principles theory
  with general application to spectroscopic analysis},}\ }\href {\doibase
  10.1103/PhysRevB.101.245119} {\bibfield  {journal} {\bibinfo  {journal}
  {Phys. Rev. B}\ }\textbf {\bibinfo {volume} {101}},\ \bibinfo {pages}
  {245119} (\bibinfo {year} {2020})}\BibitemShut {NoStop}%
\bibitem [{\citenamefont {{de Groot}}\ \emph {et~al.}(2021)\citenamefont {{de
  Groot}}, \citenamefont {Elnaggar}, \citenamefont {Frati}, \citenamefont {pan
  Wang}, \citenamefont {Delgado-Jaime}, \citenamefont {{van Veenendaal}},
  \citenamefont {Fernandez-Rodriguez}, \citenamefont {Haverkort}, \citenamefont
  {Green}, \citenamefont {{van der Laan}}, \citenamefont {Kvashnin},
  \citenamefont {Hariki}, \citenamefont {Ikeno}, \citenamefont
  {Ramanantoanina}, \citenamefont {Daul}, \citenamefont {Delley}, \citenamefont
  {Odelius}, \citenamefont {Lundberg}, \citenamefont {Kuhn}, \citenamefont
  {Bokarev}, \citenamefont {Shirley}, \citenamefont {Vinson}, \citenamefont
  {Gilmore}, \citenamefont {Stener}, \citenamefont {Fronzoni}, \citenamefont
  {Decleva}, \citenamefont {Kruger}, \citenamefont {Retegan}, \citenamefont
  {Joly}, \citenamefont {Vorwerk}, \citenamefont {Draxl}, \citenamefont
  {Rehr},\ and\ \citenamefont {Tanaka}}]{DEGROOT2021147061}%
  \BibitemOpen
  \bibfield  {author} {\bibinfo {author} {\bibfnamefont {Frank~M.F.}\
  \bibnamefont {{de Groot}}}, \bibinfo {author} {\bibfnamefont {Hebatalla}\
  \bibnamefont {Elnaggar}}, \bibinfo {author} {\bibfnamefont {Federica}\
  \bibnamefont {Frati}}, \bibinfo {author} {\bibfnamefont {Ru}~\bibnamefont
  {pan Wang}}, \bibinfo {author} {\bibfnamefont {Mario~U.}\ \bibnamefont
  {Delgado-Jaime}}, \bibinfo {author} {\bibfnamefont {Michel}\ \bibnamefont
  {{van Veenendaal}}}, \bibinfo {author} {\bibfnamefont {Javier}\ \bibnamefont
  {Fernandez-Rodriguez}}, \bibinfo {author} {\bibfnamefont {Maurits~W.}\
  \bibnamefont {Haverkort}}, \bibinfo {author} {\bibfnamefont {Robert~J.}\
  \bibnamefont {Green}}, \bibinfo {author} {\bibfnamefont {Gerrit}\
  \bibnamefont {{van der Laan}}}, \bibinfo {author} {\bibfnamefont {Yaroslav}\
  \bibnamefont {Kvashnin}}, \bibinfo {author} {\bibfnamefont {Atsushi}\
  \bibnamefont {Hariki}}, \bibinfo {author} {\bibfnamefont {Hidekazu}\
  \bibnamefont {Ikeno}}, \bibinfo {author} {\bibfnamefont {Harry}\ \bibnamefont
  {Ramanantoanina}}, \bibinfo {author} {\bibfnamefont {Claude}\ \bibnamefont
  {Daul}}, \bibinfo {author} {\bibfnamefont {Bernard}\ \bibnamefont {Delley}},
  \bibinfo {author} {\bibfnamefont {Michael}\ \bibnamefont {Odelius}}, \bibinfo
  {author} {\bibfnamefont {Marcus}\ \bibnamefont {Lundberg}}, \bibinfo {author}
  {\bibfnamefont {Oliver}\ \bibnamefont {Kuhn}}, \bibinfo {author}
  {\bibfnamefont {Sergey~I.}\ \bibnamefont {Bokarev}}, \bibinfo {author}
  {\bibfnamefont {Eric}\ \bibnamefont {Shirley}}, \bibinfo {author}
  {\bibfnamefont {John}\ \bibnamefont {Vinson}}, \bibinfo {author}
  {\bibfnamefont {Keith}\ \bibnamefont {Gilmore}}, \bibinfo {author}
  {\bibfnamefont {Mauro}\ \bibnamefont {Stener}}, \bibinfo {author}
  {\bibfnamefont {Giovanna}\ \bibnamefont {Fronzoni}}, \bibinfo {author}
  {\bibfnamefont {Piero}\ \bibnamefont {Decleva}}, \bibinfo {author}
  {\bibfnamefont {Peter}\ \bibnamefont {Kruger}}, \bibinfo {author}
  {\bibfnamefont {Marius}\ \bibnamefont {Retegan}}, \bibinfo {author}
  {\bibfnamefont {Yves}\ \bibnamefont {Joly}}, \bibinfo {author} {\bibfnamefont
  {Christian}\ \bibnamefont {Vorwerk}}, \bibinfo {author} {\bibfnamefont
  {Claudia}\ \bibnamefont {Draxl}}, \bibinfo {author} {\bibfnamefont {John}\
  \bibnamefont {Rehr}}, \ and\ \bibinfo {author} {\bibfnamefont {Arata}\
  \bibnamefont {Tanaka}},\ }\bibfield  {title} {\enquote {\bibinfo {title} {2p
  x-ray absorption spectroscopy of 3d transition metal systems},}\ }\href
  {\doibase https://doi.org/10.1016/j.elspec.2021.147061} {\bibfield  {journal}
  {\bibinfo  {journal} {Journal of Electron Spectroscopy and Related
  Phenomena}\ }\textbf {\bibinfo {volume} {249}},\ \bibinfo {pages} {147061}
  (\bibinfo {year} {2021})}\BibitemShut {NoStop}%
\bibitem [{\citenamefont {de~Groot}\ \emph {et~al.}(2009)\citenamefont
  {de~Groot}, \citenamefont {Vankó},\ and\ \citenamefont
  {Glatzel}}]{deGroot_2009}%
  \BibitemOpen
  \bibfield  {author} {\bibinfo {author} {\bibfnamefont {Frank}\ \bibnamefont
  {de~Groot}}, \bibinfo {author} {\bibfnamefont {György}\ \bibnamefont
  {Vankó}}, \ and\ \bibinfo {author} {\bibfnamefont {Pieter}\ \bibnamefont
  {Glatzel}},\ }\bibfield  {title} {\enquote {\bibinfo {title} {The 1s x-ray
  absorption pre-edge structures in transition metal oxides},}\ }\href
  {\doibase 10.1088/0953-8984/21/10/104207} {\bibfield  {journal} {\bibinfo
  {journal} {Journal of Physics: Condensed Matter}\ }\textbf {\bibinfo {volume}
  {21}},\ \bibinfo {pages} {104207} (\bibinfo {year} {2009})}\BibitemShut
  {NoStop}%
\bibitem [{\citenamefont {Vinson}\ \emph {et~al.}(2011)\citenamefont {Vinson},
  \citenamefont {Rehr}, \citenamefont {Kas},\ and\ \citenamefont
  {Shirley}}]{vinson2011bethe}%
  \BibitemOpen
  \bibfield  {author} {\bibinfo {author} {\bibfnamefont {J}~\bibnamefont
  {Vinson}}, \bibinfo {author} {\bibfnamefont {JJ}~\bibnamefont {Rehr}},
  \bibinfo {author} {\bibfnamefont {JJ}~\bibnamefont {Kas}}, \ and\ \bibinfo
  {author} {\bibfnamefont {EL}~\bibnamefont {Shirley}},\ }\bibfield  {title}
  {\enquote {\bibinfo {title} {Bethe-salpeter equation calculations of core
  excitation spectra},}\ }\href {\doibase 10.1103/PhysRevB.83.115106}
  {\bibfield  {journal} {\bibinfo  {journal} {Physical Review B}\ }\textbf
  {\bibinfo {volume} {83}},\ \bibinfo {pages} {115106} (\bibinfo {year}
  {2011})}\BibitemShut {NoStop}%
\bibitem [{\citenamefont {Vinson}(2022)}]{ocean-3}%
  \BibitemOpen
  \bibfield  {author} {\bibinfo {author} {\bibfnamefont {John}\ \bibnamefont
  {Vinson}},\ }\bibfield  {title} {\enquote {\bibinfo {title} {Advances in the
  {\sc ocean}-3 spectroscopy package},}\ }\href {\doibase 10.1039/D2CP01030E}
  {\bibfield  {journal} {\bibinfo  {journal} {Phys. Chem. Chem. Phys.}\
  }\textbf {\bibinfo {volume} {24}},\ \bibinfo {pages} {12787} (\bibinfo {year}
  {2022})}\BibitemShut {NoStop}%
\bibitem [{\citenamefont {Gulans}\ \emph {et~al.}(2014)\citenamefont {Gulans},
  \citenamefont {Kontur}, \citenamefont {Meisenbichler}, \citenamefont {Nabok},
  \citenamefont {Pavone}, \citenamefont {Rigamonti}, \citenamefont
  {Sagmeister}, \citenamefont {Werner},\ and\ \citenamefont
  {Draxl}}]{gulans2014exciting}%
  \BibitemOpen
  \bibfield  {author} {\bibinfo {author} {\bibfnamefont {Andris}\ \bibnamefont
  {Gulans}}, \bibinfo {author} {\bibfnamefont {Stefan}\ \bibnamefont {Kontur}},
  \bibinfo {author} {\bibfnamefont {Christian}\ \bibnamefont {Meisenbichler}},
  \bibinfo {author} {\bibfnamefont {Dmitrii}\ \bibnamefont {Nabok}}, \bibinfo
  {author} {\bibfnamefont {Pasquale}\ \bibnamefont {Pavone}}, \bibinfo {author}
  {\bibfnamefont {Santiago}\ \bibnamefont {Rigamonti}}, \bibinfo {author}
  {\bibfnamefont {Stephan}\ \bibnamefont {Sagmeister}}, \bibinfo {author}
  {\bibfnamefont {Ute}\ \bibnamefont {Werner}}, \ and\ \bibinfo {author}
  {\bibfnamefont {Claudia}\ \bibnamefont {Draxl}},\ }\bibfield  {title}
  {\enquote {\bibinfo {title} {Exciting: a full-potential all-electron package
  implementing density-functional theory and many-body perturbation theory},}\
  }\href {\doibase 10.1088/0953-8984/26/36/363202} {\bibfield  {journal}
  {\bibinfo  {journal} {Journal of Physics: Condensed Matter}\ }\textbf
  {\bibinfo {volume} {26}},\ \bibinfo {pages} {363202} (\bibinfo {year}
  {2014})}\BibitemShut {NoStop}%
\bibitem [{\citenamefont {Vorwerk}\ \emph {et~al.}(2017)\citenamefont
  {Vorwerk}, \citenamefont {Cocchi},\ and\ \citenamefont
  {Draxl}}]{vorwerk2017addressing}%
  \BibitemOpen
  \bibfield  {author} {\bibinfo {author} {\bibfnamefont {Christian}\
  \bibnamefont {Vorwerk}}, \bibinfo {author} {\bibfnamefont {Caterina}\
  \bibnamefont {Cocchi}}, \ and\ \bibinfo {author} {\bibfnamefont {Claudia}\
  \bibnamefont {Draxl}},\ }\bibfield  {title} {\enquote {\bibinfo {title}
  {Addressing electron-hole correlation in core excitations of solids: An
  all-electron many-body approach from first principles},}\ }\href {\doibase
  10.1103/PhysRevB.95.155121} {\bibfield  {journal} {\bibinfo  {journal}
  {Physical Review B}\ }\textbf {\bibinfo {volume} {95}},\ \bibinfo {pages}
  {155121} (\bibinfo {year} {2017})}\BibitemShut {NoStop}%
\bibitem [{\citenamefont {Vorwerk}\ \emph {et~al.}(2019)\citenamefont
  {Vorwerk}, \citenamefont {Aurich}, \citenamefont {Cocchi},\ and\
  \citenamefont {Draxl}}]{vorwerk2019implementation}%
  \BibitemOpen
  \bibfield  {author} {\bibinfo {author} {\bibfnamefont {Christian}\
  \bibnamefont {Vorwerk}}, \bibinfo {author} {\bibfnamefont {Benjamin}\
  \bibnamefont {Aurich}}, \bibinfo {author} {\bibfnamefont {Caterina}\
  \bibnamefont {Cocchi}}, \ and\ \bibinfo {author} {\bibfnamefont {Claudia}\
  \bibnamefont {Draxl}},\ }\bibfield  {title} {\enquote {\bibinfo {title}
  {Bethe-salpeter equation for absorption and scattering spectroscopy:
  Implementation in the exciting code},}\ }\href {\doibase
  10.1088/2516-1075/ab3123} {\bibfield  {journal} {\bibinfo  {journal}
  {Electronic Structure}\ }\textbf {\bibinfo {volume} {1}},\ \bibinfo {pages}
  {037001} (\bibinfo {year} {2019})}\BibitemShut {NoStop}%
\bibitem [{\citenamefont {Taillefumier}\ \emph {et~al.}(2002)\citenamefont
  {Taillefumier}, \citenamefont {Cabaret}, \citenamefont {Flank},\ and\
  \citenamefont {Mauri}}]{taillefumier2002x}%
  \BibitemOpen
  \bibfield  {author} {\bibinfo {author} {\bibfnamefont {Mathieu}\ \bibnamefont
  {Taillefumier}}, \bibinfo {author} {\bibfnamefont {Delphine}\ \bibnamefont
  {Cabaret}}, \bibinfo {author} {\bibfnamefont {Anne-Marie}\ \bibnamefont
  {Flank}}, \ and\ \bibinfo {author} {\bibfnamefont {Francesco}\ \bibnamefont
  {Mauri}},\ }\bibfield  {title} {\enquote {\bibinfo {title} {X-ray absorption
  near-edge structure calculations with the pseudopotentials: Application to
  the k edge in diamond and $\alpha$-quartz},}\ }\href {\doibase
  10.1103/PhysRevB.66.195107} {\bibfield  {journal} {\bibinfo  {journal}
  {Physical Review B}\ }\textbf {\bibinfo {volume} {66}},\ \bibinfo {pages}
  {195107} (\bibinfo {year} {2002})}\BibitemShut {NoStop}%
\bibitem [{\citenamefont {Gougoussis}\ \emph {et~al.}(2009)\citenamefont
  {Gougoussis}, \citenamefont {Calandra}, \citenamefont {Seitsonen},\ and\
  \citenamefont {Mauri}}]{gougoussis2009first}%
  \BibitemOpen
  \bibfield  {author} {\bibinfo {author} {\bibfnamefont {Christos}\
  \bibnamefont {Gougoussis}}, \bibinfo {author} {\bibfnamefont {Matteo}\
  \bibnamefont {Calandra}}, \bibinfo {author} {\bibfnamefont {Ari~P}\
  \bibnamefont {Seitsonen}}, \ and\ \bibinfo {author} {\bibfnamefont
  {Francesco}\ \bibnamefont {Mauri}},\ }\bibfield  {title} {\enquote {\bibinfo
  {title} {First-principles calculations of x-ray absorption in a scheme based
  on ultrasoft pseudopotentials: From $\alpha$-quartz to high-tc compounds},}\
  }\href {\doibase 10.1103/PhysRevB.80.075102} {\bibfield  {journal} {\bibinfo
  {journal} {Physical Review B}\ }\textbf {\bibinfo {volume} {80}},\ \bibinfo
  {pages} {075102} (\bibinfo {year} {2009})}\BibitemShut {NoStop}%
\bibitem [{\citenamefont {Puschnig}\ and\ \citenamefont
  {Ambrosch-Draxl}(2002)}]{puschnig2002optical}%
  \BibitemOpen
  \bibfield  {author} {\bibinfo {author} {\bibfnamefont {Peter}\ \bibnamefont
  {Puschnig}}\ and\ \bibinfo {author} {\bibfnamefont {Claudia}\ \bibnamefont
  {Ambrosch-Draxl}},\ }\bibfield  {title} {\enquote {\bibinfo {title} {Optical
  absorption spectra of semiconductors and insulators including electron-hole
  correlations: An ab initio study within the lapw method},}\ }\href {\doibase
  10.1103/PhysRevB.66.165105} {\bibfield  {journal} {\bibinfo  {journal}
  {Physical Review B}\ }\textbf {\bibinfo {volume} {66}},\ \bibinfo {pages}
  {165105} (\bibinfo {year} {2002})}\BibitemShut {NoStop}%
\bibitem [{\citenamefont {Vinson}\ and\ \citenamefont
  {Rehr}(2012)}]{PhysRevB.86.195135}%
  \BibitemOpen
  \bibfield  {author} {\bibinfo {author} {\bibfnamefont {J.}~\bibnamefont
  {Vinson}}\ and\ \bibinfo {author} {\bibfnamefont {J.~J.}\ \bibnamefont
  {Rehr}},\ }\bibfield  {title} {\enquote {\bibinfo {title} {Ab initio
  bethe-salpeter calculations of the x-ray absorption spectra of transition
  metals at the $l$-shell edges},}\ }\href {\doibase
  10.1103/PhysRevB.86.195135} {\bibfield  {journal} {\bibinfo  {journal} {Phys.
  Rev. B}\ }\textbf {\bibinfo {volume} {86}},\ \bibinfo {pages} {195135}
  (\bibinfo {year} {2012})}\BibitemShut {NoStop}%
\bibitem [{\citenamefont {Golze}\ \emph {et~al.}(2020)\citenamefont {Golze},
  \citenamefont {Keller},\ and\ \citenamefont {Rinke}}]{golze2020accurate}%
  \BibitemOpen
  \bibfield  {author} {\bibinfo {author} {\bibfnamefont {Dorothea}\
  \bibnamefont {Golze}}, \bibinfo {author} {\bibfnamefont {Levi}\ \bibnamefont
  {Keller}}, \ and\ \bibinfo {author} {\bibfnamefont {Patrick}\ \bibnamefont
  {Rinke}},\ }\bibfield  {title} {\enquote {\bibinfo {title} {Accurate absolute
  and relative core-level binding energies from gw},}\ }\href {\doibase
  10.1021/acs.jpclett.9b03423} {\bibfield  {journal} {\bibinfo  {journal} {The
  Journal of Physical Chemistry Letters}\ }\textbf {\bibinfo {volume} {11}},\
  \bibinfo {pages} {1840--1847} (\bibinfo {year} {2020})}\BibitemShut {NoStop}%
\bibitem [{\citenamefont {Yao}\ \emph {et~al.}(2022)\citenamefont {Yao},
  \citenamefont {Golze}, \citenamefont {Rinke}, \citenamefont {Blum},\ and\
  \citenamefont {Kanai}}]{doi:10.1021/acs.jctc.1c01180}%
  \BibitemOpen
  \bibfield  {author} {\bibinfo {author} {\bibfnamefont {Yi}~\bibnamefont
  {Yao}}, \bibinfo {author} {\bibfnamefont {Dorothea}\ \bibnamefont {Golze}},
  \bibinfo {author} {\bibfnamefont {Patrick}\ \bibnamefont {Rinke}}, \bibinfo
  {author} {\bibfnamefont {Volker}\ \bibnamefont {Blum}}, \ and\ \bibinfo
  {author} {\bibfnamefont {Yosuke}\ \bibnamefont {Kanai}},\ }\bibfield  {title}
  {\enquote {\bibinfo {title} {All-electron bse@gw method for k-edge core
  electron excitation energies},}\ }\href {\doibase 10.1021/acs.jctc.1c01180}
  {\bibfield  {journal} {\bibinfo  {journal} {Journal of Chemical Theory and
  Computation}\ }\textbf {\bibinfo {volume} {18}},\ \bibinfo {pages}
  {1569--1583} (\bibinfo {year} {2022})},\ \bibinfo {note} {pMID:
  35138865}\BibitemShut {NoStop}%
\bibitem [{\citenamefont {Rohlfing}\ and\ \citenamefont
  {Louie}(2000)}]{rohlfing2000electron}%
  \BibitemOpen
  \bibfield  {author} {\bibinfo {author} {\bibfnamefont {Michael}\ \bibnamefont
  {Rohlfing}}\ and\ \bibinfo {author} {\bibfnamefont {Steven~G}\ \bibnamefont
  {Louie}},\ }\bibfield  {title} {\enquote {\bibinfo {title} {Electron-hole
  excitations and optical spectra from first principles},}\ }\href {\doibase
  10.1103/PhysRevB.62.4927} {\bibfield  {journal} {\bibinfo  {journal}
  {Physical Review B}\ }\textbf {\bibinfo {volume} {62}},\ \bibinfo {pages}
  {4927} (\bibinfo {year} {2000})}\BibitemShut {NoStop}%
\bibitem [{\citenamefont {Giannozzi}\ \emph {et~al.}(2009)\citenamefont
  {Giannozzi}, \citenamefont {Baroni}, \citenamefont {Bonini}, \citenamefont
  {Calandra}, \citenamefont {Car}, \citenamefont {Cavazzoni}, \citenamefont
  {Ceresoli}, \citenamefont {Chiarotti}, \citenamefont {Cococcioni},
  \citenamefont {Dabo} \emph {et~al.}}]{giannozzi2009quantum}%
  \BibitemOpen
  \bibfield  {author} {\bibinfo {author} {\bibfnamefont {Paolo}\ \bibnamefont
  {Giannozzi}}, \bibinfo {author} {\bibfnamefont {Stefano}\ \bibnamefont
  {Baroni}}, \bibinfo {author} {\bibfnamefont {Nicola}\ \bibnamefont {Bonini}},
  \bibinfo {author} {\bibfnamefont {Matteo}\ \bibnamefont {Calandra}}, \bibinfo
  {author} {\bibfnamefont {Roberto}\ \bibnamefont {Car}}, \bibinfo {author}
  {\bibfnamefont {Carlo}\ \bibnamefont {Cavazzoni}}, \bibinfo {author}
  {\bibfnamefont {Davide}\ \bibnamefont {Ceresoli}}, \bibinfo {author}
  {\bibfnamefont {Guido~L}\ \bibnamefont {Chiarotti}}, \bibinfo {author}
  {\bibfnamefont {Matteo}\ \bibnamefont {Cococcioni}}, \bibinfo {author}
  {\bibfnamefont {Ismaila}\ \bibnamefont {Dabo}},  \emph {et~al.},\ }\bibfield
  {title} {\enquote {\bibinfo {title} {Quantum espresso: a modular and
  open-source software project for quantum simulations of materials},}\ }\href
  {\doibase 10.1088/0953-8984/21/39/395502} {\bibfield  {journal} {\bibinfo
  {journal} {Journal of Physics: Condensed Matter}\ }\textbf {\bibinfo {volume}
  {21}},\ \bibinfo {pages} {395502} (\bibinfo {year} {2009})}\BibitemShut
  {NoStop}%
\bibitem [{\citenamefont {Giannozzi}\ \emph {et~al.}(2017)\citenamefont
  {Giannozzi}, \citenamefont {Andreussi}, \citenamefont {Brumme}, \citenamefont
  {Bunau}, \citenamefont {Nardelli}, \citenamefont {Calandra}, \citenamefont
  {Car}, \citenamefont {Cavazzoni}, \citenamefont {Ceresoli}, \citenamefont
  {Cococcioni} \emph {et~al.}}]{giannozzi2017advanced}%
  \BibitemOpen
  \bibfield  {author} {\bibinfo {author} {\bibfnamefont {Paolo}\ \bibnamefont
  {Giannozzi}}, \bibinfo {author} {\bibfnamefont {Oliviero}\ \bibnamefont
  {Andreussi}}, \bibinfo {author} {\bibfnamefont {Thomas}\ \bibnamefont
  {Brumme}}, \bibinfo {author} {\bibfnamefont {Oana}\ \bibnamefont {Bunau}},
  \bibinfo {author} {\bibfnamefont {M~Buongiorno}\ \bibnamefont {Nardelli}},
  \bibinfo {author} {\bibfnamefont {Matteo}\ \bibnamefont {Calandra}}, \bibinfo
  {author} {\bibfnamefont {Roberto}\ \bibnamefont {Car}}, \bibinfo {author}
  {\bibfnamefont {Carlo}\ \bibnamefont {Cavazzoni}}, \bibinfo {author}
  {\bibfnamefont {Davide}\ \bibnamefont {Ceresoli}}, \bibinfo {author}
  {\bibfnamefont {Matteo}\ \bibnamefont {Cococcioni}},  \emph {et~al.},\
  }\bibfield  {title} {\enquote {\bibinfo {title} {Advanced capabilities for
  materials modelling with quantum espresso},}\ }\href {\doibase
  10.1088/1361-648X/aa8f79} {\bibfield  {journal} {\bibinfo  {journal} {Journal
  of Physics: Condensed Matter}\ }\textbf {\bibinfo {volume} {29}},\ \bibinfo
  {pages} {465901} (\bibinfo {year} {2017})}\BibitemShut {NoStop}%
\bibitem [{\citenamefont {Mo}\ and\ \citenamefont
  {Ching}(2000)}]{PhysRevB.62.7901}%
  \BibitemOpen
  \bibfield  {author} {\bibinfo {author} {\bibfnamefont {Shang-Di}\
  \bibnamefont {Mo}}\ and\ \bibinfo {author} {\bibfnamefont {W.~Y.}\
  \bibnamefont {Ching}},\ }\bibfield  {title} {\enquote {\bibinfo {title} {Ab
  initio calculation of the core-hole effect in the electron energy-loss
  near-edge structure},}\ }\href {\doibase 10.1103/PhysRevB.62.7901} {\bibfield
   {journal} {\bibinfo  {journal} {Phys. Rev. B}\ }\textbf {\bibinfo {volume}
  {62}},\ \bibinfo {pages} {7901--7907} (\bibinfo {year} {2000})}\BibitemShut
  {NoStop}%
\bibitem [{\citenamefont {Jackson}\ and\ \citenamefont
  {Pederson}(1991)}]{PhysRevLett.67.2521}%
  \BibitemOpen
  \bibfield  {author} {\bibinfo {author} {\bibfnamefont {Koblar~A.}\
  \bibnamefont {Jackson}}\ and\ \bibinfo {author} {\bibfnamefont {Mark~R.}\
  \bibnamefont {Pederson}},\ }\bibfield  {title} {\enquote {\bibinfo {title}
  {New theoretical model for the diamond 1s core exciton},}\ }\href {\doibase
  10.1103/PhysRevLett.67.2521} {\bibfield  {journal} {\bibinfo  {journal}
  {Phys. Rev. Lett.}\ }\textbf {\bibinfo {volume} {67}},\ \bibinfo {pages}
  {2521--2524} (\bibinfo {year} {1991})}\BibitemShut {NoStop}%
\bibitem [{\citenamefont {Lanczos}(1950)}]{lanczos1950iteration}%
  \BibitemOpen
  \bibfield  {author} {\bibinfo {author} {\bibfnamefont {Cornelius}\
  \bibnamefont {Lanczos}},\ }\bibfield  {title} {\enquote {\bibinfo {title} {An
  iteration method for the solution of the eigenvalue problem of linear
  differential and integral operators},}\ }\href@noop {} {\bibfield  {journal}
  {\bibinfo  {journal} {Journal of Research of the National Bureau of
  Standards}\ }\textbf {\bibinfo {volume} {45}},\ \bibinfo {pages} {255}
  (\bibinfo {year} {1950})}\BibitemShut {NoStop}%
\bibitem [{\citenamefont {Lanczos}(1952)}]{lanczos1952solution}%
  \BibitemOpen
  \bibfield  {author} {\bibinfo {author} {\bibfnamefont {Cornelius}\
  \bibnamefont {Lanczos}},\ }\bibfield  {title} {\enquote {\bibinfo {title}
  {Solution of systems of linear equations by},}\ }\href@noop {} {\bibfield
  {journal} {\bibinfo  {journal} {Journal of Research of the National Bureau of
  Standards}\ }\textbf {\bibinfo {volume} {49}},\ \bibinfo {pages} {33}
  (\bibinfo {year} {1952})}\BibitemShut {NoStop}%
\bibitem [{\citenamefont {Vinson}\ and\ \citenamefont
  {Shirley}(2021)}]{PhysRevB.103.245143}%
  \BibitemOpen
  \bibfield  {author} {\bibinfo {author} {\bibfnamefont {John}\ \bibnamefont
  {Vinson}}\ and\ \bibinfo {author} {\bibfnamefont {Eric~L.}\ \bibnamefont
  {Shirley}},\ }\bibfield  {title} {\enquote {\bibinfo {title} {Fast,
  efficient, and accurate dielectric screening using a local real-space
  approach},}\ }\href {\doibase 10.1103/PhysRevB.103.245143} {\bibfield
  {journal} {\bibinfo  {journal} {Phys. Rev. B}\ }\textbf {\bibinfo {volume}
  {103}},\ \bibinfo {pages} {245143} (\bibinfo {year} {2021})}\BibitemShut
  {NoStop}%
\bibitem [{\citenamefont {Jain}\ \emph {et~al.}(2013)\citenamefont {Jain},
  \citenamefont {Ong}, \citenamefont {Hautier}, \citenamefont {Chen},
  \citenamefont {Richards}, \citenamefont {Dacek}, \citenamefont {Cholia},
  \citenamefont {Gunter}, \citenamefont {Skinner}, \citenamefont {Ceder} \emph
  {et~al.}}]{jain2013commentary}%
  \BibitemOpen
  \bibfield  {author} {\bibinfo {author} {\bibfnamefont {Anubhav}\ \bibnamefont
  {Jain}}, \bibinfo {author} {\bibfnamefont {Shyue~Ping}\ \bibnamefont {Ong}},
  \bibinfo {author} {\bibfnamefont {Geoffroy}\ \bibnamefont {Hautier}},
  \bibinfo {author} {\bibfnamefont {Wei}\ \bibnamefont {Chen}}, \bibinfo
  {author} {\bibfnamefont {William~Davidson}\ \bibnamefont {Richards}},
  \bibinfo {author} {\bibfnamefont {Stephen}\ \bibnamefont {Dacek}}, \bibinfo
  {author} {\bibfnamefont {Shreyas}\ \bibnamefont {Cholia}}, \bibinfo {author}
  {\bibfnamefont {Dan}\ \bibnamefont {Gunter}}, \bibinfo {author}
  {\bibfnamefont {David}\ \bibnamefont {Skinner}}, \bibinfo {author}
  {\bibfnamefont {Gerbrand}\ \bibnamefont {Ceder}},  \emph {et~al.},\
  }\bibfield  {title} {\enquote {\bibinfo {title} {Commentary: The materials
  project: A materials genome approach to accelerating materials innovation},}\
  }\href {\doibase 10.1063/1.4812323} {\bibfield  {journal} {\bibinfo
  {journal} {APL materials}\ }\textbf {\bibinfo {volume} {1}},\ \bibinfo
  {pages} {011002} (\bibinfo {year} {2013})}\BibitemShut {NoStop}%
\bibitem [{\citenamefont {Petousis}\ \emph {et~al.}(2017)\citenamefont
  {Petousis}, \citenamefont {Mrdjenovich}, \citenamefont {Ballouz},
  \citenamefont {Liu}, \citenamefont {Winston}, \citenamefont {Chen},
  \citenamefont {Graf}, \citenamefont {Schladt}, \citenamefont {Persson},\ and\
  \citenamefont {Prinz}}]{petousis2017high}%
  \BibitemOpen
  \bibfield  {author} {\bibinfo {author} {\bibfnamefont {Ioannis}\ \bibnamefont
  {Petousis}}, \bibinfo {author} {\bibfnamefont {David}\ \bibnamefont
  {Mrdjenovich}}, \bibinfo {author} {\bibfnamefont {Eric}\ \bibnamefont
  {Ballouz}}, \bibinfo {author} {\bibfnamefont {Miao}\ \bibnamefont {Liu}},
  \bibinfo {author} {\bibfnamefont {Donald}\ \bibnamefont {Winston}}, \bibinfo
  {author} {\bibfnamefont {Wei}\ \bibnamefont {Chen}}, \bibinfo {author}
  {\bibfnamefont {Tanja}\ \bibnamefont {Graf}}, \bibinfo {author}
  {\bibfnamefont {Thomas~D}\ \bibnamefont {Schladt}}, \bibinfo {author}
  {\bibfnamefont {Kristin~A}\ \bibnamefont {Persson}}, \ and\ \bibinfo {author}
  {\bibfnamefont {Fritz~B}\ \bibnamefont {Prinz}},\ }\bibfield  {title}
  {\enquote {\bibinfo {title} {High-throughput screening of inorganic compounds
  for the discovery of novel dielectric and optical materials},}\ }\href
  {\doibase 10.1038/sdata.2016.134} {\bibfield  {journal} {\bibinfo  {journal}
  {Scientific Data}\ }\textbf {\bibinfo {volume} {4}},\ \bibinfo {pages}
  {1--12} (\bibinfo {year} {2017})}\BibitemShut {NoStop}%
\bibitem [{\citenamefont {Bergengren}(1920)}]{bergengren1920rontgenabsorption}%
  \BibitemOpen
  \bibfield  {author} {\bibinfo {author} {\bibfnamefont {J}~\bibnamefont
  {Bergengren}},\ }\bibfield  {title} {\enquote {\bibinfo {title} {{\"U}ber die
  r{\"o}ntgenabsorption des phosphors},}\ }\href@noop {} {\bibfield  {journal}
  {\bibinfo  {journal} {Zeitschrift f{\"u}r Physik}\ }\textbf {\bibinfo
  {volume} {3}},\ \bibinfo {pages} {247--249} (\bibinfo {year}
  {1920})}\BibitemShut {NoStop}%
\bibitem [{\citenamefont {Lindh}(1921)}]{lindh1921kenntnis}%
  \BibitemOpen
  \bibfield  {author} {\bibinfo {author} {\bibfnamefont {Axel~E}\ \bibnamefont
  {Lindh}},\ }\bibfield  {title} {\enquote {\bibinfo {title} {Zur kenntnis des
  r{\"o}ntgenabsorptionsspektrums von chlor},}\ }\href {\doibase
  10.1007/BF01327991} {\bibfield  {journal} {\bibinfo  {journal} {Zeitschrift
  f{\"u}r Physik}\ }\textbf {\bibinfo {volume} {6}},\ \bibinfo {pages}
  {303--310} (\bibinfo {year} {1921})}\BibitemShut {NoStop}%
\bibitem [{\citenamefont {Stelling}(1930)}]{stelling1930ztschr}%
  \BibitemOpen
  \bibfield  {author} {\bibinfo {author} {\bibfnamefont {O}~\bibnamefont
  {Stelling}},\ }\href@noop {} {\bibfield  {journal} {\bibinfo  {journal} {Z.
  Phys. Chem., B}\ }\textbf {\bibinfo {volume} {7}},\ \bibinfo {pages} {210}
  (\bibinfo {year} {1930})}\BibitemShut {NoStop}%
\bibitem [{\citenamefont {Kunzl}(1932)}]{kunzl1932linear}%
  \BibitemOpen
  \bibfield  {author} {\bibinfo {author} {\bibfnamefont {V}~\bibnamefont
  {Kunzl}},\ }\bibfield  {title} {\enquote {\bibinfo {title} {A linear
  dependence of energy levels on the valency of elements},}\ }\href {\doibase
  10.1135/cccc19320213} {\bibfield  {journal} {\bibinfo  {journal} {Collection
  of Czechoslovak Chemical Communications}\ }\textbf {\bibinfo {volume} {4}},\
  \bibinfo {pages} {213--224} (\bibinfo {year} {1932})}\BibitemShut {NoStop}%
\bibitem [{\citenamefont {Farges}(2009)}]{Farges2009}%
  \BibitemOpen
  \bibfield  {author} {\bibinfo {author} {\bibfnamefont {F.}~\bibnamefont
  {Farges}},\ }\bibfield  {title} {\enquote {\bibinfo {title} {Chromium
  speciation in oxide-type compounds: application to minerals, gems, aqueous
  solutions and silicate glasses},}\ }\href {\doibase
  https://doi.org/10.1007/s00269-009-0293-3} {\bibfield  {journal} {\bibinfo
  {journal} {Phys Chem Minerals}\ }\textbf {\bibinfo {volume} {36}},\ \bibinfo
  {pages} {463–481} (\bibinfo {year} {2009})}\BibitemShut {NoStop}%
\bibitem [{\citenamefont {Tromp}\ \emph {et~al.}(2007)\citenamefont {Tromp},
  \citenamefont {Moulin}, \citenamefont {Reid},\ and\ \citenamefont
  {Evans}}]{Tromp2007}%
  \BibitemOpen
  \bibfield  {author} {\bibinfo {author} {\bibfnamefont {Moniek}\ \bibnamefont
  {Tromp}}, \bibinfo {author} {\bibfnamefont {Jerome}\ \bibnamefont {Moulin}},
  \bibinfo {author} {\bibfnamefont {Gillian}\ \bibnamefont {Reid}}, \ and\
  \bibinfo {author} {\bibfnamefont {John}\ \bibnamefont {Evans}},\ }\bibfield
  {title} {\enquote {\bibinfo {title} {Cr k‐edge xanes spectroscopy: Ligand
  and oxidation state dependence — what is oxidation state?}}\ }\href
  {\doibase 10.1063/1.2644637} {\bibfield  {journal} {\bibinfo  {journal} {AIP
  Conference Proceedings}\ }\textbf {\bibinfo {volume} {882}},\ \bibinfo
  {pages} {699--701} (\bibinfo {year} {2007})}\BibitemShut {NoStop}%
\bibitem [{\citenamefont {Suchet}(1965)}]{suchet1965chemical}%
  \BibitemOpen
  \bibfield  {author} {\bibinfo {author} {\bibfnamefont {Jacques~Paul}\
  \bibnamefont {Suchet}},\ }\href@noop {} {\emph {\bibinfo {title} {Chemical
  physics of semiconductors}}}\ (\bibinfo  {publisher} {Van Nostrand},\
  \bibinfo {year} {1965})\BibitemShut {NoStop}%
\bibitem [{\citenamefont {van Setten}\ \emph
  {et~al.}(2018{\natexlab{a}})\citenamefont {van Setten}, \citenamefont
  {Costa}, \citenamefont {Vi{\~n}es},\ and\ \citenamefont
  {Illas}}]{doi:10.1021/acs.jctc.7b01192}%
  \BibitemOpen
  \bibfield  {author} {\bibinfo {author} {\bibfnamefont {Michiel~J.}\
  \bibnamefont {van Setten}}, \bibinfo {author} {\bibfnamefont {Ramon}\
  \bibnamefont {Costa}}, \bibinfo {author} {\bibfnamefont {Francesc}\
  \bibnamefont {Vi{\~n}es}}, \ and\ \bibinfo {author} {\bibfnamefont
  {Francesc}\ \bibnamefont {Illas}},\ }\bibfield  {title} {\enquote {\bibinfo
  {title} {Assessing gw approaches for predicting core level binding
  energies},}\ }\href {\doibase 10.1021/acs.jctc.7b01192} {\bibfield  {journal}
  {\bibinfo  {journal} {Journal of Chemical Theory and Computation}\ }\textbf
  {\bibinfo {volume} {14}},\ \bibinfo {pages} {877--883} (\bibinfo {year}
  {2018}{\natexlab{a}})},\ \bibinfo {note} {pMID: 29320628}\BibitemShut
  {NoStop}%
\bibitem [{\citenamefont {Golze}\ \emph {et~al.}(2018)\citenamefont {Golze},
  \citenamefont {Wilhelm}, \citenamefont {van Setten},\ and\ \citenamefont
  {Rinke}}]{doi:10.1021/acs.jctc.8b00458}%
  \BibitemOpen
  \bibfield  {author} {\bibinfo {author} {\bibfnamefont {Dorothea}\
  \bibnamefont {Golze}}, \bibinfo {author} {\bibfnamefont {Jan}\ \bibnamefont
  {Wilhelm}}, \bibinfo {author} {\bibfnamefont {Michiel~J.}\ \bibnamefont {van
  Setten}}, \ and\ \bibinfo {author} {\bibfnamefont {Patrick}\ \bibnamefont
  {Rinke}},\ }\bibfield  {title} {\enquote {\bibinfo {title} {Core-level
  binding energies from gw: An efficient full-frequency approach within a
  localized basis},}\ }\href {\doibase 10.1021/acs.jctc.8b00458} {\bibfield
  {journal} {\bibinfo  {journal} {Journal of Chemical Theory and Computation}\
  }\textbf {\bibinfo {volume} {14}},\ \bibinfo {pages} {4856--4869} (\bibinfo
  {year} {2018})}\BibitemShut {NoStop}%
\bibitem [{vxc()}]{vxc}%
  \BibitemOpen
  \href@noop {} {}\bibinfo {note} {Prior to version 3.1, {\sc ocean}
  incorrectly did not remove the $V_{xc}$ contribution to the core-hole energy.
  Removing $V_{xc}$ gives an almost system-independent shift of 42.58~eV
  $\pm0.03$~eV, suggesting that the errors in relative alignment from this bug
  are minor.}\BibitemShut {Stop}%
\bibitem [{\citenamefont {Carbone}\ \emph {et~al.}(2023)\citenamefont
  {Carbone}, \citenamefont {Meng}, \citenamefont {Vorwerk}, \citenamefont
  {Maurer}, \citenamefont {Peschel}, \citenamefont {Qu}, \citenamefont
  {Stavitski}, \citenamefont {Draxl}, \citenamefont {Vinson},\ and\
  \citenamefont {Lu}}]{carbone2022lightshow}%
  \BibitemOpen
  \bibfield  {author} {\bibinfo {author} {\bibfnamefont {Matthew~R}\
  \bibnamefont {Carbone}}, \bibinfo {author} {\bibfnamefont {Fanchen}\
  \bibnamefont {Meng}}, \bibinfo {author} {\bibfnamefont {Christian}\
  \bibnamefont {Vorwerk}}, \bibinfo {author} {\bibfnamefont {Benedikt}\
  \bibnamefont {Maurer}}, \bibinfo {author} {\bibfnamefont {Fabian}\
  \bibnamefont {Peschel}}, \bibinfo {author} {\bibfnamefont {Xiaohui}\
  \bibnamefont {Qu}}, \bibinfo {author} {\bibfnamefont {Eli}\ \bibnamefont
  {Stavitski}}, \bibinfo {author} {\bibfnamefont {Claudia}\ \bibnamefont
  {Draxl}}, \bibinfo {author} {\bibfnamefont {John}\ \bibnamefont {Vinson}}, \
  and\ \bibinfo {author} {\bibfnamefont {Deyu}\ \bibnamefont {Lu}},\ }\bibfield
   {title} {\enquote {\bibinfo {title} {Lightshow: a python package for
  generating computational x-ray absorption spectroscopy input files},}\ }\href
  {\doibase 10.21105/joss.05182} {\bibfield  {journal} {\bibinfo  {journal}
  {Journal of Open Source Software}\ }\textbf {\bibinfo {volume} {8}},\
  \bibinfo {pages} {5182} (\bibinfo {year} {2023})}\BibitemShut {NoStop}%
\bibitem [{\citenamefont {van Setten}\ \emph
  {et~al.}(2018{\natexlab{b}})\citenamefont {van Setten}, \citenamefont
  {Giantomassi}, \citenamefont {Bousquet}, \citenamefont {Verstraete},
  \citenamefont {Hamann}, \citenamefont {Gonze},\ and\ \citenamefont
  {Rignanese}}]{van2018pseudodojo}%
  \BibitemOpen
  \bibfield  {author} {\bibinfo {author} {\bibfnamefont {Michiel~J}\
  \bibnamefont {van Setten}}, \bibinfo {author} {\bibfnamefont {Matteo}\
  \bibnamefont {Giantomassi}}, \bibinfo {author} {\bibfnamefont {Eric}\
  \bibnamefont {Bousquet}}, \bibinfo {author} {\bibfnamefont {Matthieu~J}\
  \bibnamefont {Verstraete}}, \bibinfo {author} {\bibfnamefont {Don~R}\
  \bibnamefont {Hamann}}, \bibinfo {author} {\bibfnamefont {Xavier}\
  \bibnamefont {Gonze}}, \ and\ \bibinfo {author} {\bibfnamefont {G-M}\
  \bibnamefont {Rignanese}},\ }\bibfield  {title} {\enquote {\bibinfo {title}
  {The pseudodojo: Training and grading a 85 element optimized norm-conserving
  pseudopotential table},}\ }\href {\doibase 10.1016/j.cpc.2018.01.012}
  {\bibfield  {journal} {\bibinfo  {journal} {Computer Physics Communications}\
  }\textbf {\bibinfo {volume} {226}},\ \bibinfo {pages} {39--54} (\bibinfo
  {year} {2018}{\natexlab{b}})}\BibitemShut {NoStop}%
\bibitem [{\citenamefont {Prandini}\ \emph {et~al.}(2018)\citenamefont
  {Prandini}, \citenamefont {Marrazzo}, \citenamefont {Castelli}, \citenamefont
  {Mounet},\ and\ \citenamefont {Marzari}}]{prandini2018precision}%
  \BibitemOpen
  \bibfield  {author} {\bibinfo {author} {\bibfnamefont {Gianluca}\
  \bibnamefont {Prandini}}, \bibinfo {author} {\bibfnamefont {Antimo}\
  \bibnamefont {Marrazzo}}, \bibinfo {author} {\bibfnamefont {Ivano~E}\
  \bibnamefont {Castelli}}, \bibinfo {author} {\bibfnamefont {Nicolas}\
  \bibnamefont {Mounet}}, \ and\ \bibinfo {author} {\bibfnamefont {Nicola}\
  \bibnamefont {Marzari}},\ }\bibfield  {title} {\enquote {\bibinfo {title}
  {Precision and efficiency in solid-state pseudopotential calculations},}\
  }\href {\doibase 10.1038/s41524-018-0127-2} {\bibfield  {journal} {\bibinfo
  {journal} {npj Computational Materials}\ }\textbf {\bibinfo {volume} {4}},\
  \bibinfo {pages} {1--13} (\bibinfo {year} {2018})}\BibitemShut {NoStop}%
\bibitem [{\citenamefont {Perdew}\ \emph {et~al.}(1996)\citenamefont {Perdew},
  \citenamefont {Burke},\ and\ \citenamefont
  {Ernzerhof}}]{perdew1996generalized}%
  \BibitemOpen
  \bibfield  {author} {\bibinfo {author} {\bibfnamefont {John~P}\ \bibnamefont
  {Perdew}}, \bibinfo {author} {\bibfnamefont {Kieron}\ \bibnamefont {Burke}},
  \ and\ \bibinfo {author} {\bibfnamefont {Matthias}\ \bibnamefont
  {Ernzerhof}},\ }\bibfield  {title} {\enquote {\bibinfo {title} {Generalized
  gradient approximation made simple},}\ }\href {\doibase
  10.1103/PhysRevLett.77.3865} {\bibfield  {journal} {\bibinfo  {journal}
  {Physical Review Letters}\ }\textbf {\bibinfo {volume} {77}},\ \bibinfo
  {pages} {3865} (\bibinfo {year} {1996})}\BibitemShut {NoStop}%
\bibitem [{\citenamefont {H{\'e}bert}(2007)}]{HEBERT200712}%
  \BibitemOpen
  \bibfield  {author} {\bibinfo {author} {\bibfnamefont {C.}~\bibnamefont
  {H{\'e}bert}},\ }\bibfield  {title} {\enquote {\bibinfo {title} {Practical
  aspects of running the wien2k code for electron spectroscopy},}\ }\href
  {\doibase https://doi.org/10.1016/j.micron.2006.03.010} {\bibfield  {journal}
  {\bibinfo  {journal} {Micron}\ }\textbf {\bibinfo {volume} {38}},\ \bibinfo
  {pages} {12--28} (\bibinfo {year} {2007})}\BibitemShut {NoStop}%
\bibitem [{\citenamefont {Spearman}(1904)}]{spearman1904}%
  \BibitemOpen
  \bibfield  {author} {\bibinfo {author} {\bibfnamefont {Charles}\ \bibnamefont
  {Spearman}},\ }\bibfield  {title} {\enquote {\bibinfo {title} {The proof and
  measurement of association between two things.}}\ }\href {\doibase
  10.1037/11491-005} {\bibfield  {journal} {\bibinfo  {journal} {Am. J.
  Psychol.}\ }\textbf {\bibinfo {volume} {15}},\ \bibinfo {pages} {72--101}
  (\bibinfo {year} {1904})}\BibitemShut {NoStop}%
\bibitem [{SI()}]{SI}%
  \BibitemOpen
  \href@noop {} {}\bibinfo {note} {See Supplemental Material at [URL will be
  inserted by publisher]}\BibitemShut {NoStop}%
\bibitem [{\citenamefont {Rehr}\ \emph
  {et~al.}(2005{\natexlab{b}})\citenamefont {Rehr}, \citenamefont {Soininen},\
  and\ \citenamefont {Shirley}}]{Rehr_2005}%
  \BibitemOpen
  \bibfield  {author} {\bibinfo {author} {\bibfnamefont {J~J}\ \bibnamefont
  {Rehr}}, \bibinfo {author} {\bibfnamefont {J~A}\ \bibnamefont {Soininen}}, \
  and\ \bibinfo {author} {\bibfnamefont {E~L}\ \bibnamefont {Shirley}},\
  }\bibfield  {title} {\enquote {\bibinfo {title} {Final-state rule vs the
  bethe-salpeter equation for deep-core x-ray absorption spectra},}\ }\href
  {\doibase 10.1238/Physica.Topical.115a00207} {\bibfield  {journal} {\bibinfo
  {journal} {Physica Scripta}\ }\textbf {\bibinfo {volume} {2005}},\ \bibinfo
  {pages} {207} (\bibinfo {year} {2005}{\natexlab{b}})}\BibitemShut {NoStop}%
\bibitem [{\citenamefont {Ohtaka}\ and\ \citenamefont
  {Tanabe}(1990)}]{RevModPhys.62.929}%
  \BibitemOpen
  \bibfield  {author} {\bibinfo {author} {\bibfnamefont {K.}~\bibnamefont
  {Ohtaka}}\ and\ \bibinfo {author} {\bibfnamefont {Y.}~\bibnamefont
  {Tanabe}},\ }\bibfield  {title} {\enquote {\bibinfo {title} {Theory of the
  soft-x-ray edge problem in simple metals: historical survey and recent
  developments},}\ }\href {\doibase 10.1103/RevModPhys.62.929} {\bibfield
  {journal} {\bibinfo  {journal} {Rev. Mod. Phys.}\ }\textbf {\bibinfo {volume}
  {62}},\ \bibinfo {pages} {929--991} (\bibinfo {year} {1990})}\BibitemShut
  {NoStop}%
\bibitem [{\citenamefont {Roychoudhury}\ and\ \citenamefont
  {Prendergast}(2023)}]{PhysRevB.107.035146}%
  \BibitemOpen
  \bibfield  {author} {\bibinfo {author} {\bibfnamefont {Subhayan}\
  \bibnamefont {Roychoudhury}}\ and\ \bibinfo {author} {\bibfnamefont {David}\
  \bibnamefont {Prendergast}},\ }\bibfield  {title} {\enquote {\bibinfo {title}
  {Efficient core-excited state orbital perspective on calculating x-ray
  absorption transitions in determinant framework},}\ }\href {\doibase
  10.1103/PhysRevB.107.035146} {\bibfield  {journal} {\bibinfo  {journal}
  {Phys. Rev. B}\ }\textbf {\bibinfo {volume} {107}},\ \bibinfo {pages}
  {035146} (\bibinfo {year} {2023})}\BibitemShut {NoStop}%
\bibitem [{\citenamefont {Liang}\ and\ \citenamefont
  {Prendergast}(2018)}]{PhysRevB.97.205127}%
  \BibitemOpen
  \bibfield  {author} {\bibinfo {author} {\bibfnamefont {Yufeng}\ \bibnamefont
  {Liang}}\ and\ \bibinfo {author} {\bibfnamefont {David}\ \bibnamefont
  {Prendergast}},\ }\bibfield  {title} {\enquote {\bibinfo {title} {Quantum
  many-body effects in x-ray spectra efficiently computed using a basic graph
  algorithm},}\ }\href {\doibase 10.1103/PhysRevB.97.205127} {\bibfield
  {journal} {\bibinfo  {journal} {Phys. Rev. B}\ }\textbf {\bibinfo {volume}
  {97}},\ \bibinfo {pages} {205127} (\bibinfo {year} {2018})}\BibitemShut
  {NoStop}%
\bibitem [{\citenamefont {Onida}\ \emph {et~al.}(2002)\citenamefont {Onida},
  \citenamefont {Reining},\ and\ \citenamefont {Rubio}}]{RevModPhys.74.601}%
  \BibitemOpen
  \bibfield  {author} {\bibinfo {author} {\bibfnamefont {Giovanni}\
  \bibnamefont {Onida}}, \bibinfo {author} {\bibfnamefont {Lucia}\ \bibnamefont
  {Reining}}, \ and\ \bibinfo {author} {\bibfnamefont {Angel}\ \bibnamefont
  {Rubio}},\ }\bibfield  {title} {\enquote {\bibinfo {title} {Electronic
  excitations: density-functional versus many-body green's-function
  approaches},}\ }\href {\doibase 10.1103/RevModPhys.74.601} {\bibfield
  {journal} {\bibinfo  {journal} {Rev. Mod. Phys.}\ }\textbf {\bibinfo {volume}
  {74}},\ \bibinfo {pages} {601--659} (\bibinfo {year} {2002})}\BibitemShut
  {NoStop}%
\bibitem [{\citenamefont {Strinati}(1984)}]{PhysRevB.29.5718}%
  \BibitemOpen
  \bibfield  {author} {\bibinfo {author} {\bibfnamefont {G.}~\bibnamefont
  {Strinati}},\ }\bibfield  {title} {\enquote {\bibinfo {title} {Effects of
  dynamical screening on resonances at inner-shell thresholds in
  semiconductors},}\ }\href {\doibase 10.1103/PhysRevB.29.5718} {\bibfield
  {journal} {\bibinfo  {journal} {Phys. Rev. B}\ }\textbf {\bibinfo {volume}
  {29}},\ \bibinfo {pages} {5718--5726} (\bibinfo {year} {1984})}\BibitemShut
  {NoStop}%
\bibitem [{\citenamefont {Ankudinov}\ \emph {et~al.}(2003)\citenamefont
  {Ankudinov}, \citenamefont {Nesvizhskii},\ and\ \citenamefont
  {Rehr}}]{PhysRevB.67.115120}%
  \BibitemOpen
  \bibfield  {author} {\bibinfo {author} {\bibfnamefont {A.~L.}\ \bibnamefont
  {Ankudinov}}, \bibinfo {author} {\bibfnamefont {A.~I.}\ \bibnamefont
  {Nesvizhskii}}, \ and\ \bibinfo {author} {\bibfnamefont {J.~J.}\ \bibnamefont
  {Rehr}},\ }\bibfield  {title} {\enquote {\bibinfo {title} {Dynamic screening
  effects in x-ray absorption spectra},}\ }\href {\doibase
  10.1103/PhysRevB.67.115120} {\bibfield  {journal} {\bibinfo  {journal} {Phys.
  Rev. B}\ }\textbf {\bibinfo {volume} {67}},\ \bibinfo {pages} {115120}
  (\bibinfo {year} {2003})}\BibitemShut {NoStop}%
\bibitem [{\citenamefont {Hybertsen}\ and\ \citenamefont
  {Louie}(1986)}]{PhysRevB.34.5390}%
  \BibitemOpen
  \bibfield  {author} {\bibinfo {author} {\bibfnamefont {Mark~S.}\ \bibnamefont
  {Hybertsen}}\ and\ \bibinfo {author} {\bibfnamefont {Steven~G.}\ \bibnamefont
  {Louie}},\ }\bibfield  {title} {\enquote {\bibinfo {title} {Electron
  correlation in semiconductors and insulators: Band gaps and quasiparticle
  energies},}\ }\href {\doibase 10.1103/PhysRevB.34.5390} {\bibfield  {journal}
  {\bibinfo  {journal} {Phys. Rev. B}\ }\textbf {\bibinfo {volume} {34}},\
  \bibinfo {pages} {5390--5413} (\bibinfo {year} {1986})}\BibitemShut {NoStop}%
\bibitem [{\citenamefont {Fleszar}\ and\ \citenamefont
  {Resta}(1985)}]{PhysRevB.31.5305}%
  \BibitemOpen
  \bibfield  {author} {\bibinfo {author} {\bibfnamefont {Andrzej}\ \bibnamefont
  {Fleszar}}\ and\ \bibinfo {author} {\bibfnamefont {Raffaele}\ \bibnamefont
  {Resta}},\ }\bibfield  {title} {\enquote {\bibinfo {title} {Dielectric
  matrices in semiconductors: A direct approach},}\ }\href {\doibase
  10.1103/PhysRevB.31.5305} {\bibfield  {journal} {\bibinfo  {journal} {Phys.
  Rev. B}\ }\textbf {\bibinfo {volume} {31}},\ \bibinfo {pages} {5305--5310}
  (\bibinfo {year} {1985})}\BibitemShut {NoStop}%
\bibitem [{\citenamefont {Kunc}\ and\ \citenamefont
  {Tosatti}(1984)}]{PhysRevB.29.7045}%
  \BibitemOpen
  \bibfield  {author} {\bibinfo {author} {\bibfnamefont {K.}~\bibnamefont
  {Kunc}}\ and\ \bibinfo {author} {\bibfnamefont {E.}~\bibnamefont {Tosatti}},\
  }\bibfield  {title} {\enquote {\bibinfo {title} {"direct" evaluation of the
  inverse dielectric matrix in semiconductors},}\ }\href {\doibase
  10.1103/PhysRevB.29.7045} {\bibfield  {journal} {\bibinfo  {journal} {Phys.
  Rev. B}\ }\textbf {\bibinfo {volume} {29}},\ \bibinfo {pages} {7045--7047}
  (\bibinfo {year} {1984})}\BibitemShut {NoStop}%
\bibitem [{\citenamefont {Li}\ \emph {et~al.}(2020)\citenamefont {Li},
  \citenamefont {Govind}, \citenamefont {Isborn}, \citenamefont {DePrince},\
  and\ \citenamefont {Lopata}}]{doi:10.1021/acs.chemrev.0c00223}%
  \BibitemOpen
  \bibfield  {author} {\bibinfo {author} {\bibfnamefont {Xiaosong}\
  \bibnamefont {Li}}, \bibinfo {author} {\bibfnamefont {Niranjan}\ \bibnamefont
  {Govind}}, \bibinfo {author} {\bibfnamefont {Christine}\ \bibnamefont
  {Isborn}}, \bibinfo {author} {\bibfnamefont {A.~Eugene~III}\ \bibnamefont
  {DePrince}}, \ and\ \bibinfo {author} {\bibfnamefont {Kenneth}\ \bibnamefont
  {Lopata}},\ }\bibfield  {title} {\enquote {\bibinfo {title} {Real-time
  time-dependent electronic structure theory},}\ }\href {\doibase
  10.1021/acs.chemrev.0c00223} {\bibfield  {journal} {\bibinfo  {journal}
  {Chemical Reviews}\ }\textbf {\bibinfo {volume} {120}},\ \bibinfo {pages}
  {9951--9993} (\bibinfo {year} {2020})},\ \bibinfo {note} {pMID:
  32813506}\BibitemShut {NoStop}%
\bibitem [{\citenamefont {Shirley}(2005)}]{SHIRLEY2005}%
  \BibitemOpen
  \bibfield  {author} {\bibinfo {author} {\bibfnamefont {Eric~L.}\ \bibnamefont
  {Shirley}},\ }\bibfield  {title} {\enquote {\bibinfo {title} {Bethe--salpeter
  treatment of x-ray absorption including core-hole multiplet effects},}\
  }\href {\doibase https://doi.org/10.1016/j.elspec.2005.01.191} {\bibfield
  {journal} {\bibinfo  {journal} {Journal of Electron Spectroscopy and Related
  Phenomena}\ }\textbf {\bibinfo {volume} {144-147}},\ \bibinfo {pages}
  {1187--1190} (\bibinfo {year} {2005})},\ \bibinfo {note} {{P}roceeding of the
  Fourteenth International Conference on Vacuum Ultraviolet Radiation
  Physics}\BibitemShut {NoStop}%
\bibitem [{\citenamefont {Olovsson}\ \emph {et~al.}(2009)\citenamefont
  {Olovsson}, \citenamefont {Tanaka}, \citenamefont {Mizoguchi}, \citenamefont
  {Puschnig},\ and\ \citenamefont {Ambrosch-Draxl}}]{olovsson2009all}%
  \BibitemOpen
  \bibfield  {author} {\bibinfo {author} {\bibfnamefont {W}~\bibnamefont
  {Olovsson}}, \bibinfo {author} {\bibfnamefont {I}~\bibnamefont {Tanaka}},
  \bibinfo {author} {\bibfnamefont {T}~\bibnamefont {Mizoguchi}}, \bibinfo
  {author} {\bibfnamefont {P}~\bibnamefont {Puschnig}}, \ and\ \bibinfo
  {author} {\bibfnamefont {C}~\bibnamefont {Ambrosch-Draxl}},\ }\bibfield
  {title} {\enquote {\bibinfo {title} {All-electron bethe-salpeter calculations
  for shallow-core x-ray absorption near-edge structures},}\ }\href@noop {}
  {\bibfield  {journal} {\bibinfo  {journal} {Physical Review B}\ }\textbf
  {\bibinfo {volume} {79}},\ \bibinfo {pages} {041102} (\bibinfo {year}
  {2009})}\BibitemShut {NoStop}%
\bibitem [{\citenamefont {Mauchamp}\ \emph {et~al.}(2009)\citenamefont
  {Mauchamp}, \citenamefont {Jaouen},\ and\ \citenamefont
  {Schattschneider}}]{PhysRevB.79.235106}%
  \BibitemOpen
  \bibfield  {author} {\bibinfo {author} {\bibfnamefont {Vincent}\ \bibnamefont
  {Mauchamp}}, \bibinfo {author} {\bibfnamefont {Michel}\ \bibnamefont
  {Jaouen}}, \ and\ \bibinfo {author} {\bibfnamefont {Peter}\ \bibnamefont
  {Schattschneider}},\ }\bibfield  {title} {\enquote {\bibinfo {title}
  {Core-hole effect in the one-particle approximation revisited from density
  functional theory},}\ }\href {\doibase 10.1103/PhysRevB.79.235106} {\bibfield
   {journal} {\bibinfo  {journal} {Phys. Rev. B}\ }\textbf {\bibinfo {volume}
  {79}},\ \bibinfo {pages} {235106} (\bibinfo {year} {2009})}\BibitemShut
  {NoStop}%
\bibitem [{\citenamefont {Carta}\ \emph {et~al.}(2015)\citenamefont {Carta},
  \citenamefont {Mountjoy}, \citenamefont {Regoutz}, \citenamefont {Khiat},
  \citenamefont {Serb},\ and\ \citenamefont {Prodromakis}}]{carta2015x}%
  \BibitemOpen
  \bibfield  {author} {\bibinfo {author} {\bibfnamefont {Daniela}\ \bibnamefont
  {Carta}}, \bibinfo {author} {\bibfnamefont {Gavin}\ \bibnamefont {Mountjoy}},
  \bibinfo {author} {\bibfnamefont {Anna}\ \bibnamefont {Regoutz}}, \bibinfo
  {author} {\bibfnamefont {Ali}\ \bibnamefont {Khiat}}, \bibinfo {author}
  {\bibfnamefont {Alexantrou}\ \bibnamefont {Serb}}, \ and\ \bibinfo {author}
  {\bibfnamefont {Themistoklis}\ \bibnamefont {Prodromakis}},\ }\bibfield
  {title} {\enquote {\bibinfo {title} {X-ray absorption spectroscopy study of
  tio2--x thin films for memory applications},}\ }\href@noop {} {\bibfield
  {journal} {\bibinfo  {journal} {The Journal of Physical Chemistry C}\
  }\textbf {\bibinfo {volume} {119}},\ \bibinfo {pages} {4362--4370} (\bibinfo
  {year} {2015})}\BibitemShut {NoStop}%
\bibitem [{\citenamefont {Hariki}\ \emph {et~al.}(2022)\citenamefont {Hariki},
  \citenamefont {Higashi}, \citenamefont {Yamaguchi}, \citenamefont {Li},
  \citenamefont {Kalha}, \citenamefont {Mascheck}, \citenamefont {Eriksson},
  \citenamefont {Wiell}, \citenamefont {de~Groot},\ and\ \citenamefont
  {Regoutz}}]{Kariki2022}%
  \BibitemOpen
  \bibfield  {author} {\bibinfo {author} {\bibfnamefont {Atsushi}\ \bibnamefont
  {Hariki}}, \bibinfo {author} {\bibfnamefont {Keisuke}\ \bibnamefont
  {Higashi}}, \bibinfo {author} {\bibfnamefont {Tatsuya}\ \bibnamefont
  {Yamaguchi}}, \bibinfo {author} {\bibfnamefont {Jiebin}\ \bibnamefont {Li}},
  \bibinfo {author} {\bibfnamefont {Curran}\ \bibnamefont {Kalha}}, \bibinfo
  {author} {\bibfnamefont {Manfred}\ \bibnamefont {Mascheck}}, \bibinfo
  {author} {\bibfnamefont {Susanna~K.}\ \bibnamefont {Eriksson}}, \bibinfo
  {author} {\bibfnamefont {Tomas}\ \bibnamefont {Wiell}}, \bibinfo {author}
  {\bibfnamefont {Frank M.~F.}\ \bibnamefont {de~Groot}}, \ and\ \bibinfo
  {author} {\bibfnamefont {Anna}\ \bibnamefont {Regoutz}},\ }\bibfield  {title}
  {\enquote {\bibinfo {title} {Satellites in the {T}i $1s$ core level spectra
  of {SrTiO}$_3$ and {TiO}$_2$},}\ }\href {\doibase
  10.1103/PhysRevB.106.205138} {\bibfield  {journal} {\bibinfo  {journal}
  {Phys. Rev. B}\ }\textbf {\bibinfo {volume} {106}},\ \bibinfo {pages}
  {205138} (\bibinfo {year} {2022})}\BibitemShut {NoStop}%
\bibitem [{\citenamefont {Brouder}(1990)}]{Brouder_1990}%
  \BibitemOpen
  \bibfield  {author} {\bibinfo {author} {\bibfnamefont {C}~\bibnamefont
  {Brouder}},\ }\bibfield  {title} {\enquote {\bibinfo {title} {Angular
  dependence of x-ray absorption spectra},}\ }\href {\doibase
  10.1088/0953-8984/2/3/018} {\bibfield  {journal} {\bibinfo  {journal} {J.
  Phys.: Condens. Matter}\ }\textbf {\bibinfo {volume} {2}},\ \bibinfo {pages}
  {701} (\bibinfo {year} {1990})}\BibitemShut {NoStop}%
\bibitem [{\citenamefont {Kas}\ \emph {et~al.}(2015)\citenamefont {Kas},
  \citenamefont {Vila}, \citenamefont {Rehr},\ and\ \citenamefont
  {Chambers}}]{kas2015real}%
  \BibitemOpen
  \bibfield  {author} {\bibinfo {author} {\bibfnamefont {J.~J.}\ \bibnamefont
  {Kas}}, \bibinfo {author} {\bibfnamefont {F.~D.}\ \bibnamefont {Vila}},
  \bibinfo {author} {\bibfnamefont {J.~J.}\ \bibnamefont {Rehr}}, \ and\
  \bibinfo {author} {\bibfnamefont {S.~A.}\ \bibnamefont {Chambers}},\
  }\bibfield  {title} {\enquote {\bibinfo {title} {Real-time cumulant approach
  for charge-transfer satellites in x-ray photoemission spectra},}\ }\href
  {\doibase 10.1103/PhysRevB.91.121112} {\bibfield  {journal} {\bibinfo
  {journal} {Phys. Rev. B}\ }\textbf {\bibinfo {volume} {91}},\ \bibinfo
  {pages} {121112} (\bibinfo {year} {2015})}\BibitemShut {NoStop}%
\bibitem [{\citenamefont {Uozumi}\ \emph {et~al.}(1992)\citenamefont {Uozumi},
  \citenamefont {Okada}, \citenamefont {Kotani}, \citenamefont {Durmeyer},
  \citenamefont {Kappler}, \citenamefont {Beaurepaire},\ and\ \citenamefont
  {Parlebas}}]{uozumi1992experimental}%
  \BibitemOpen
  \bibfield  {author} {\bibinfo {author} {\bibfnamefont {T}~\bibnamefont
  {Uozumi}}, \bibinfo {author} {\bibfnamefont {K}~\bibnamefont {Okada}},
  \bibinfo {author} {\bibfnamefont {A}~\bibnamefont {Kotani}}, \bibinfo
  {author} {\bibfnamefont {O}~\bibnamefont {Durmeyer}}, \bibinfo {author}
  {\bibfnamefont {JP}~\bibnamefont {Kappler}}, \bibinfo {author} {\bibfnamefont
  {E}~\bibnamefont {Beaurepaire}}, \ and\ \bibinfo {author} {\bibfnamefont
  {JC}~\bibnamefont {Parlebas}},\ }\bibfield  {title} {\enquote {\bibinfo
  {title} {Experimental and theoretical investigation of the pre-peaks at the
  ti k-edge absorption spectra in tio2},}\ }\href@noop {} {\bibfield  {journal}
  {\bibinfo  {journal} {Europhysics Letters}\ }\textbf {\bibinfo {volume}
  {18}},\ \bibinfo {pages} {85} (\bibinfo {year} {1992})}\BibitemShut {NoStop}%
\bibitem [{\citenamefont {Vedrinskii}\ \emph {et~al.}(1998)\citenamefont
  {Vedrinskii}, \citenamefont {Kraizman}, \citenamefont {Novakovich},
  \citenamefont {Demekhin},\ and\ \citenamefont
  {Urazhdin}}]{vedrinskii1998pre}%
  \BibitemOpen
  \bibfield  {author} {\bibinfo {author} {\bibfnamefont {RV}~\bibnamefont
  {Vedrinskii}}, \bibinfo {author} {\bibfnamefont {VL}~\bibnamefont
  {Kraizman}}, \bibinfo {author} {\bibfnamefont {AA}~\bibnamefont
  {Novakovich}}, \bibinfo {author} {\bibfnamefont {Ph~V}\ \bibnamefont
  {Demekhin}}, \ and\ \bibinfo {author} {\bibfnamefont {SV}~\bibnamefont
  {Urazhdin}},\ }\bibfield  {title} {\enquote {\bibinfo {title} {Pre-edge fine
  structure of the 3d atom k x-ray absorption spectra and quantitative atomic
  structure determinations for ferroelectric perovskite structure crystals},}\
  }\href@noop {} {\bibfield  {journal} {\bibinfo  {journal} {Journal of
  Physics: condensed matter}\ }\textbf {\bibinfo {volume} {10}},\ \bibinfo
  {pages} {9561} (\bibinfo {year} {1998})}\BibitemShut {NoStop}%
\bibitem [{\citenamefont {Cabaret}\ \emph {et~al.}(1999)\citenamefont
  {Cabaret}, \citenamefont {Joly}, \citenamefont {Renevier} \emph
  {et~al.}}]{cabaret1999pre}%
  \BibitemOpen
  \bibfield  {author} {\bibinfo {author} {\bibfnamefont {D}~\bibnamefont
  {Cabaret}}, \bibinfo {author} {\bibfnamefont {Y}~\bibnamefont {Joly}},
  \bibinfo {author} {\bibfnamefont {H}~\bibnamefont {Renevier}},  \emph
  {et~al.},\ }\bibfield  {title} {\enquote {\bibinfo {title} {Pre-edge
  structure analysis of ti k-edge polarized x-ray absorption spectra in tio2 by
  full-potential xanes calculations},}\ }\href@noop {} {\bibfield  {journal}
  {\bibinfo  {journal} {Journal of Synchrotron Radiation}\ }\textbf {\bibinfo
  {volume} {6}},\ \bibinfo {pages} {258--260} (\bibinfo {year}
  {1999})}\BibitemShut {NoStop}%
\bibitem [{\citenamefont {Joly}\ \emph {et~al.}(1999)\citenamefont {Joly},
  \citenamefont {Cabaret}, \citenamefont {Renevier},\ and\ \citenamefont
  {Natoli}}]{PhysRevLett.82.2398}%
  \BibitemOpen
  \bibfield  {author} {\bibinfo {author} {\bibfnamefont {Yves}\ \bibnamefont
  {Joly}}, \bibinfo {author} {\bibfnamefont {Delphine}\ \bibnamefont
  {Cabaret}}, \bibinfo {author} {\bibfnamefont {Hubert}\ \bibnamefont
  {Renevier}}, \ and\ \bibinfo {author} {\bibfnamefont {Calogero~R.}\
  \bibnamefont {Natoli}},\ }\bibfield  {title} {\enquote {\bibinfo {title}
  {Electron population analysis by full-potential x-ray absorption
  simulations},}\ }\href {\doibase 10.1103/PhysRevLett.82.2398} {\bibfield
  {journal} {\bibinfo  {journal} {Phys. Rev. Lett.}\ }\textbf {\bibinfo
  {volume} {82}},\ \bibinfo {pages} {2398--2401} (\bibinfo {year}
  {1999})}\BibitemShut {NoStop}%
\bibitem [{\citenamefont {Danger}\ \emph {et~al.}(2002)\citenamefont {Danger},
  \citenamefont {Le~F{\`e}vre}, \citenamefont {Magnan}, \citenamefont
  {Chandesris}, \citenamefont {Bourgeois}, \citenamefont {Jupille},
  \citenamefont {Eickhoff},\ and\ \citenamefont
  {Drube}}]{danger2002quadrupolar}%
  \BibitemOpen
  \bibfield  {author} {\bibinfo {author} {\bibfnamefont {J}~\bibnamefont
  {Danger}}, \bibinfo {author} {\bibfnamefont {P}~\bibnamefont {Le~F{\`e}vre}},
  \bibinfo {author} {\bibfnamefont {H}~\bibnamefont {Magnan}}, \bibinfo
  {author} {\bibfnamefont {D}~\bibnamefont {Chandesris}}, \bibinfo {author}
  {\bibfnamefont {S}~\bibnamefont {Bourgeois}}, \bibinfo {author}
  {\bibfnamefont {J}~\bibnamefont {Jupille}}, \bibinfo {author} {\bibfnamefont
  {T}~\bibnamefont {Eickhoff}}, \ and\ \bibinfo {author} {\bibfnamefont
  {W}~\bibnamefont {Drube}},\ }\bibfield  {title} {\enquote {\bibinfo {title}
  {Quadrupolar transitions evidenced by resonant auger spectroscopy},}\
  }\href@noop {} {\bibfield  {journal} {\bibinfo  {journal} {Physical review
  letters}\ }\textbf {\bibinfo {volume} {88}},\ \bibinfo {pages} {243001}
  (\bibinfo {year} {2002})}\BibitemShut {NoStop}%
\bibitem [{\citenamefont {Yamamoto}\ \emph {et~al.}(2005)\citenamefont
  {Yamamoto}, \citenamefont {Mizoguchi},\ and\ \citenamefont
  {Tanaka}}]{yamamoto2005core}%
  \BibitemOpen
  \bibfield  {author} {\bibinfo {author} {\bibfnamefont {Tomoyuki}\
  \bibnamefont {Yamamoto}}, \bibinfo {author} {\bibfnamefont {Teruyasu}\
  \bibnamefont {Mizoguchi}}, \ and\ \bibinfo {author} {\bibfnamefont {Isao}\
  \bibnamefont {Tanaka}},\ }\bibfield  {title} {\enquote {\bibinfo {title}
  {Core-hole effect on dipolar and quadrupolar transitions of {SrTiO}$_{3}$ and
  {BaTiO}$_{3}$ at {Ti} {K} edge},}\ }\href@noop {} {\bibfield  {journal}
  {\bibinfo  {journal} {Physical Review B}\ }\textbf {\bibinfo {volume} {71}},\
  \bibinfo {pages} {245113} (\bibinfo {year} {2005})}\BibitemShut {NoStop}%
\bibitem [{\citenamefont {Woicik}\ \emph {et~al.}(2007)\citenamefont {Woicik},
  \citenamefont {Shirley}, \citenamefont {Hellberg}, \citenamefont {Andersen},
  \citenamefont {Sambasivan}, \citenamefont {Fischer}, \citenamefont {Chapman},
  \citenamefont {Stern}, \citenamefont {Ryan}, \citenamefont {Ederer} \emph
  {et~al.}}]{woicik2007ferroelectric}%
  \BibitemOpen
  \bibfield  {author} {\bibinfo {author} {\bibfnamefont {JC}~\bibnamefont
  {Woicik}}, \bibinfo {author} {\bibfnamefont {Eric~L}\ \bibnamefont
  {Shirley}}, \bibinfo {author} {\bibfnamefont {CS}~\bibnamefont {Hellberg}},
  \bibinfo {author} {\bibfnamefont {KE}~\bibnamefont {Andersen}}, \bibinfo
  {author} {\bibfnamefont {S}~\bibnamefont {Sambasivan}}, \bibinfo {author}
  {\bibfnamefont {DA}~\bibnamefont {Fischer}}, \bibinfo {author} {\bibfnamefont
  {BD}~\bibnamefont {Chapman}}, \bibinfo {author} {\bibfnamefont
  {EA}~\bibnamefont {Stern}}, \bibinfo {author} {\bibfnamefont {P}~\bibnamefont
  {Ryan}}, \bibinfo {author} {\bibfnamefont {DL}~\bibnamefont {Ederer}},  \emph
  {et~al.},\ }\bibfield  {title} {\enquote {\bibinfo {title} {Ferroelectric
  distortion in srtio 3 thin films on si (001) by x-ray absorption fine
  structure spectroscopy: Experiment and first-principles calculations},}\
  }\href@noop {} {\bibfield  {journal} {\bibinfo  {journal} {Physical Review
  B}\ }\textbf {\bibinfo {volume} {75}},\ \bibinfo {pages} {140103} (\bibinfo
  {year} {2007})}\BibitemShut {NoStop}%
\bibitem [{\citenamefont {Wu}\ \emph {et~al.}(2004)\citenamefont {Wu},
  \citenamefont {Xian}, \citenamefont {Hu}, \citenamefont {Xie}, \citenamefont
  {Tao}, \citenamefont {Natoli}, \citenamefont {Paris},\ and\ \citenamefont
  {Marcelli}}]{wu2004quadrupolar}%
  \BibitemOpen
  \bibfield  {author} {\bibinfo {author} {\bibfnamefont {ZY}~\bibnamefont
  {Wu}}, \bibinfo {author} {\bibfnamefont {DC}~\bibnamefont {Xian}}, \bibinfo
  {author} {\bibfnamefont {TD}~\bibnamefont {Hu}}, \bibinfo {author}
  {\bibfnamefont {YN}~\bibnamefont {Xie}}, \bibinfo {author} {\bibfnamefont
  {Y}~\bibnamefont {Tao}}, \bibinfo {author} {\bibfnamefont {CR}~\bibnamefont
  {Natoli}}, \bibinfo {author} {\bibfnamefont {E}~\bibnamefont {Paris}}, \ and\
  \bibinfo {author} {\bibfnamefont {A}~\bibnamefont {Marcelli}},\ }\bibfield
  {title} {\enquote {\bibinfo {title} {Quadrupolar transitions and
  medium-range-order effects in metal k-edge x-ray absorption spectra of 3 d
  transition-metal compounds},}\ }\href@noop {} {\bibfield  {journal} {\bibinfo
   {journal} {Physical Review B}\ }\textbf {\bibinfo {volume} {70}},\ \bibinfo
  {pages} {033104} (\bibinfo {year} {2004})}\BibitemShut {NoStop}%
\bibitem [{\citenamefont {De~Groot}\ \emph {et~al.}(1990)\citenamefont
  {De~Groot}, \citenamefont {Fuggle}, \citenamefont {Thole},\ and\
  \citenamefont {Sawatzky}}]{de19902}%
  \BibitemOpen
  \bibfield  {author} {\bibinfo {author} {\bibfnamefont {FMF}\ \bibnamefont
  {De~Groot}}, \bibinfo {author} {\bibfnamefont {JC}~\bibnamefont {Fuggle}},
  \bibinfo {author} {\bibfnamefont {BT}~\bibnamefont {Thole}}, \ and\ \bibinfo
  {author} {\bibfnamefont {GA}~\bibnamefont {Sawatzky}},\ }\bibfield  {title}
  {\enquote {\bibinfo {title} {{\emph{L}}$_{2,3}$ x-ray-absorption edges of
  $d^0$ compounds: {K}$^+$, {Ca}$^{2+}$, {Sc}$^{3+}$, and {Ti}$^{4+}$ in
  {O}$_h$ (octahedral) symmetry},}\ }\href@noop {} {\bibfield  {journal}
  {\bibinfo  {journal} {Physical Review B}\ }\textbf {\bibinfo {volume} {41}},\
  \bibinfo {pages} {928} (\bibinfo {year} {1990})}\BibitemShut {NoStop}%
\bibitem [{\citenamefont {Cabaret}\ \emph {et~al.}(2010)\citenamefont
  {Cabaret}, \citenamefont {Bordage}, \citenamefont {Juhin}, \citenamefont
  {Arfaoui},\ and\ \citenamefont {Gaudry}}]{B926499J}%
  \BibitemOpen
  \bibfield  {author} {\bibinfo {author} {\bibfnamefont {Delphine}\
  \bibnamefont {Cabaret}}, \bibinfo {author} {\bibfnamefont {Amélie}\
  \bibnamefont {Bordage}}, \bibinfo {author} {\bibfnamefont {Amélie}\
  \bibnamefont {Juhin}}, \bibinfo {author} {\bibfnamefont {Mounir}\
  \bibnamefont {Arfaoui}}, \ and\ \bibinfo {author} {\bibfnamefont {Emilie}\
  \bibnamefont {Gaudry}},\ }\bibfield  {title} {\enquote {\bibinfo {title}
  {First-principles calculations of x-ray absorption spectra at the k-edge of
  3d transition metals: an electronic structure analysis of the pre-edge},}\
  }\href {\doibase 10.1039/B926499J} {\bibfield  {journal} {\bibinfo  {journal}
  {Phys. Chem. Chem. Phys.}\ }\textbf {\bibinfo {volume} {12}},\ \bibinfo
  {pages} {5619--5633} (\bibinfo {year} {2010})}\BibitemShut {NoStop}%
\bibitem [{\citenamefont {Rossi}\ \emph {et~al.}(2019)\citenamefont {Rossi},
  \citenamefont {Grolimund}, \citenamefont {Nachtegaal}, \citenamefont
  {Cannelli}, \citenamefont {Mancini}, \citenamefont {Bacellar}, \citenamefont
  {Kinschel}, \citenamefont {Rouxel}, \citenamefont {Ohannessian},
  \citenamefont {Pergolesi}, \citenamefont {Lippert},\ and\ \citenamefont
  {Chergui}}]{PhysRevB.100.245207}%
  \BibitemOpen
  \bibfield  {author} {\bibinfo {author} {\bibfnamefont {T.~C.}\ \bibnamefont
  {Rossi}}, \bibinfo {author} {\bibfnamefont {D.}~\bibnamefont {Grolimund}},
  \bibinfo {author} {\bibfnamefont {M.}~\bibnamefont {Nachtegaal}}, \bibinfo
  {author} {\bibfnamefont {O.}~\bibnamefont {Cannelli}}, \bibinfo {author}
  {\bibfnamefont {G.~F.}\ \bibnamefont {Mancini}}, \bibinfo {author}
  {\bibfnamefont {C.}~\bibnamefont {Bacellar}}, \bibinfo {author}
  {\bibfnamefont {D.}~\bibnamefont {Kinschel}}, \bibinfo {author}
  {\bibfnamefont {J.~R.}\ \bibnamefont {Rouxel}}, \bibinfo {author}
  {\bibfnamefont {N.}~\bibnamefont {Ohannessian}}, \bibinfo {author}
  {\bibfnamefont {D.}~\bibnamefont {Pergolesi}}, \bibinfo {author}
  {\bibfnamefont {T.}~\bibnamefont {Lippert}}, \ and\ \bibinfo {author}
  {\bibfnamefont {M.}~\bibnamefont {Chergui}},\ }\bibfield  {title} {\enquote
  {\bibinfo {title} {X-ray absorption linear dichroism at the ti $k$ edge of
  anatase ${\mathrm{tio}}_{2}$ single crystals},}\ }\href {\doibase
  10.1103/PhysRevB.100.245207} {\bibfield  {journal} {\bibinfo  {journal}
  {Phys. Rev. B}\ }\textbf {\bibinfo {volume} {100}},\ \bibinfo {pages}
  {245207} (\bibinfo {year} {2019})}\BibitemShut {NoStop}%
\bibitem [{\citenamefont {Rossi}\ \emph {et~al.}(2020)\citenamefont {Rossi},
  \citenamefont {Grolimund}, \citenamefont {Cannelli}, \citenamefont {Mancini},
  \citenamefont {Bacellar}, \citenamefont {Kinschel}, \citenamefont {Rouxel},
  \citenamefont {Ohannessian}, \citenamefont {Pergolesi},\ and\ \citenamefont
  {Chergui}}]{Rossi:gb5100}%
  \BibitemOpen
  \bibfield  {author} {\bibinfo {author} {\bibfnamefont {T.~C.}\ \bibnamefont
  {Rossi}}, \bibinfo {author} {\bibfnamefont {D.}~\bibnamefont {Grolimund}},
  \bibinfo {author} {\bibfnamefont {O.}~\bibnamefont {Cannelli}}, \bibinfo
  {author} {\bibfnamefont {G.~F.}\ \bibnamefont {Mancini}}, \bibinfo {author}
  {\bibfnamefont {C.}~\bibnamefont {Bacellar}}, \bibinfo {author}
  {\bibfnamefont {D.}~\bibnamefont {Kinschel}}, \bibinfo {author}
  {\bibfnamefont {J.~R.}\ \bibnamefont {Rouxel}}, \bibinfo {author}
  {\bibfnamefont {N.}~\bibnamefont {Ohannessian}}, \bibinfo {author}
  {\bibfnamefont {D.}~\bibnamefont {Pergolesi}}, \ and\ \bibinfo {author}
  {\bibfnamefont {M.}~\bibnamefont {Chergui}},\ }\bibfield  {title} {\enquote
  {\bibinfo {title} {{X-ray absorption linear dichroism at the Ti {\it K}-edge
  of rutile (001) TiO${\sb 2}$ single crystal}},}\ }\href {\doibase
  10.1107/S160057752000051X} {\bibfield  {journal} {\bibinfo  {journal}
  {Journal of Synchrotron Radiation}\ }\textbf {\bibinfo {volume} {27}},\
  \bibinfo {pages} {425--435} (\bibinfo {year} {2020})}\BibitemShut {NoStop}%
\bibitem [{\citenamefont {Durmeyer}\ \emph {et~al.}(2010)\citenamefont
  {Durmeyer}, \citenamefont {Beaurepaire}, \citenamefont {Kappler},
  \citenamefont {Brouder},\ and\ \citenamefont {Baudelet}}]{Durmeyer_2010}%
  \BibitemOpen
  \bibfield  {author} {\bibinfo {author} {\bibfnamefont {O}~\bibnamefont
  {Durmeyer}}, \bibinfo {author} {\bibfnamefont {E}~\bibnamefont
  {Beaurepaire}}, \bibinfo {author} {\bibfnamefont {J-P}\ \bibnamefont
  {Kappler}}, \bibinfo {author} {\bibfnamefont {Ch}~\bibnamefont {Brouder}}, \
  and\ \bibinfo {author} {\bibfnamefont {F}~\bibnamefont {Baudelet}},\
  }\bibfield  {title} {\enquote {\bibinfo {title} {Temperature dependence of
  the pre-edge structure in the ti k-edge x-ray absorption spectrum of
  rutile},}\ }\href {\doibase 10.1088/0953-8984/22/12/125504} {\bibfield
  {journal} {\bibinfo  {journal} {Journal of Physics: Condensed Matter}\
  }\textbf {\bibinfo {volume} {22}},\ \bibinfo {pages} {125504} (\bibinfo
  {year} {2010})}\BibitemShut {NoStop}%
\bibitem [{\citenamefont {Tinte}\ and\ \citenamefont
  {Shirley}(2008)}]{Tinte_2008}%
  \BibitemOpen
  \bibfield  {author} {\bibinfo {author} {\bibfnamefont {Silvia}\ \bibnamefont
  {Tinte}}\ and\ \bibinfo {author} {\bibfnamefont {Eric~L}\ \bibnamefont
  {Shirley}},\ }\bibfield  {title} {\enquote {\bibinfo {title} {Vibrational
  effects on srtio3 ti 1s absorption spectra studied using first-principles
  methods},}\ }\href {\doibase 10.1088/0953-8984/20/36/365221} {\bibfield
  {journal} {\bibinfo  {journal} {Journal of Physics: Condensed Matter}\
  }\textbf {\bibinfo {volume} {20}},\ \bibinfo {pages} {365221} (\bibinfo
  {year} {2008})}\BibitemShut {NoStop}%
\bibitem [{\citenamefont {Jiang}\ \emph {et~al.}(2007)\citenamefont {Jiang},
  \citenamefont {Su},\ and\ \citenamefont {Spence}}]{PhysRevB.76.214117}%
  \BibitemOpen
  \bibfield  {author} {\bibinfo {author} {\bibfnamefont {N.}~\bibnamefont
  {Jiang}}, \bibinfo {author} {\bibfnamefont {D.}~\bibnamefont {Su}}, \ and\
  \bibinfo {author} {\bibfnamefont {J.~C.~H.}\ \bibnamefont {Spence}},\
  }\bibfield  {title} {\enquote {\bibinfo {title} {Determination of ti
  coordination from pre-edge peaks in ti $k$-edge xanes},}\ }\href {\doibase
  10.1103/PhysRevB.76.214117} {\bibfield  {journal} {\bibinfo  {journal} {Phys.
  Rev. B}\ }\textbf {\bibinfo {volume} {76}},\ \bibinfo {pages} {214117}
  (\bibinfo {year} {2007})}\BibitemShut {NoStop}%
\bibitem [{nom()}]{nomad-doi}%
  \BibitemOpen
  \href@noop {} {}\bibinfo {note} {Data can be downloaded at the following
  link: \url{https/doi.org/10.17172/NOMAD/2023.03.24-1}}\BibitemShut {NoStop}%
\bibitem [{\citenamefont {Draxl}\ and\ \citenamefont
  {Scheffler}(2019)}]{draxl2019nomad}%
  \BibitemOpen
  \bibfield  {author} {\bibinfo {author} {\bibfnamefont {Claudia}\ \bibnamefont
  {Draxl}}\ and\ \bibinfo {author} {\bibfnamefont {Matthias}\ \bibnamefont
  {Scheffler}},\ }\bibfield  {title} {\enquote {\bibinfo {title} {The nomad
  laboratory: from data sharing to artificial intelligence},}\ }\href@noop {}
  {\bibfield  {journal} {\bibinfo  {journal} {Journal of Physics: Materials}\
  }\textbf {\bibinfo {volume} {2}},\ \bibinfo {pages} {036001} (\bibinfo {year}
  {2019})}\BibitemShut {NoStop}%
\end{thebibliography}
\end{document}